\documentclass[prd, nobibnotes, nofootinbib, eqsecnum, onecolumn,
10pt, a4paper, superscriptaddress, nofootinbib, showpacs,
byrevtex]{revtex4}

\usepackage{psfig}

\newcommand{\fin}{4}
\newcommand{\www}{w}
\newcommand{\Lperp}{L^\perp}

\newcommand{\pe}{\partial_\eta}
\newcommand{\pezB}{\partial_u}
\newcommand{\ppzN}{\partial_n}

\newcommand{\HcB}{{\cal H}}
\newcommand{\IcB}{{\cal I}}
\newcommand{\UcB}{{\cal U}}

\newcommand{\HpN}{H}
\newcommand{\IpN}{I}
\newcommand{\UpN}{U}

\newcommand{\Hconf}{\tilde \HcB}
\newcommand{\Iconf}{\tilde \IcB}
\newcommand{\Uconf}{\tilde \UcB}
\newcommand{\Hperp}{\tilde \HpN}
\newcommand{\Iperp}{\tilde \IpN}
\newcommand{\Uperp}{\tilde \UpN}

\newcommand{\COURB}{{\rm k}}
\newcommand{\KzAA}{K}
\newcommand{\DELTA}{\nabla^2}
\newcommand{\DzAA}{\Delta}

\newcommand{\IzA}{\frac{1}{\IA}}
\newcommand{\IzB}{\frac{1}{\IB}}
\newcommand{\IzN}{\frac{1}{\IN}}
\newcommand{\IAA}{\IA^2}
\newcommand{\IBB}{\IB^2}
\newcommand{\INN}{\IN^2}
\newcommand{\IAB}{\IA \IB}
\newcommand{\IBN}{\IB \IN}
\newcommand{\INA}{\IN \IA}
\newcommand{\IAAA}{\IA^3}
\newcommand{\IAAAA}{\IA^4}
\newcommand{\IzAA}{\frac{1}{\IA^2}}
\newcommand{\IzBB}{\frac{1}{\IB^2}}
\newcommand{\IzNN}{\frac{1}{\IN^2}}
\newcommand{\IzAB}{\frac{1}{\IA \IB}}
\newcommand{\IzBN}{\frac{1}{\IB \IN}}
\newcommand{\IzNA}{\frac{1}{\IN \IA}}
\newcommand{\IAzB}{\frac{\IA}{\IB}}
\newcommand{\IAzN}{\frac{\IA}{\IN}}
\newcommand{\IBzA}{\frac{\IB}{\IA}}
\newcommand{\IBzN}{\frac{\IB}{\IN}}
\newcommand{\INzA}{\frac{\IN}{\IA}}
\newcommand{\INzB}{\frac{\IN}{\IB}}
\newcommand{\IABzN}{\frac{\IAB}{\IN}}

\newcommand{\IBBzN}{\frac{\IBB}{\IN}}
\newcommand{\INNzA}{\frac{\INN}{\IA}}
\newcommand{\INNzB}{\frac{\INN}{\IB}}
\newcommand{\IAAzB}{\frac{\IAA}{\IB}}
\newcommand{\IAAzN}{\frac{\IAA}{\IN}}
\newcommand{\IAAzBB}{\frac{\IAA}{\IBB}}
\newcommand{\IAAzNN}{\frac{\IAA}{\INN}}
\newcommand{\IBBzAA}{\frac{\IBB}{\IAA}}
\newcommand{\IBBzNN}{\frac{\IBB}{\INN}}
\newcommand{\INNzAA}{\frac{\INN}{\IAA}}
\newcommand{\INNzBB}{\frac{\INN}{\IBB}}
\newcommand{\IAAzBN}{\frac{\IAA}{\IB \IN}}
\newcommand{\IBNzAA}{\frac{\IB \IN}{\IAA}}
\newcommand{\IBBzNA}{\frac{\IBB}{\IN \IA}}
\newcommand{\INAzBB}{\frac{\IN \IA}{\IBB}}
\newcommand{\INNzAB}{\frac{\INN}{\IA \IB}}
\newcommand{\IABzNN}{\frac{\IA \IB}{\INN}}

\newcommand{\IAAAzB}{\frac{\IAAA}{\IB}}
\newcommand{\IAAAzN}{\frac{\IAAA}{\IN}}
\newcommand{\IAABzN}{\frac{\IAA \IB}{\IN}}
\newcommand{\IABBzN}{\frac{\IA \IBB}{\IN}}
\newcommand{\IBBNzA}{\frac{\IBB \IN}{\IA}}
\newcommand{\IBNNzA}{\frac{\IB \INN}{\IA}}
\newcommand{\INAAzB}{\frac{\IN \IAA}{\IB}}
\newcommand{\INNAzB}{\frac{\INN \IA}{\IB}}

\newcommand{\IAAAAzBB}{\frac{\IAAAA}{\IBB}}
\newcommand{\IAAAAzBN}{\frac{\IAAAA}{\IB \IN}}
\newcommand{\IAAAAzNN}{\frac{\IAAAA}{\INN}}
\newcommand{\IAAABzNN}{\frac{\IAAA \IB}{\INN}}
\newcommand{\IAABBzNN}{\frac{\IAA \IBB}{\INN}}
\newcommand{\IBBNNzAA}{\frac{\IBB \INN}{\IAA}}
\newcommand{\INNAAzBB}{\frac{\INN \IAA}{\IBB}}
\newcommand{\INAAAzBB}{\frac{\IN \IAAA}{\IBB}}

\newcommand{\BET}{\beta}
\newcommand{\GAM}{\gamma}
\newcommand{\BTT}{\BET^2}
\newcommand{\GMM}{\GAM^2}
\newcommand{\BGM}{\BET \GAM}
\newcommand{\BGG}{\BET \GAM^2}

\newcommand{\REST}[1]{{#1}_0}
\newcommand{\RVU}[1]{{#1}^0}
\newcommand{\RVD}[1]{{#1}_0}
\newcommand{\VVRVU}[1]{\VV{#1}^0}

\newcommand{\QINT}{{\cal Q}}

\newcommand{\WN}{\BRN{\cal E}}
\newcommand{\DWN}{\BRN {\delta {\cal E}}}
\newcommand{\XZ}{X^{(0)}}
\newcommand{\XP}{X^{(\perp)}}
\newcommand{\VP}{X^{(*)}}
\newcommand{\EZA}{E^{(01)}}
\newcommand{\EZB}{E^{(02)}}
\newcommand{\EPA}{E^{(\perp 1)}}
\newcommand{\EPB}{E^{(\perp 2)}}

\newcommand{\DIR}{{\rm D}}

\newcommand{\BRNXI}{{\BRN{\Xi}}}
\newcommand{\BRUPS}{{\BRN{\Upsilon}}}

\newcommand{\IMET}{\chi}
\newcommand{\INDMET}{\BRN{\IMET}{}}
\newcommand{\DINDMET}{\BRN{\delta \IMET}{}}

\newcommand{\NVEC}{\perp}
\newcommand{\NORMVEC}{\BRN{\NVEC}{}}
\newcommand{\DNORMVEC}{\BRN{\delta \NVEC}{}}

\newcommand{\FFF}{\BRN{q}{}}
\newcommand{\DFFF}{\BRN{\delta q}{}}

\newcommand{\BRNkappa}{\kappa_4}

\newcommand{\BRNPhi}{\BRN{\Phi}}
\newcommand{\BRNPsi}{\BRN{\Psi}}

\newcommand{\UDT}[1]{\overline {#1}}
\newcommand{\BRN}[1]{\underline {#1}}
\newcommand{\SBLK}{{\rm B}}
\newcommand{\SBRN}{{\rm b}}

\newcommand{\CEX}{\BRN{\cal K}{}}
\newcommand{\STT}{\BRN{{\cal S}}{}}
\newcommand{\DCEX}{\BRN{\delta {\cal K}}{}}
\newcommand{\DSTT}{\BRN{\delta {\cal S}}{}}
\newcommand{\ND}[2]{{}^{#1}{#2}}

\newcommand{\CONT}[1]{\left<{#1} \right>}
\newcommand{\DISC}[1]{\left[{#1} \right]}

\newcommand{\ABBE}{\left(\IAzB B + \IAAzBB \dot E\right)}
\newcommand{\ANEE}{\left(\IAzN E_\perp + \IAAzNN E'\right)}
\newcommand{\BABBE}{\left(\IA B + \IAAzB \dot E\right)}
\newcommand{\NANEE}{\left(\IA E_\perp + \IAAzN E'\right)}
\newcommand{\VV}[1]{\bar {#1}}
\newcommand{\TT}[1]{\bar {\bar {#1}}}

\newcommand{\TDEMI}{{\textstyle \frac{1}{2}}}
\newcommand{\DDEMI}{{\displaystyle \frac{1}{2}}}
\newcommand{\ddd}{{\rm d}}

\newcommand{\XXE}{X^{\scriptscriptstyle(\eta)}}
\newcommand{\XXP}{X^{\scriptscriptstyle(\perp)}}

\newcommand{\EQREF}[3]{{#1} \{{#2}\} {#3}}
\newcommand{\TEQREFA}[4]{
\begin{array}{p{1cm}l}
   \mbox{$#1$} &
   \mbox{$\EQREF{#2}{#3}{#4} $}
\end{array}}
\newcommand{\TEQREFB}[4]{
\begin{array}{p{2cm}l}
   \mbox{$#1$} &
   \mbox{$\EQREF{#2}{#3}{#4} $}
\end{array}}
\newcommand{\TEQREFC}[4]{
\begin{array}{p{4cm}l}
   \mbox{$#1$} &
   \mbox{$\EQREF{#2}{#3}{#4} $}
\end{array}}
\newcommand{\TEQREFD}[4]{
\begin{array}{p{6cm}l}
   \mbox{$#1$} &
   \mbox{$\EQREF{#2}{#3}{#4} $}
\end{array}}

\newcounter{APPsection}
\newcounter{subAPPsection}[APPsection]
\renewcommand{\theAPPsection}{\Alph{APPsection}}

\newcommand{\thessubAPPsection}{\arabic{subAPPsection}}

\newcommand{\newAPP}{
\renewcommand{\theequation}{\theAPPsection\arabic{equation}}}

\newcommand{\APPsection}[1]{
\refstepcounter{APPsection}
\setcounter{equation}{0}
\begin{center} 
\vskip 1cm
\bf{\small APPENDIX \theAPPsection:\hskip 0.4cm \uppercase{#1}}
\end{center}}
                            
\newcommand{\subAPPsection}[1]{
\refstepcounter{subAPPsection}
\vskip 0.3cm
\begin{center} 
\bf{\small \thessubAPPsection.\hskip 0.4cm #1} 
\end{center}}

\newcommand{\fd}{y}
\newcommand{\pp}{\partial_y}
\newcommand{\POSBRN}{{\rm y}_\SBRN{}}

\newcommand{\finN}{{\scriptscriptstyle M}}

\newcommand{\SfinN}{{\scriptstyle M}}

\newcommand{\SSfinN}{M}

\newcommand{\Bperp}{B_\perp}
\newcommand{\Eperp}{E_\perp}
\newcommand{\epi}{\Eperp{}}
\newcommand{\eip}{\Eperp}
\newcommand{\Vepi}{\VV{E}_{\scriptscriptstyle(\perp)}{}}
\newcommand{\epp}{E_{\perp \perp}}

\newcommand{\ea}{E_{(A)}}
\newcommand{\vea}{\VV{E}_{(A)}{}}
\newcommand{\eab}{E_{(AB)}}

\newcommand{\IA}{a}
\newcommand{\IB}{n}
\newcommand{\IN}{b}

\newcommand{\KAPPAN}{\kappa_{\scriptscriptstyle N + 2}}
\newcommand{\KAPPACQ}{\kappa_5}

\newcommand{\PRESY}{Y}

\newcommand{\WEYL}{{\cal Z}}

\newcommand{\hpp}{h}
\newcommand{\BRNhpp}{\BRN{h}}
\newcommand{\vp}{\Sigma}
\newcommand{\BRNvp}{\BRN{\Sigma}}
\newcommand{\hci}{\VV{h}}
\newcommand{\hic}{\VV{h}}
\newcommand{\vi}{\VV{\Sigma}}

\newcommand{\dni}{f}

\newcommand{\UPS}{\epsilon}
\newcommand{\ZETA}{\zeta}

\begin{document}

\title{Gauge invariant cosmological perturbation theory for braneworlds}

\author{Alain Riazuelo}
\email{riazuelo@spht.saclay.cea.fr}
\affiliation{Service de Physique Th\'eorique, CEA/DSM/SPhT, \\
Unit\'e de recherche associ\'ee au CNRS, CEA/Saclay, F--91191
Gif-sur-Yvette c\'edex, France}
\affiliation{D\'epartement de Physique Th\'eorique, Universit\'e de
Gen\`eve, 24, Quai Ernest Ansermet, 1211 Gen\`eve 4, Switzerland}

\author{Filippo Vernizzi}
\email{vernizzi@amorgos.unige.ch}
\affiliation{D\'epartement de Physique Th\'eorique, Universit\'e de
Gen\`eve, 24, Quai Ernest Ansermet, 1211 Gen\`eve 4, Switzerland}

\author{Dani\`ele Steer}
\email{steer@th.u-psud.fr}
\affiliation{Laboratoire de Physique Th\'eorique, B\^at. 210,
Universit\'e Paris XI, F--91405 Orsay Cedex, France}
\affiliation{D\'epartement de Physique Th\'eorique, Universit\'e de
Gen\`eve, 24, Quai Ernest Ansermet, 1211 Gen\`eve 4, Switzerland}

\author{Ruth Durrer}
\email{ruth.durrer@physics.unige.ch}
\affiliation{D\'epartement de Physique Th\'eorique, Universit\'e de
Gen\`eve, 24, Quai Ernest Ansermet, 1211 Gen\`eve 4, Switzerland}

\date{\today}

\begin{abstract}
We derive the gauge invariant perturbation equations for a
$5$-dimensional bulk spacetime in the presence of a brane.  The
equations are derived in full generality, without specifying a
particular energy content of the bulk or the brane. We do not assume
$Z_2$ symmetry, and show that the degree of freedom associated with
brane motion plays a crucial role. Our formalism may also be used in
the $Z_2$ symmetric case where it simplifies considerably.

\vspace*{0.5cm}
\noindent
{\footnotesize Preprint numbers: SPhT-Saclay T02/040; ORSAY-LPT-02-36;
{\tt hep-th/0205???} }
\end{abstract}


\maketitle

\tableofcontents

\section{Introduction}

The idea that our $4$-dimensional observed universe may be a
hypersurface or ``brane'' in a higher dimensional spacetime is
motivated by string- and $M$-theory~\cite{Polch,HW1,HW2}. In
particular, $5$-dimensional braneworld scenarios, in which our
universe represents the boundary of a $5$-dimensional spacetime, have
recently received considerable
attention~\cite{ebp1,ebp2,ebp3,ebp4,ebp5,ebp6,RS1,RS2,Bine1,Bine2,Muko,gcc1,gcc2,gcc3,Lorenzo,Deruelle,Deruelle2,Deruelle3,Tanaka,GKR,fields,bulksf,DDPV00,perkins00,Carter1,Carter2,JP1,JP2,JP3,Fterm,SMS,Roy,Carsten,bp1,bp2a,bp2b,bp2c,bp4,bp5,bp6,bp7,bp8,bp9,bp10,Bozza,seto,giovannini,BarrowRoy,Khoury,Anisotropicstress,Dick,Christos,Kraus,Ida,Mukoya}.
The case in which the bulk spacetime is Anti~de~Sitter space and
orbifold compactification is realized with the brane as fixed point
has been particularly studied. In this situation, it has been shown
that $4$-dimensional gravity is recovered on the
brane~\cite{Deruelle2,Tanaka,RS2,GKR} on energy scales much lower than
the brane tension and/or bulk curvature, and late time cosmology is
not changed if the brane tension is sufficiently
high~\cite{Bine1,Bine2,gcc1,gcc2,gcc3,Lorenzo}. In an attempt to solve
the fine-tuning problem between the bulk cosmological constant and the
brane tension, more complex models in which the bulk or the brane are
filled with several species (such as scalar fields) have recently been
proposed (see for example~\cite{fields} and references therein).

In these models $Z_2$ symmetry is often assumed, and this is
particularly convenient when considering boundary conditions on the
brane. If $Z_2$ symmetry is dropped, brane motion in the bulk must be
taken into account and involved calculations are required in order to
determine the boundary conditions on the brane. Whilst $Z_2$ symmetry
is motivated by $M$-theory and is required for a supersymmetric brane
configuration, such as a BPS state~\cite{Polch}, there exist
situations in which $Z_2$ symmetry is broken.  This occurs, for
example, when the brane is charged and couples to a $4$-form field in
the bulk~\cite{JP1}.  Cosmological asymmetric brane models have been
studied
in~\cite{DDPV00,perkins00,Carter1,Carter2,JP1,JP2,JP3,Deruelle3}.

These developments have prompted us to derive gauge invariant
perturbation theory for brane cosmology with one codimension. Our aim
is develop a formalism which may then be applied to any situation of
cosmological interest.  Previously, perturbations in braneworld
cosmology have been extensively studied in the literature mostly for
the case of $Z_2$
symmetry~\cite{Roy,Carsten,bp1,bp2a,bp2b,bp2c,bp4,bp5,bp6,bp7,bp8,bp9,bp10,Bozza,seto,giovannini,BarrowRoy}.
Here we consider the most general situation in which the spatial
background geometry on the brane has maximal symmetry and thus
represents a space of constant curvature $\COURB$.  We do not assume
$Z_2$ symmetry, and the boundary conditions on the brane are
discussed. Also, no particular gauge choice for the metric component
$g_{\fin \fin}$ is made.  The perturbation equations in the bulk and
on the brane are derived for general bulk and brane stress-energy
tensors. This makes our formalism particularly convenient when
analyzing situations in which different bulk components (such as
several scalar fields) are also considered. The formalism can be used
to study phenomena which have important observational consequences,
the most important of them being the calculation of the anisotropies
of the cosmic microwave background~\cite{bp10,BarrowRoy}.  Since one
must in general first determine the behaviour of perturbations in the
bulk before being able to determine their behaviour on the
brane~\cite{SMS}, we pay particular attention to the relation between
bulk and brane gauge invariant perturbation variables. These become
more subtle when the position of the brane is displaced.  Indeed we
define a set of gauge invariant variables in which the perturbation
equations on the brane become similar to the usual $4$-dimensional
equations. We then study the new terms arising in braneworlds.

Since we assume very general background spacetimes and no $Z_2$
symmetry, some of our equations are extremely cumbersome.  In order to
guide the reader through the rest of the paper, we now give a general
overview of the methods we use, the variables we introduce, and the
equations we derive in this paper.

The basic setup is one of a $3+1$-dimensional brane where the
$3$-space of constant time is maximally symmetric (a space of constant
curvature), embedded in a $4+1$-dimensional bulk.  As $Z_2$ symmetry
is not assumed, the bulk spacetimes on each side of the brane will
generally differ. Both the brane and the bulk may contain arbitrary
matter. Our notation is as follows:
\begin{itemize}

\item $x^\alpha, \; \alpha = 0$, $1$, $2$, $3$, $\fin$~: spacetime
coordinates (Greek indices), with metric $g_{\alpha \beta}$ and
covariant derivative $D_\alpha$,

\item $x^i, \; i = 1$, $2$, $3$~: coordinates on the maximally
symmetric $3$-space (second part of Latin alphabet) with metric
$\gamma_{i j}$ and covariant derivative $\nabla_i$,

\item $\sigma^a, \; a = 0$, $1$, $2$, $3$~: brane-worldsheet
coordinates (first part of Latin alphabet),

\item $X^\alpha (\sigma^a)$~: brane position in target-space.

\item A Roman subscript ${\rm b}$ indicates ``brane'' whilst ${\rm
B}$ denotes ``bulk''.

\item Certain variables such as the brane matter content ($\BRN{P}
(\sigma^a)$, $\BRN{\rho} (\sigma^a)$, etc) are only defined {\em on}
the brane. Other variables such as the normal vector to the brane
$\NORMVEC^\alpha$ or the extrinsic curvature $\CEX_{\alpha \beta}$ are
also defined at the brane position, but since they describe the
embedding of the brane in the bulk, they may take different values
{\em on either side} of the brane. All these brane-related variables
are underlined.

\end{itemize}

The action for the system is
\begin{eqnarray}
\label{action}
S & = & S_{\rm EH} + S^m_\SBLK + \BRN{S}^m_\SBRN + S_{\rm GH} 
\nonumber \\
  & = &   \int \ddd^5 x \sqrt{|g|} \left(  \frac{1}{2 \KAPPACQ} R
                                         + {\cal L}_\SBLK^m \right)
        + \int \ddd^4 \sigma \sqrt{|\INDMET|} \BRN{\cal L}_\SBRN^m
        + S_{\rm GH} .
\end{eqnarray}
Here $S_{\rm GH}$ is the Gibbons-Hawking boundary term required to
consistently derive the Israel junction conditions~\cite{Dick}, and
$\KAPPACQ$ is the fundamental $5$-dimensional Newton constant (related
to the $5$-dimensional Planck mass $M_5$ by $\KAPPACQ = 6 \pi^2
M_5^3$).  Furthermore, $R$ is the bulk scalar curvature, $g_{\alpha
\beta}$ and $\INDMET_{a b}$ are the bulk metric and the induced metric
on the brane respectively, and ${\cal L}_\SBLK^m$ and $\BRN{\cal
L}_\SBRN^m$ are respectively the Lagrangians for arbitrary matter in
the bulk and matter confined on the brane.  They may also contain a
cosmological constant or brane tension.  The induced metric on the
brane is (see for example~\cite{induced})
\begin{equation}
\INDMET_{a b} (\sigma)
 = g_{\mu \nu}(X) \frac{\partial X^\mu}{\partial \sigma^a}
                  \frac{\partial X^\nu}{\partial \sigma^b} ,
\end{equation}
and the Einstein equations resulting from action~(\ref{action}) are
\begin{equation}
\label{Einstein}
G_{\alpha \beta}
 = \KAPPACQ \left(  T_{\alpha \beta}
                  + \DIR \BRN{T}_{\alpha \beta} \right) ,
\end{equation}
where $\DIR$ is a covariant Dirac $\delta$-function specifying the
position of the brane (see Section~\ref{ssec_ee}), and
\begin{eqnarray}
\label{T1}
T_{\alpha \beta}
 & = & \frac{2}{\sqrt{|g|}}
       \left( \frac{\delta S_\SBLK^m}{\delta g^{\alpha \beta}} \right) , \\
\label{T2}
\BRN{T}^{\alpha \beta}
 & = &  \frac{2}{\sqrt{|\INDMET|}}
    \frac{\partial X^\alpha}{\partial \sigma^a}
         \frac{\partial X^\beta}{\partial \sigma^b}
         \left( \frac{\delta \BRN{S}_\SBRN^m}{\delta \INDMET_{a b}} \right) .
\end{eqnarray}
As noted above, we underline $\BRN{T}^{\alpha \beta}$ and $\INDMET_{a
b}$ to emphasize that they are only defined on the brane (see
Section~\ref{secIII}).

\begin{figure}
\centerline{\bf ~ \hskip 0.4cm EINSTEIN EQUATIONS \hskip 4cm
CONSERVATION EQUATIONS}
\centerline{\psfig{file=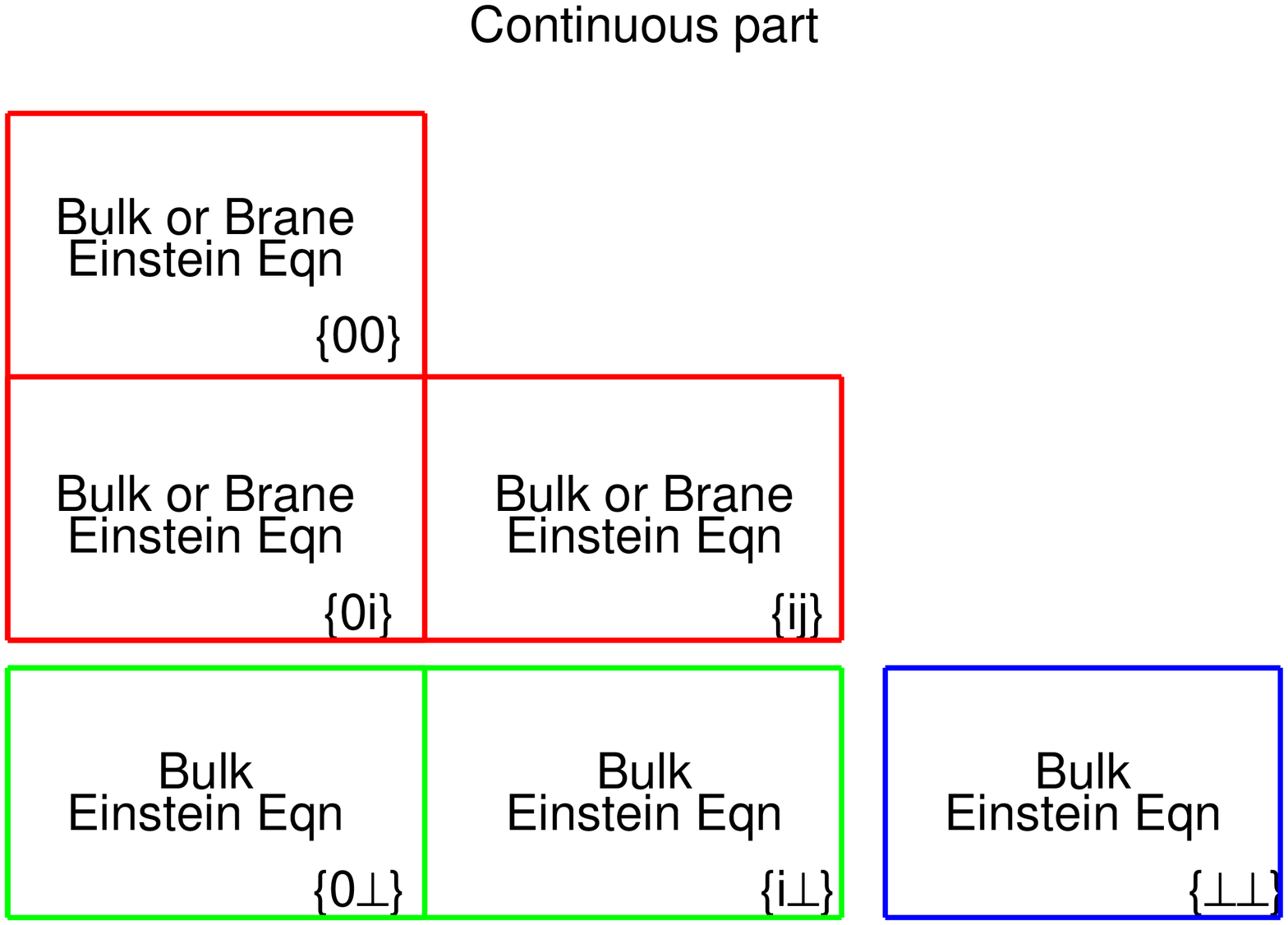,angle=270,width=3.5in}
            \psfig{file=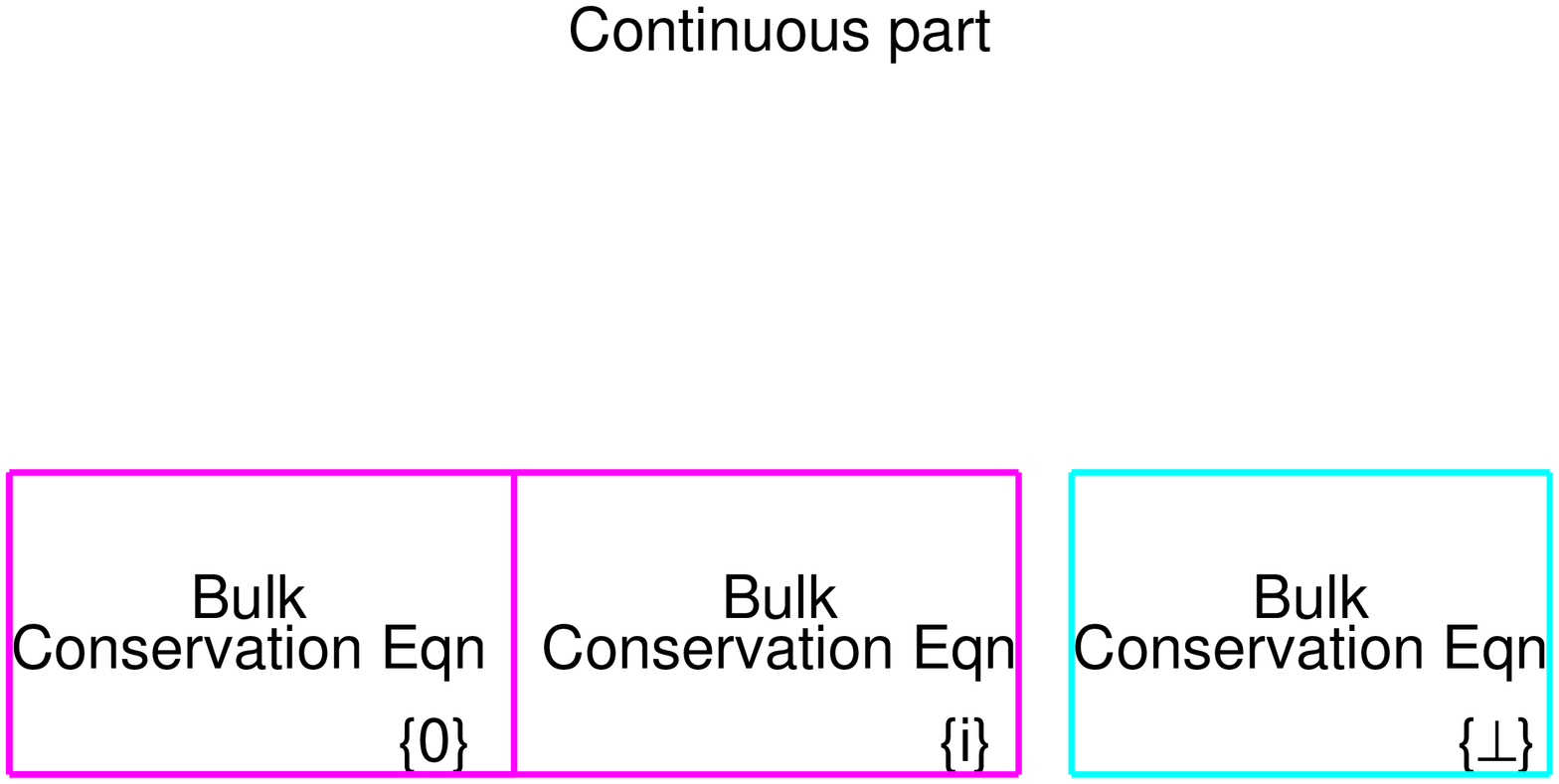,angle=270,width=3.5in}}
\centerline{\psfig{file=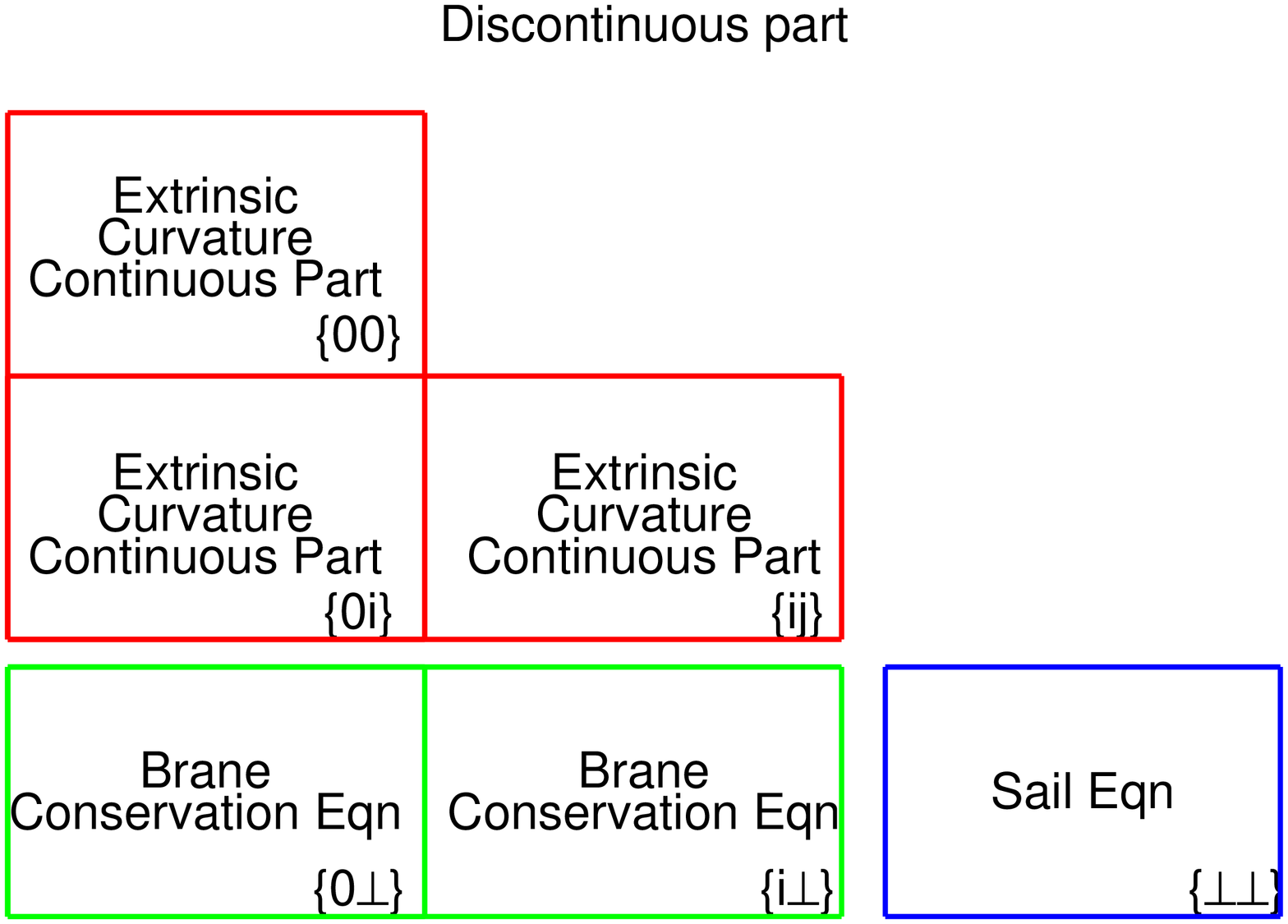,angle=270,width=3.5in}
            \psfig{file=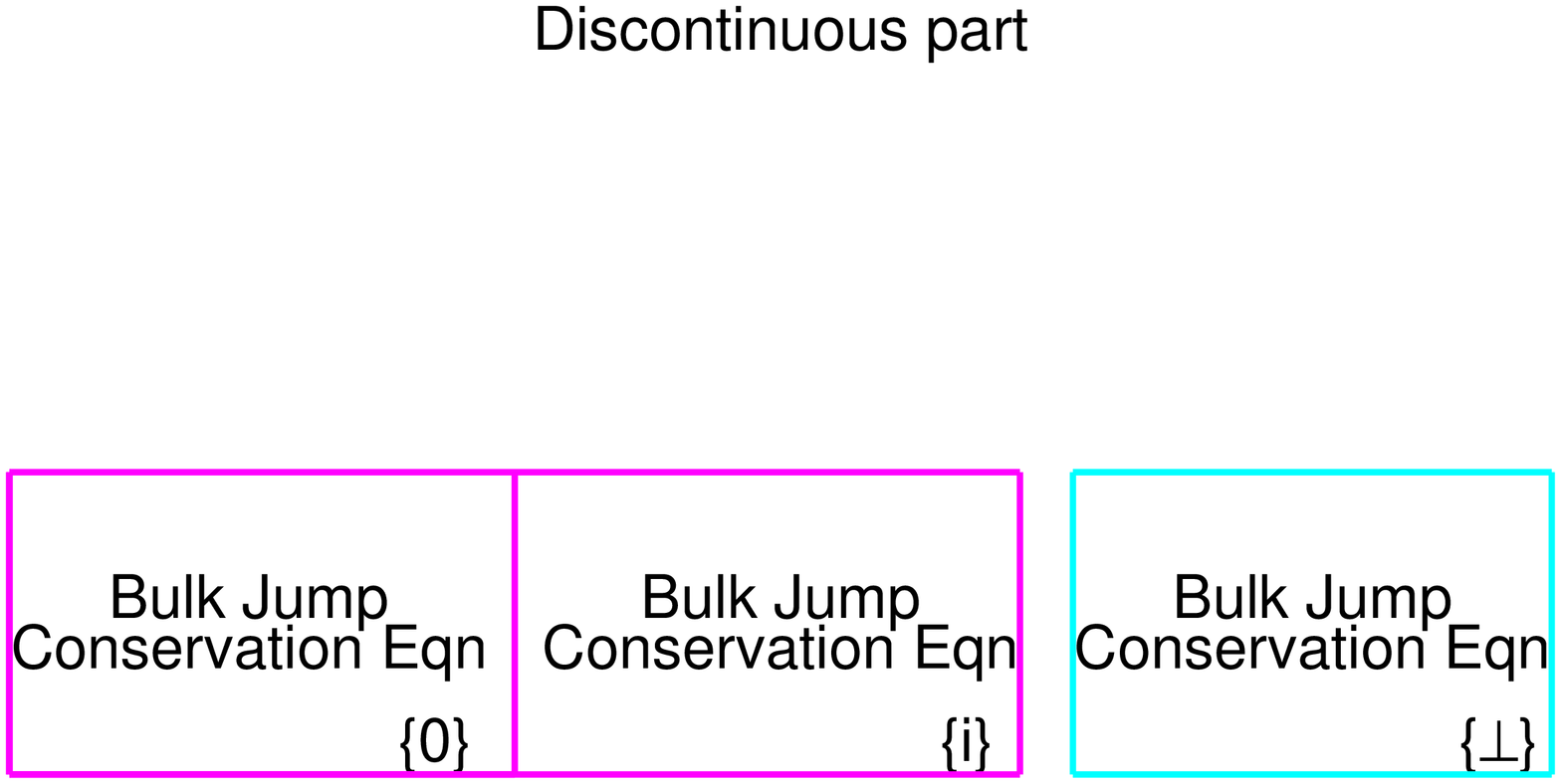,angle=270,width=3.5in}}
\centerline{\psfig{file=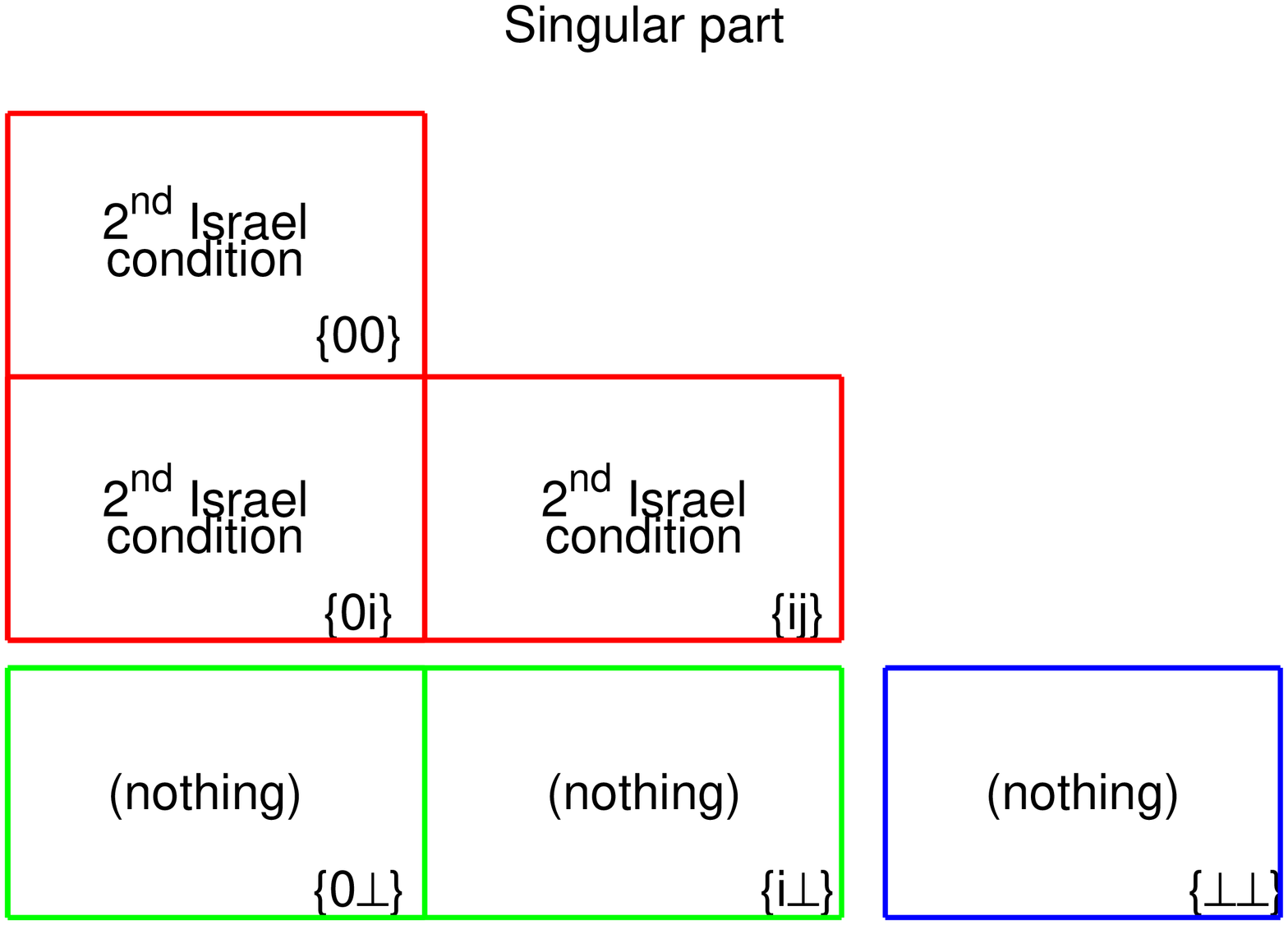,angle=270,width=3.5in}
            \psfig{file=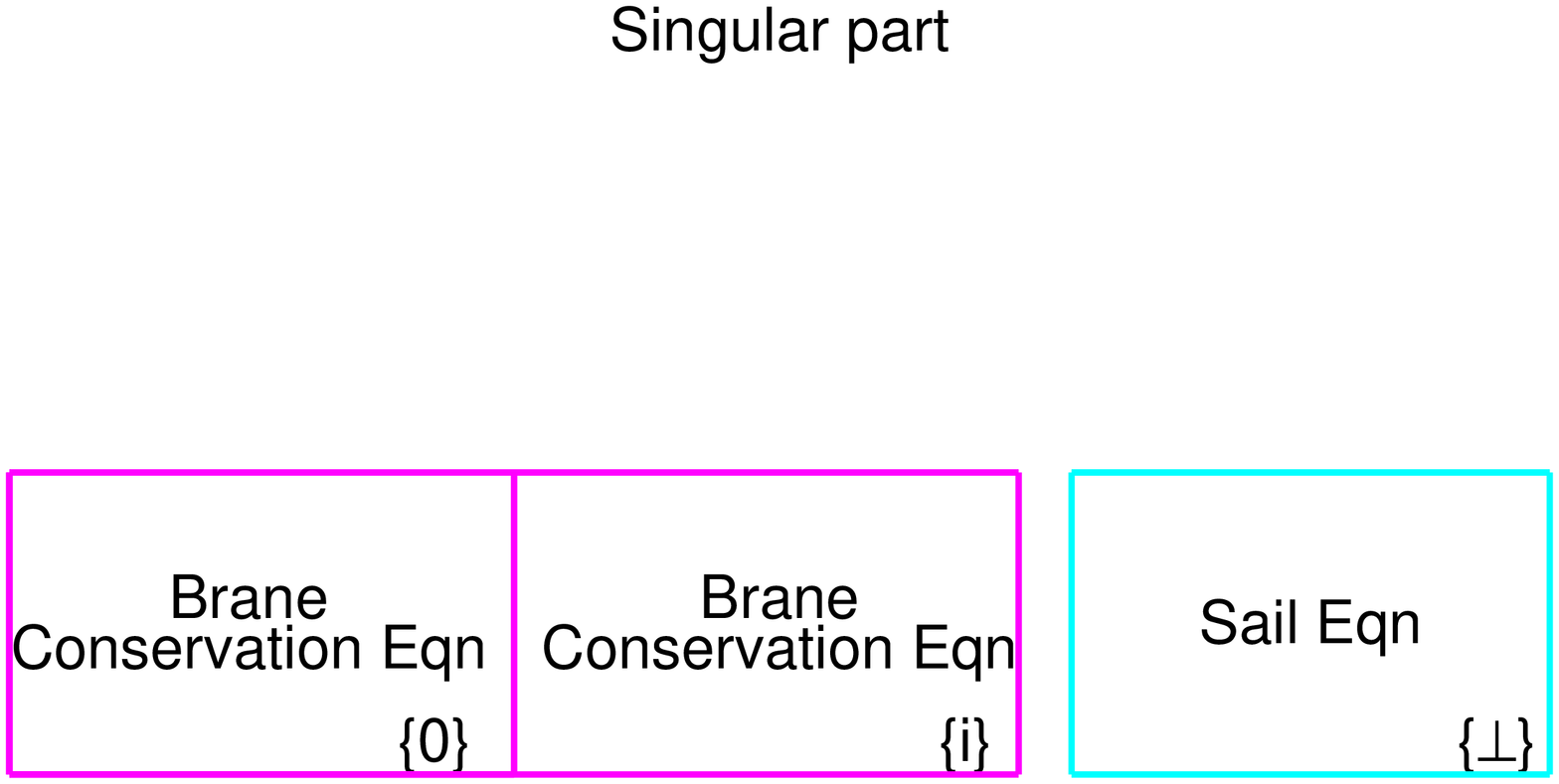,angle=270,width=3.5in}}
\caption{Structure of the Einstein equations and of the energy momentum
conservation equations in the coordinate system~(\ref{metricback}),
where coordinates $0$, $i$ are also brane coordinates and $\fin$
represents the direction orthogonal to the brane. In components, the
Einstein equations can be split into three parts: $\{\mu \nu\}$,
$\{\mu \fin\}$, and $\{\fin \fin\}$, where $\mu$, $\nu$ run on indices
$0$, $i$.  These three parts possess a continuous part (defined
everywhere in the bulk) and a jump at the brane position. Part $\{\mu
\nu\}$ also exhibits a singular term at the brane position. The role
played by all these terms is shown in the above diagrams.}
\label{fig1}
\end{figure}

We consider these Einstein equations~(\ref{Einstein}) for a
homogeneous and isotropic brane and bulk background with first
order perturbations. As is summarised schematically in the left
hand panels of Fig.~\ref{fig1}, these equations contain three
parts: one is continuous; the second is discontinuous across the
brane; and the third part is singular at the brane position
(proportional to $\DIR$).  The coefficients of each of the
individual parts must be equated, leading to a number of different
equations. The continuous part gives the Einstein equations in the
bulk and, via the Gauss-Codacci equation, they also determine the
$4$-dimensional Einstein tensor on the brane (see
Section~\ref{secIV}). The discontinuous (but non singular) part is
only non-trivial when $Z_2$ symmetry is not assumed. It then gives
equations for the continuous part of the extrinsic curvature, and
it describes the energy and momentum exchange between the brane
and bulk, leading to the equation of motion for the brane --- the
so-called ``sail equation''~\cite{JP1,JP2,JP3}. Finally, the
singular part represents the second junction condition which
relates the bulk geometries on each side of the brane through the
brane geometry and matter content.

We also discuss the so-called conservation equations for the
stress-energy tensors given in Eqns~(\ref{T1},\ref{T2}). Again,
these contain a discontinuous, continuous and singular part.  As
is summarized schematically in the right hand panels of
Fig.~\ref{fig1}, the continuous part gives the bulk energy
momentum conservation, the discontinuity simply describes the
conservation of the jump of the bulk stress-energy on the brane,
and the singular part leads to energy momentum conservation of the
brane with a possible contribution from the bulk, and to the sail
equation.

When discussing the perturbations of these equations, we will make use
of the maximal symmetry of the $3$-dimensional subspaces parallel to
the brane.  Our geometrical quantities will be decomposed into scalar,
vector and tensor degrees of freedom (with respect to these
$3$-spaces).  This decomposition is not identical to the more physical
one containing density modes, vorticity modes, and $5$-dimensional
gravitational waves. The relationship between these two approaches is
given in Section~\ref{ssecSVTdvg}. Finally, in order to set up a
consistent gauge-invariant formalism for the evolution of these
perturbations, we will see that it is crucial to take fully into
account the perturbed brane motion (which can be written in a gauge
invariant manner). This degree of freedom will be central to our
analysis.

The outline of the paper is the following.  In the next section
(Section~\ref{secII}) we discuss the unperturbed (or background)
$5$-dimensional bulk: we allow a foliation (with two codimensions)
into maximally symmetric $3$-spaces, and do not specify the presence
of the brane. The Einstein and conservation equations for the bulk
background are derived.  In Section~\ref{secIII} we introduce the
brane and we discuss the boundary conditions at the brane position for
the unperturbed spacetime without imposing $Z_2$ symmetry.  In
Section~\ref{secIV} we derive the background equations for an observer
on the brane.  In Section~\ref{secVI}, we perturb the background. We
introduce gauge invariant variables and derive the perturbed Einstein
and conservation equations in terms of these variables. The perturbed
brane including the perturbation of the brane position is discussed in
Section~\ref{secVII}.  In
Section~\ref{secIX} we reformulate the perturbation theory from the
point of view of an observer confined to the brane, and in the last
section we draw some conclusions.

Finally, we also provide an extensive and highly technical Appendix
where we present all the relevant intermediate steps required to
obtain the results presented in the text.  (Examples are, for
instance, the perturbed Christoffel symbols and the components of the
perturbed Riemann and Weyl tensors.)  The Appendix is, in fact, more
general than the main text since there we consider an
$N+1$-dimensional brane (with an $N$-dimensional maximally symmetric
subspace) embedded in a $N+2$-dimensional bulk: in the text we have
set $N = 3$.  Furthermore, whilst the text presents the perturbation
equations in full generality, some specific examples such as a bulk
scalar field are discussed briefly in the Appendix.

\section{Bulk background}
\label{secII}

In this section we describe the bulk background geometry and energy
content without introducing a brane.  We assume that the space
orthogonal to the fifth dimension is maximally symmetric so that a
homogeneous and isotropic brane can be accommodated, as discussed in
the next section.  We consider the most general stress-energy tensor
which satisfies these symmetry conditions, and then derive the
Einstein equations and the conservation equations.

\subsection{Metric and notation}

We consider a $5$-dimensional spacetime with one timelike coordinate
$x^0 \equiv \eta$, and four spacelike coordinates $\{x^1, x^2, x^3,
x^\fin\}$, where $x^\fin \equiv y$.  We assume that the constant time
hypersurfaces are locally of the form ${\cal M} \times {\bf R}$, where
${\cal M}$ is a $3$-dimensional maximally symmetric space, i.e., a
$3$-space of constant curvature, parameterized by the coordinates $
\{x^1, x^2, x^3\}$, with spatial metric $\IA^2 (\eta, \fd)
\gamma_{i j}$. The curvature of this space will be denoted by
$\COURB$. For example, we may choose the coordinates $\{x^1, x^2,
x^3\}$ such that
\begin{equation}
\label{cartsymmax}
\gamma_{i j}
 =   \delta_{i j}
   + \frac{\COURB x^i x^j}{1 - \COURB \delta_{p q} x^p x^q} ,
\end{equation}
where $\delta_{i j}$ is the Kronecker symbol.  The last spacelike
coordinate $\fd$ (the ``extra dimension'') is orthogonal to the
maximally symmetric space.  The metric has the signature $+----$.  The
line element of the metric can therefore be written as
\begin{equation}
\label{metricback}
\ddd s^2 =   \IBB \ddd \eta^2
           - \IAA \gamma_{i j} \ddd x^i \ddd x^j
           - \INN \ddd \fd{}^2 .
\end{equation}

An overdot will denote derivation with respect to $\eta$, and a prime
derivation with respect to $\fd$. In addition, we shall define
\begin{eqnarray}
\label{defperpu}
\pezB & \equiv & \IzB \pe , \\
\label{defperpn}
\ppzN & \equiv & \IzN \pp .
\end{eqnarray}
Covariant derivatives with respect to the full metric will be denoted
by $D_\alpha$, and those with respect to $\gamma_{i j}$ by
$\nabla_i$. For convenience we also define
\begin{eqnarray}
\HcB \equiv \IzB \frac{\dot \IA}{\IA}
 \quad , \qquad
\IcB & \equiv & \IzB \frac{\dot \IB}{\IB}
 \quad , \qquad
\UcB \equiv \IzB \frac{\dot \IN}{\IN} , \\
\HpN \equiv \IzN \frac{\IA'}{\IA}
 \quad , \qquad
\IpN & \equiv & \IzN \frac{\IB'}{\IB}
 \quad , \qquad
\UpN \equiv \IzN \frac{\IN'}{\IN} ,
\end{eqnarray}
as well as
\begin{equation}
\DELTA \equiv \nabla_i \nabla^i
 \quad , \qquad
\DzAA  \equiv  \frac{\DELTA}{\IAA}
 \quad , \qquad
\nabla_{i j} \equiv \nabla_i\nabla_j
 \quad , \qquad
\KzAA \equiv \frac{\COURB}{\IAA} .
\end{equation}
Notice that $\pezB$ and $\ppzN$ do not commute since $\pezB
\ppzN - \ppzN \pezB = \IpN \pezB - \UcB \ppzN$.

\subsection{Stress-energy tensor}

We now consider the bulk stress-energy tensor $T_{\alpha \beta}$ whose
energy flux need not be at rest with respect to our $(\eta, \fd)$
coordinates. Let $U^\alpha$ be the normalized timelike eigenvector of
$T_\alpha^\beta$ with eigenvalue $\REST{\rho}$. Correspondingly let
$N^\alpha$ be the normalized eigenvector orthogonal to both $U^\alpha$
and to the maximally symmetric $3$-spaces, with eigenvalue
$\REST{\PRESY}$. Finally, let $\REST{P}$ be the eigenvalue of the
three eigenvectors parallel to the maximally symmetric $3$-dimensional
slices. Note that any symmetric tensor can be decomposed in this way,
and that the symmetry requires that the eigenvectors parallel to the
symmetric $3$-spaces are degenerate.  In terms of these variables, the
bulk stress-energy tensor can be written as
\begin{eqnarray}
\label{Tdef}
T_{\alpha \beta}
 & = &   (\REST{P} + \REST{\rho}) U_\alpha U_\beta
       - (\REST{P} - \REST{\PRESY}) N_\alpha N_\beta
       - \REST{P} g_{\alpha \beta} ,
\end{eqnarray}
where the vectors $U^\alpha$ and $N^\alpha$ are given by
\begin{eqnarray}
\label{Udef}
U^\alpha & = & \left(\IzB \GAM, {\bf 0}, \IzN \BGM \right)
 \quad , \qquad
U_\mu U^\mu = 1 , \\
\label{Ndef}
N^\alpha & = & \left(- \IzB \BGM, {\bf 0}, - \IzN \GAM \right)
 \quad , \qquad
N_\mu N^\mu = -1 .
\end{eqnarray}
Here $\BET$ represents the Lorentz boost which must be performed along
the $\fd$ axis in order to be in the fluid's rest frame.  When one is
not in the rest frame of the fluid, both its energy density and its
pressure along the extra dimension are modified, and the fluid
exhibits a flux through an $\fd = {\rm constant}$ hypersurface. As
usual, $\GAM = 1 / \sqrt{1 - \BET^2}$.

Below it will be more convenient to use a different definition for the
stress-energy tensor components --- a definition which is less adapted
to the fluid, but better adapted to our coordinates.  To derive it,
let us denote by $u^\alpha$ the $5$-velocity of a bulk observer who
is at rest with respect to our coordinate system,
\begin{equation}
\label{uuu}
u^\alpha = \left(\IzB, {\bf 0}, 0 \right)
 \quad , \qquad
u_\mu u^\mu = 1 ,
\end{equation}
and by $n^\alpha$ the spacelike unit vector orthogonal to both
$u^\alpha$ and ${\cal M}$,
\begin{equation}
\label{NNN}
n^\alpha = \left(0, {\bf 0}, -\IzN \right)
 \quad , \qquad
n_\mu n^\mu = - 1
 \quad , \qquad
n_\mu u^\mu = 0 .
\end{equation}
Note that neither $u^\alpha$ nor $n^\alpha$ are geodesic vector
fields, but their orthogonality and normalization is conserved.  The
most general form of the bulk stress-energy tensor satisfying the
required symmetry with respect to translations and rotations in ${\cal
M}$ can be written as
\begin{equation}
\label{emt}
T_{\alpha \beta}
 =   (P + \rho) u_\alpha u_\beta
   - (P - \PRESY) n_\alpha n_\beta
   - P g_{\alpha \beta}
   - 2 F u_{(\alpha} n_{\beta)} ,
\end{equation}
where $f_{(\alpha} g_{\beta)} \equiv \TDEMI (f_\alpha g_\beta +
g_\alpha f_\beta)$ denotes symmetrization. In components this gives
\begin{eqnarray}
T_{0 0} & = & \IBB \rho , \\
T_{i j} & = & \IAA P \gamma_{ij} , \\
T_{0 \fin} & = & - \IBN F , \\
T_{\fin \fin} & = & \INN \PRESY .
\end{eqnarray}
Thus $\rho = T_{\mu \nu} u^\mu u^\nu$ is the bulk energy density as
measured by an observer with $5$-velocity $u^\mu$, $F = T_{\mu \nu}
u^\mu n^\nu$ is the energy flux transverse to ${\cal M}$, and $P$,
$\PRESY$ are the pressure along the directions $x^i$, $\fd$,
respectively.  The new variables $\rho$, $\PRESY$, $P$ and $F$ are
related to the old ones by
\begin{eqnarray}
\label{cordtointdeb}
\rho & =  & \GMM (\REST{\rho} + \BTT \REST{\PRESY}) , \\
\PRESY & = & \GMM (\REST{\PRESY} + \BTT \REST{\rho}) , \\
P & = & \REST{P} , \\
\label{cordtointfin}
F & = & \BGG (\REST{\rho} + \REST{\PRESY}) .
\end{eqnarray}
The last relation again shows that $F$ represents the energy flux in
$\fd$ direction. This flux, as well as the energy density $\rho$ and
pressure $\PRESY$ along the $\fd$ direction measured by an observer at
rest with respect to the coordinate system, are obtained from the
components of the stress-energy tensor in a frame at rest with respect
to the fluid simply by a Lorentz transformation. Somewhat more
complicated but equally straightforward expressions express the old
variables in terms of the new ones (see
Appendix~\ref{ssec_app_set}). Note that in~(\ref{Udef}) the bulk
velocity of the fluid is $\BET = n_\mu U^\mu / u_\nu U^\nu$. When
$\BET = 0$, $\GAM = 1$, we recover the case in which $U^\alpha =
u^\alpha$ and $N^\alpha = n^\alpha$, so that the rest frame of the
bulk matter and the coordinate system coincide.

\subsection{Einstein equations}

The Christoffel symbols, the Riemann, Ricci, Einstein and Weyl
tensors for the metric~(\ref{metricback}) are given in
Appendices~\ref{ssec_app_christ}, \ref{ssec_app_ric},
 \ref{ssec_app_ein}, \ref{ssec_app_riem} and \ref{ssec_app_weyl}
respectively, and the background bulk Einstein equations are
\begin{equation}
\label{eback}
G_{\alpha \beta} = \KAPPACQ T_{\alpha \beta} .
\end{equation}
With the stress-energy tensor~(\ref{emt}) and the Einstein tensor from
Appendix~\ref{ssec_app_ein}, Eq.~(\ref{eback}) becomes
\begin{eqnarray}
\label{einback00q}
   3 \KzAA
 + 3 \HcB \left( \HcB + \UcB \right)
 - 3 \left(\ppzN + 2 \HpN \right) \HpN
 & = & \TEQREFA{\KAPPACQ \rho}{}{00}{} , \\
\label{einbackijq}
 - \KzAA
 - 3 \left(\HcB^2 - \HpN^2 \right)
 - \left(\pezB + \UcB \right) \left(\UcB + 2 \HcB \right)
 + \left(\ppzN + \IpN \right) \left(\IpN + 2 \HpN \right)
 & = & \TEQREFA{\KAPPACQ P}{}{ij}{} , \\
\label{einback05q}
3 \left(\pezB \HpN + \HcB \HpN - \HcB \IpN \right)
 & = & \TEQREFA{\KAPPACQ F}{}{0\fin}{} , \\
\label{einback55q}
 - 3 \KzAA
 - 3 \left(\pezB + 2 \HcB \right) \HcB
 + 3 \left(\HpN +  \IpN \right) \HpN
 & = & \TEQREFA{\KAPPACQ  \PRESY}{}{\fin\fin}{} ,
\end{eqnarray}
where we have indicated in braces on the right hand side from which
component of the Einstein tensor these bulk Einstein equations are
derived.  Equations~(\ref{einback00q},\ref{einback55q}) were first
discussed in~\cite{Bine1}, and integrated with respect to the fifth
dimension in~\cite{Bine2}, for the case of a negative bulk
cosmological constant, $P = \PRESY = - \rho = \Lambda$.

The first and the third of these equations
(\ref{einback00q},\ref{einback05q}) are constraints (i.e., they do not
involve second derivatives with respect to time). The other two are
dynamical equations. In fact, there are only two independent dynamical
variables which can be written as a combination of the scale factors
$\IB$, $\IA$, and $\IN$.  One can choose coordinates to remove this
ambiguity: for example, in Gaussian coordinates $\IN = 1$ as
in~\cite{Bine2}, and in conformal coordinates $\IN =
\IB$~\cite{Deruelle,Carsten}.  Of course other choices of coordinates
are also allowed. We shall, however, keep $\IN$ undetermined, so that
any useful choice for $\IN$ can be made at the end.

\subsection{Conservation equations}

The Bianchi identities lead to the so-called conservation equations for
the stress-energy tensor,
\begin{equation}
D_\mu T^{\mu \alpha} = 0 .
\end{equation}
Only for $\alpha = 0$ and $\alpha = \fin$ are there non-trivial
relations,
\begin{eqnarray}
\label{backbianchi1}
   \pezB \rho + 3 \HcB (P + \rho) + \UcB (\PRESY + \rho)
 + \left(\ppzN + 3 \HpN + 2 \IpN \right) F & = & \TEQREFA{0}{}{0}{} , \\
\label{backbianchi2}
   \left(\pezB + 3 \HcB + 2 \UcB \right) F
 + \ppzN \PRESY + 3 \HpN (\PRESY - P) + \IpN (\PRESY + \rho)
 & = & \TEQREFA{0}{}{\fin}{} .
\end{eqnarray}
These are the conservation equations for the energy density and the
energy flux of the bulk components, respectively. The generalisation
to several components is straightforward (see
Appendix~\ref{ssec_app_cons}). Written in term of the intrinsic fluid
quantities, they give an equation of evolution for the energy density
$\REST{\rho}$ and for fluid bulk velocity $\BET$.

\section{Bulk background with a brane}
\label{secIII}

We now consider a homogeneous and isotropic $3$-brane orthogonal to
$\fd$ (lying in the space of maximal symmetry) as a singular source,
with intrinsic stress-energy tensor $\BRN{T}_{\alpha \beta}$.

\subsection{Brane position, induced metric and first fundamental form}

Let us choose the intrinsic brane coordinates $(\sigma^0, \sigma^i) =
(\eta, x^i)$, and embed the brane according to
\begin{eqnarray}
\label{embdeb}
X^0 & = & \eta , \\
X^i & = & x^i , \\
\label{embfin}
X^\fin & = & \POSBRN = {\rm constant} .
\end{eqnarray}
Note that it is always possible to choose the background coordinate
$\fd$ such that the unperturbed brane is at rest: this is the only
coordinate choice made in this paper.

As we shall see, the presence of the brane will introduce
discontinuities at $\fd = \POSBRN$ in several variables. For that
reason, it is useful to decompose a given function $f$ as
\begin{equation}
f =   \DISC{f} \left( \theta (\fd - \POSBRN) - \TDEMI \right)
    + \CONT{f} (\fd) ,
\end{equation}
where $\theta$ is the Heaviside function.  This equation defines the
continuous function $\CONT{f} (\fd)$, whilst the discontinuity or jump
of $f$ when going from one side to the other side of the brane is
given by
\begin{equation}
\DISC{f} =  \lim_{\varepsilon \to 0^+}
            \left(f(\POSBRN + \varepsilon) - f(\POSBRN - \varepsilon) \right)
    \equiv f^+ - f^- .
\end{equation}
Notice that we have the two product relations
\begin{eqnarray}
\label{pr1}
\CONT{f g} & = & \CONT{f} \CONT{g} + \frac{1}{4} \DISC{f} \DISC{g} , \\
\label{pr2}
\DISC{f g} & = & \CONT{f}\DISC{g} + \DISC{f} \CONT{g} .
\end{eqnarray}
For later convenience, and when considering a continuous function
$\CONT{f}$, we will also define the continuous part and the jump of
its derivative by
\begin{eqnarray}
\CONT{\ppzN} \CONT{f} & \equiv & \CONT{\ppzN \CONT{f}} , \\
\DISC{\ppzN} \CONT{f} & \equiv & \DISC{\ppzN \CONT{f}} .
\end{eqnarray}

Sometimes we shall also need $\DISC{\BRN{f}}$ for variables $\BRN{f}$
describing the embedding of the brane, and thus which may take
different values, $\BRN{f}^+$, $\BRN{f}^-$, on either side of the
brane. The quantities $\DISC{\BRN{f}}$ and $\CONT{\BRN{f}}$ are
defined by
\begin{eqnarray}
\label{discbrn}
\DISC{\BRN{f}} & \equiv & \BRN{f}^+ - \BRN{f}^- , \\
\label{contbrn} 
\CONT{\BRN{f}} & \equiv & \TDEMI \left(\BRN{f}^+ +
\BRN{f}^- \right) .
\end{eqnarray}

The normal unit vector to the brane, $\NORMVEC_\alpha$, is given by
\begin{equation}
\NORMVEC_\mu \frac{\partial X^\mu}{\partial \sigma^a} = 0
\quad , \qquad
\NORMVEC_\mu \NORMVEC^\mu = - 1 .
\end{equation}
One obtains (up to an overall sign)
\begin{equation}
\label{sign_perp}
\NORMVEC^\alpha = \left(0, {\bf 0}, - \IzN \right) .
\end{equation}
As we shall see, $\IN$ can be discontinuous on the brane and
$\NORMVEC^\alpha$ can have different values {\em on either side} of
the brane.

From the induced metric one can define the first fundamental
form~\cite{fundform} $\FFF_{\alpha \beta} = g_{\alpha \beta} +
\NORMVEC_\alpha \NORMVEC_\beta$, where $g_{\alpha \beta}$ is again
evaluated on (either side of) the brane, and we have $\FFF_{\alpha
\mu} \NORMVEC^\mu = 0$. On the brane, $\FFF_{\alpha \beta}$ is related
to $\INDMET_{a b} (\sigma)$ by
\begin{equation}
\label{def_q}
\FFF^{\alpha \beta} (X)
 = \frac{\partial X^\alpha}{\partial \sigma^a} 
   \frac{\partial X^\beta}{\partial \sigma^b} 
   \INDMET^{a b} (\sigma) .
\end{equation}

We can decompose the stress-energy tensor {\em on} the brane,
$\BRN{T}_{\alpha \beta} (X)$, as
\begin{equation}
\label{bemt}
\BRN{T}_{\alpha \beta}
 =   (\BRN{P} + \BRN{\rho}) \BRN{u}_\alpha \BRN{u}_\beta
   - \BRN{P} \FFF_{\alpha \beta} ,
\end{equation}
where $\BRN{u}_\alpha$ is the 4-vector of the energy flux on the brane
matter,
\begin{equation}
\BRN{u}^\alpha =\left(\IzB, {\bf 0}, 0 \right)
\quad , \qquad
\BRN{u}_\mu \BRN{u}^\mu = 1 .
\end{equation}
Note that $\BRN{T}_{\alpha \mu} \NORMVEC^\mu = \BRN{u}_\mu
\NORMVEC^\mu = 0$.  This is the most generic stress-energy tensor
compatible with a homogeneous and isotropic brane.

\subsection{Einstein equation}
\label{ssec_ee}

In the presence of the brane, the $5$-dimensional Einstein equations
become
\begin{equation}
\label{Einsteinbulk}
G_{\alpha \beta}= \KAPPACQ \left( T_{\alpha \beta} + \DIR
 \BRN{T}_{\alpha \beta} \right) ,
\end{equation}
where, from Eqn~(\ref{action}), the ``covariant Dirac function''
$\DIR$ is
\begin{equation}
\DIR = \frac{\sqrt{|\FFF|}}{\sqrt{|g|}} \delta(\fd - \POSBRN) .
\end{equation}
Here $g$ and $\FFF$ are the determinants of the metric $g_{\alpha
\beta}$ and first fundamental form $\FFF_{\alpha \beta}$ respectively,
evaluated at the brane position.  Written in components, the Einstein
equations with the brane, Eq.~(\ref{Einsteinbulk}), become
\begin{eqnarray}
\label{backein00}
   3 \KzAA
 + 3 \HcB \left( \HcB + \UcB \right)
 - 3 \left(\ppzN + 2 \HpN \right) \HpN
 & = & \TEQREFB{\KAPPACQ \left(\rho + \DIR \BRN{\rho} \right)}{}{00}{} , \\
\label{backeinij}
 - \KzAA
 - 3 \left(\HcB^2 - \HpN^2 \right)
 - \left(\pezB + \UcB \right) \left(\UcB + 2 \HcB \right)
 + \left(\ppzN + \IpN \right) \left(\IpN + 2 \HpN \right)
 & = & \TEQREFB{\KAPPACQ \left(P + \DIR \BRN{P} \right)}{}{ij}{} , \\
\label{backein05}
3 \left(\pezB \HpN + \HcB \HpN - \HcB \IpN \right)
 & = & \TEQREFB{\KAPPACQ F}{}{0\fin}{} , \\
\label{backein55}
 - 3 \KzAA
 - 3 \left(\pezB + 2 \HcB \right) \HcB
 + 3 \left(\HpN +  \IpN \right) \HpN
 & = & \TEQREFB{\KAPPACQ  \PRESY}{}{\fin\fin}{} .
\end{eqnarray}
A global solution to these equations has been derived
in~\cite{Bine1,Muko} with the assumption of a pure negative
cosmological constant in the bulk, and using Gaussian coordinates.
The right hand sides of Eqns~(\ref{backein00},\ref{backeinij}) contain
a singular term proportional to $\DIR$ due to the presence of the
brane.  As we will see below, although the first fundamental form is
continuous on the brane, its first derivative with respect to the
fifth dimension $\fd$ (i.e., the terms $\HpN$ and $\IpN$) may jump and
its second derivative ($\ppzN \HpN$ and $\ppzN \IpN$) can be
singular. Thus the Einstein tensor contains a singular part which must
be matched with the singular part of the stress-energy tensor. We now
turn to the problem of relating these terms to the brane matter
content.

\subsection{Israel junction conditions}
\label{ssec_Israel_and}

\begin{figure}
\centerline{\psfig{file=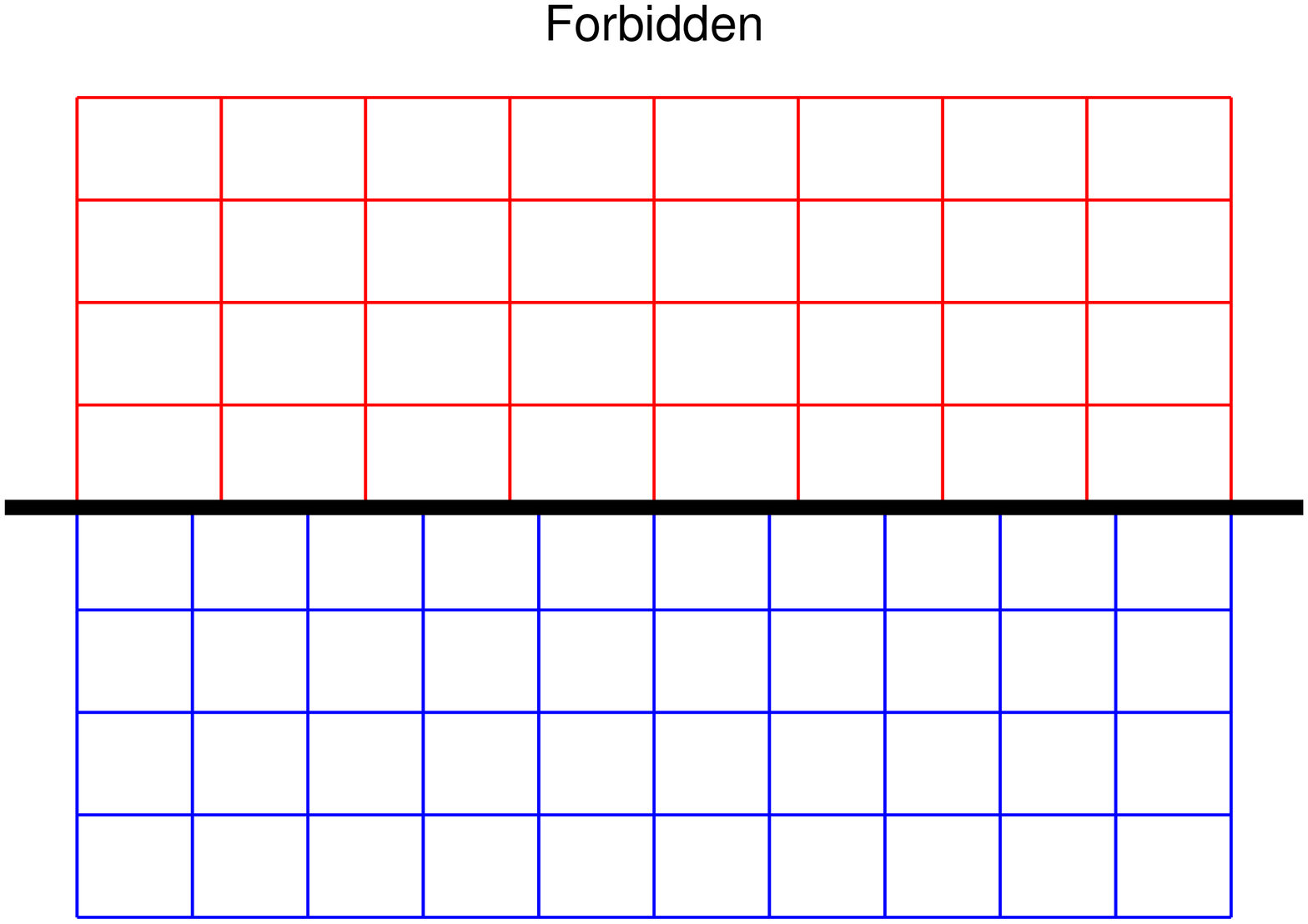,angle=270,width=3.5in}
            \psfig{file=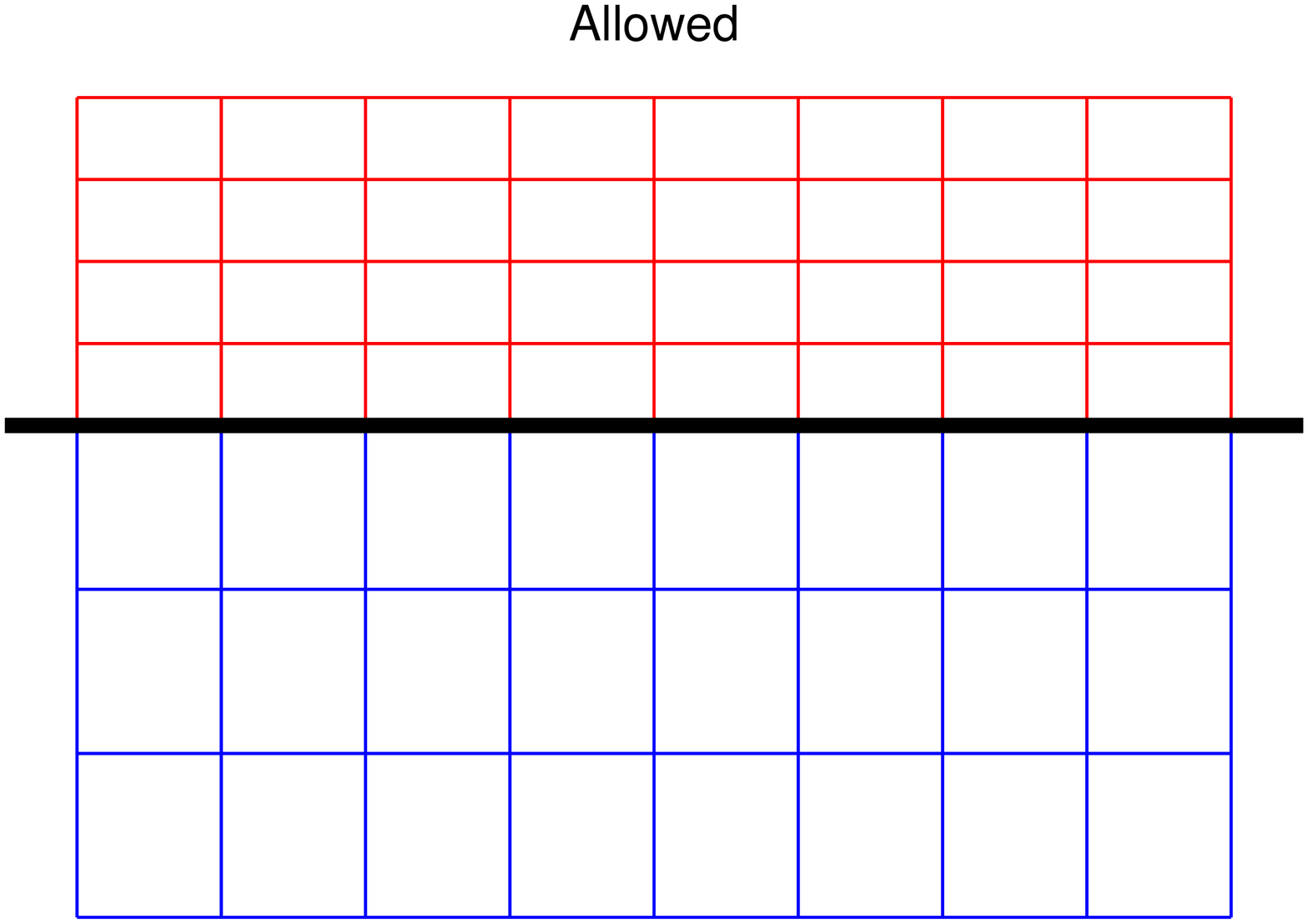,angle=270,width=3.5in}}
\caption{Schematic illustration of the first Israel condition. We
have embedded in a Minkowskian space a $2$-dimensional spacelike bulk
of metric $\ddd s^2 = \IAA \ddd x^2 + \INN \ddd \fd{}^2$.  The brane
is the thick horizontal line in the middle of both panels, and the
grids represent the metric coefficients so that the grid spacing is
proportional to $\IN$ and $\IA$ along the vertical and the horizontal
directions respectively. In the left panel, $\DISC{\IA} \neq 0$,
$\DISC{\IN} = 0$, in the right one, $\DISC{\IN} \neq 0$, $\DISC{\IA} =
0$.  The first Israel condition states that when considering a line of
constant $x$, there must not be any discontinuity when crossing the
brane: this is obviously not the case in the left panel. On the
contrary, nothing is said about how the spacing of the horizontal
lines evolves across the brane.  This translates into the fact that
$\IN$ is allowed to be discontinuous (right panel).}
\label{fig2}
\end{figure}

The extrinsic curvature formalism is a useful tool in the analysis of
junction conditions on a singular surface~\cite{MTW}.  The {\em first
Israel condition}~\cite{Ijc} imposes the continuity of the first
fundamental form,
\begin{equation}
\label{impose1}
\DISC{\FFF_{\alpha \beta}} = 0 .
\end{equation}
Hence $\FFF_{\alpha \beta}$ is well-defined {\em on} the brane. Since
we have $\FFF_{0 0} = \IBB(X)$ and $\FFF_{i j} = - \IAA(X) \gamma_{i
j}$, this condition implies the continuity of the scale factors $\IB$
and $\IA$: $\DISC{\IB} = \DISC{\IA} = 0$ (see Figure~\ref{fig2}). Note
that the continuity of the metric function $\IN$ is not required by
the junction conditions and will not be assumed in what
follows\footnote{Allowing $\IN$ to be discontinuous makes the
covariant Dirac function $\DIR$ ill-defined. This is not a serious
problem, as all the terms involving this function can be grouped
together to give the second Israel condition. Therefore, we shall
continue to use the notation $\DIR$ and suppose that when $\IN$ is not
continuous, it corresponds to a regularized and mathematically
consistent expression.} (see also
Appendix~\ref{sec_app_pert_bulk_geom}).

\begin{figure}
\centerline{\psfig{file=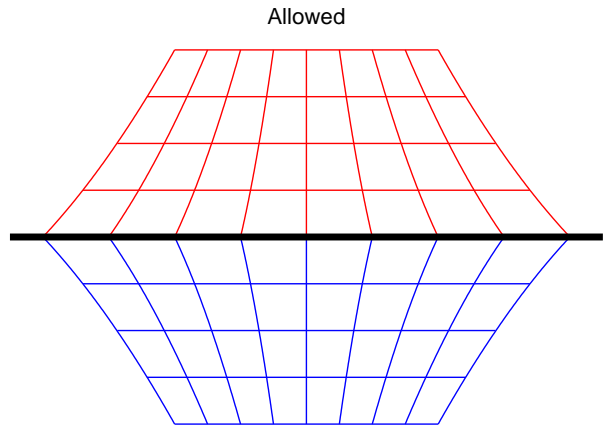,angle=270,width=3.5in}}
\caption{Illustration of the second Israel condition. With the
same conventions as in fig.~\ref{fig2}, we show an example where
$\DISC{\HpN} \neq 0$. This is possible if the energy density on the
brane is non zero (it is positive in this illustration).}
\label{fig3}
\end{figure}

Nevertheless, the first derivative with respect to $\fd$ of $\IA$ and
$\IB$ (which are proportional to $\IpN$ and $\HpN$), are allowed to
jump. In order to study the behaviour of these quantities on the brane
we consider the extrinsic curvature tensor (or second fundamental
form) with respect to the brane, namely
\begin{equation}
\label{ext}
\CEX_{\alpha \beta}
 = \FFF_{(\alpha}^\mu  D_\mu \NORMVEC_{\beta)} .
\end{equation}
For the background metric, the components of the
extrinsic curvature are
\begin{eqnarray}
\label{K00}
\CEX_{0 0} & = & - \IBB \IpN , \\
\label{Kij}
\CEX_{i j} & = &   \IAA \HpN \gamma_{i j} .
\end{eqnarray}
Let us define the surface ``stress tensor'' $\STT_{\alpha \beta}$ on
the brane by
\begin{equation}
\STT_{\alpha \beta} =   \BRN{T}_{\alpha \beta}
                      - \frac{1}{3} \BRN{T} \FFF_{\alpha \beta} .
\end{equation}
Then the {\em second Israel condition}~\cite{Ijc} relates the jump
in the extrinsic curvature with the energy content on the brane
and requires that
\begin{equation} 
\label{juncgen}
\DISC{\CEX_{\alpha \beta}} = - \KAPPACQ \STT_{\alpha \beta} .
\end{equation}
(Note that the choice of the sign here is consistent with our choice
for the sign of $\NORMVEC_\alpha$ in Eq.~(\ref{sign_perp}).)  For our
background this condition can be written as (see
Appendix~\ref{ssec_app_isr2})
\begin{eqnarray}
\label{bjcn}
 \DISC{\IpN}
 & = & \KAPPACQ \left(\frac{2}{3} \BRN{\rho} + \BRN{P} \right) , \\
\label{bjca} 
\DISC{\HpN}
 & = & - \KAPPACQ \frac{1}{3} \BRN{\rho} .
\end{eqnarray}
(See also Figure~\ref{fig3}.)  Alternatively,
Eqns~(\ref{bjcn},\ref{bjca}) can be obtained directly from the
singular part of the Einstein
equations~(\ref{backein00},\ref{backeinij}).

\subsection{Boundary conditions in the bulk}

We now comment briefly on the question of boundary conditions at the
brane. Consider first the bulk Einstein equations
(\ref{backein00}--\ref{backein55}).  They form a system of second
order partial differential equations in $\eta$ and $\fd$, and in order
to solve them we must specify initial conditions on a spacelike Cauchy
hypersurface, boundary conditions far from our braneworld (at infinity
in a one brane scenario, or on another brane), and boundary conditions
at our brane.  The Israel junction conditions impose the continuity of
$\IA$ and $\IB$, and fix the jump in their normal derivatives at the
brane. Since the Einstein equations represent a set of second order
partial differential equations, these junction conditions are
sufficient to allow us to solve the Einstein equations everywhere in
the bulk.

We now turn to the Einstein equations on the brane.

\section{The brane point of view}
\label{secIV}

An observer on the brane will not see $4$-dimensional Einstein
gravity. This may, however, be recovered in particular situations at
low energy. The $4$-dimensional Einstein tensor in general depends on
bulk quantities and is quadratic in the brane stress-energy tensor.

Here we discuss the $4$-dimensional ``Einstein equations'' which lead
to the modified Friedmann equations, and also the conservation
equation on the brane. Finally we interpret the deviation from the
$4$-dimensional theory in terms of the $5$-dimensional one.

\subsection{Einstein gravity on the brane}
\label{ssec_bc}

Through the Gauss-Codacci equations, we can write the $4$-dimensional
Einstein tensor $\ND{(4)}{\BRN{G}}_{\alpha \beta}$ in terms of the
bulk stress-energy tensor $T_{\alpha \beta}$, the extrinsic curvature
$\CEX_{\alpha \beta}$ and the projected Weyl tensor $\WN_{\alpha
\beta}$.  The details of the calculation can be found in~\cite{SMS},
and the result is
\begin{eqnarray}
\label{genein}
\ND{(4)}{\BRN{G}}_{\alpha \beta}
 & = & \frac{2}{3}
       \left(   G_{\mu \nu} \FFF_\alpha^\mu \FFF_\beta^\nu
             - \left(  G_{\mu \nu} \NORMVEC^\mu \NORMVEC^\nu
                     + \frac{1}{4} G \right) \FFF_{\alpha \beta} \right)
\nonumber \\ 
 &   & - \CEX \CEX_{\alpha \beta} + \CEX_\alpha^\mu \CEX_{\mu \beta}
       + \frac{1}{2} \FFF_{\alpha \beta}
         (\CEX^2 - \CEX^{\mu \nu} \CEX_{\mu \nu})
       + \WN_{\alpha \beta} .
\end{eqnarray}
The projected Weyl tensor $\WN_{\alpha \beta}$ on the brane is
obtained from the bulk Weyl tensor $C_{\alpha \beta \gamma \delta}$ as
follows
\begin{eqnarray}
C_{\alpha \beta \gamma \delta}
 & = &   R_{\alpha \beta \gamma \delta}
       + \frac{2}{3} (  g_{\alpha [\delta} R_{\gamma] \beta}
                      + g_{\beta [\gamma} R_{\delta] \alpha})
       + \frac{1}{6} R g_{\alpha [\gamma} g_{\delta] \beta} , \\
\WN_{\alpha \beta}
 & = & C_{\alpha \mu \beta \nu} \NORMVEC^\mu \NORMVEC^\nu .
\end{eqnarray}
Here $f_{[\alpha} g_{\beta]} \equiv \TDEMI (f_\alpha g_\beta -
g_\alpha f_\beta)$ denotes antisymmetrization.  This projection
represents the contribution of the free gravity in the bulk to the
gravity on the brane. In components, we have
\begin{eqnarray}
\WN_{0 0} & = & \frac{1}{2} \IBB \WEYL , \\
\WN_{i j} & = & \frac{1}{6} \IAA \WEYL \gamma_{i j} , \\
\label{weylback}
\WEYL
 & = &   \KzAA
       + \left(\pezB + \UcB \right) \left(\UcB - \HcB \right)
       - \left(\ppzN + \IpN \right) \left(\IpN - \HpN \right) .
\end{eqnarray}
There is only one independent component in the Weyl tensor (as well as
in its projection on the brane). This is related to the fact that this
spacetime is the $5$-dimensional analog of a $4$-dimensional type-$D$
spacetime in Petrov's classification~\cite{petrov}.

\subsection{Friedmann equations on the brane}

Since the tensor $\ND{(4)}{\BRN{G}}_{\alpha \beta}$ contains only
derivatives of the continuous first fundamental form with respect to
$\eta$ and $x^i$, it is continuous. Hence, on taking the continuous
part of the right hand side of Eq.~(\ref{genein}) (and applying the
product relation~(\ref{pr1})), we find the projected $4$-dimensional
Einstein equation on the brane,
\begin{eqnarray}
\label{generale}
\ND{(4)}{\BRN{G}}_{\alpha \beta}
 & = & \frac{2}{3} \KAPPACQ
       \left(  \CONT{T_{\mu \nu}} \FFF_\alpha^\mu \FFF_\beta^\nu
             - \left(  \CONT{T_{\mu \nu} \NORMVEC^\mu \NORMVEC^\nu}
                  + \frac{1}{4} \CONT{T} \right) \FFF_{\alpha \beta} \right)
\nonumber \\
 &   & - \CONT{\CEX} \CONT{\CEX_{\alpha \beta}}
       + \CONT{\CEX_\alpha^\mu} \CONT{\CEX_{\mu \beta}}
       + \frac{1}{2} \FFF_{\alpha \beta}
         (\CONT{\CEX}^2 - \CONT{\CEX^{\mu \nu}} \CONT{\CEX_{\mu \nu}})
\nonumber \\ 
 &   & + \frac{1}{4}
         \left(- \DISC{\CEX} \DISC{\CEX_{\alpha \beta}}
               + \DISC{\CEX_\alpha^\mu} \DISC{\CEX_{\mu \beta}}
               + \frac{1}{2} \FFF_{\alpha \beta}
                 (\DISC{\CEX}^2 - \DISC{\CEX^{\mu \nu}} \DISC{\CEX_{\mu \nu}} )
         \right)
       + \CONT{\WN_{\alpha \beta}} .
\end{eqnarray}
The right hand side of equation~(\ref{generale}) can be split into
four parts. The first depends on the average of the bulk stress-energy
tensor, $\CONT{T_{\alpha \beta}}$.  The second is given by four terms
quadratic in the average of the extrinsic curvature
$\CONT{\CEX_{\alpha \beta}}$. These terms are known once the bulk
Einstein equations have been solved. They vanish when $Z_2$ symmetry
is assumed --- we return to this point below. Then there is a third
part which contains four terms quadratic in the jump of the extrinsic
curvature $\DISC{\CEX_{\alpha \beta}}$.  We have already determined
these through the second junction condition~(\ref{bjcn},\ref{bjca}),
and they are related to the brane stress-energy tensor: these terms
will be responsible for the non-standard $\BRN{\rho}^2$ contribution
in the brane Friedmann equations. Finally there is a fourth part, the
average of the projected bulk Weyl tensor on the brane, describing the
effect from the free gravity in the bulk.

In components, we obtain the modified Friedmann equations which
contain a dynamical equation and a constraint:
\begin{eqnarray}
\label{bf1}
3 \left(\HcB^2 + \KzAA \right)
 & = &   \DDEMI \KAPPACQ \CONT{\rho + P - \PRESY}
       + \frac{\KAPPACQ^2}{12} \BRN{\rho}^2
       + 3 \CONT{\HpN}^2 + \frac{1}{2} \CONT{\WEYL} , \\
\label{bf2}
 - 2 \pezB \HcB - 3 \HcB^2 - \KzAA 
 & = &   \frac{1}{6} \KAPPACQ \CONT{\rho + P + 3 \PRESY}
       + \frac{\KAPPACQ^2}{12} (\BRN{\rho}+ 2 \BRN{P}) \BRN{\rho}
       - \CONT{\HpN} \CONT{2 \IpN + \HpN}
       + \frac{1}{6} \CONT{\WEYL} ,
\end{eqnarray}
where we have isolated the continuous part of the projection of the
bulk Weyl tensor,
\begin{equation}
\label{contweyl}
\CONT{\WEYL} =
   \KzAA - \pezB \HcB
 + \CONT{(\pezB + \UcB - \HcB) \UcB}
 - \CONT{\ppzN + \IpN} \CONT{\IpN - \HpN}
 - \frac{\KAPPACQ^2}{4} \left(\frac{2}{3} \BRN{\rho} + \BRN{P} \right)
   (\BRN{\rho} + \BRN{P}) .
\end{equation}
(These equations could alternatively have been obtained from the
continuous part of the Einstein equations,
Eqns~(\ref{backein00}--\ref{backein55}).) 

The cosmological consequences of these equations have been studied
in~\cite{Bine2,gcc1,gcc2,gcc3,Lorenzo} with assumption of $Z_2$
symmetry, in which case $\CONT{\CEX_{\alpha \beta}} = \CONT{\HpN} =
\CONT{\IpN} = 0$.  These authors considered a negative
cosmological constant in the bulk and assumed that the brane
stress-energy tensor consists of a rigid part --- the brane
tension --- and a fluid,
\begin{equation}
\label{brnlambda}
\BRN{T}_{\alpha \beta}
 = \lambda \FFF_{\alpha \beta} + \BRN{T}_{\alpha \beta}^f .
\end{equation}
In this case the bulk stress-energy tensor can be tuned to the
brane tension in such a way that deviations from standard Friedmann equations
are effective only at energies of order  $\lambda$ and higher.

If we assume $F = \CONT{H} = \dot \PRESY = 0$, integrate the sum of
Eq.~(\ref{backein05}) and (\ref{backein55}) once with respect to time,
and compare the result with~(\ref{bf1}), one obtains an expression for
the sum of the continuous part of the projected bulk Weyl tensor and
the continuous part of the bulk energy density and pressure: this
behaves as a radiation term,
\begin{equation}
\CONT{\WEYL} + \KAPPACQ \CONT{\rho + P} = {\cal C} / \IA^4 ,
\end{equation}
where ${\cal C}$ is an integration constant (see
Appendix~\ref{ssec_app_newfried}).
In~\cite{Bine2,gcc1,gcc2,gcc3,Lorenzo} this is discussed for the case
$\rho + P = 0$ so that $\CONT{\WEYL} \propto \IA^{- 4}$.

Furthermore, notice that Eq.~(\ref{brnlambda}) is
not a necessary requirement in order for the $5$-dimensional Friedmann
equations on the brane to reduce to the standard $4$-dimensional
Friedmann equations at low energy. It suffices that the brane
stress-energy tensor is dominated by a term which is almost a
cosmological constant today (i.e., it can be a slow-rolling scalar
field, see for example~\cite{bulksf}). In this case, one has
\begin{eqnarray}
\BRN{\rho} & = & \BRN{\rho}_\phi + \BRN{\rho}^f , \\
\BRN{P} & = & \BRN{P}_\phi + \BRN{P}^f , \\
\BRN{\rho}_\phi \simeq - \BRN{P}_\phi & \simeq & \lambda ,
\end{eqnarray}
where $\BRN{\rho}^f$, $\BRN{P}^f$ are the energy density and pressure
of the  ordinary matter content on the brane. Now, if in addition
\begin{eqnarray}
|\rho + P| & \ll & |\PRESY| , \\
\PRESY & \simeq & \Lambda , \\
\Lambda & \simeq & \frac{1}{6} \KAPPACQ \lambda^2 ,
\end{eqnarray}
the first two terms of the right hand side of the above Friedmann
equations on the brane are proportional to $\BRNkappa \BRN{\rho}^f$,
$\BRNkappa \BRN{P}^f$, with
\begin{equation}
\BRNkappa = \frac{1}{6} \KAPPACQ^2 \lambda
          = \KAPPACQ \frac{\Lambda}{\lambda} ,
\end{equation}
and one re-obtains a term linear in the brane matter content. Note
also that there are no extra $\lambda^2$ appearing in the projected
Weyl tensor~(\ref{contweyl}) since with these conditions $\BRN{P} +
\BRN{\rho}$ is $\lambda$-independent. This was also noted in
Ref.~\cite{Khoury}.

\subsection{Closing the system when {\protect $Z_2$} symmetry is
broken}
\label{subsecBound}

When solving the Einstein equations on the brane~(\ref{generale}), we
need the continuous part of the extrinsic curvature
$\CONT{\CEX_{\alpha \beta}}$.  If $Z_2$ symmetry is assumed --- as
motivated by $M$-theory --- the evolution of the bulk is the same on
both the sides of the brane.  In this case the Israel conditions
determine $\CEX_{\alpha \beta}$ entirely: it is always possible to
choose a coordinate system in which $\IB (\POSBRN + \fd) = \IB
(\POSBRN - \fd)$ and $\IA (\POSBRN + \fd) = \IA (\POSBRN - \fd)$,
i.e., where $\IB$ and $\IA$ are even.  Thus $\IpN$ and $\HpN$ are odd
and the continuous value of $\CEX_{\alpha \beta}$ across the brane
vanishes,
\begin{equation}
\label{contK}
\CONT{\CEX_{\alpha \beta}} = 0
 \qquad \qquad
(Z_2 \mbox{ symmetry}) .
\end{equation}
This implies
\begin{eqnarray}
  \CONT{\IpN} & = & \CONT{\HpN} = 0 ,  \qquad \qquad
(Z_2 \mbox{ symmetry}) ,
\nonumber \\  
\IpN^+ & = & - \IpN^- = \TDEMI \DISC{\IpN} ,
\nonumber \\  
\HpN^+ & = & - \HpN^- = \TDEMI \DISC{\HpN} .
\end{eqnarray}
If $Z_2$ symmetry is not assumed, as in this paper, the evolution on
either side will in general be different and $\CONT{\CEX_{\alpha
\beta}}$ no longer vanishes,
\begin{equation}
\label{contKnonZ2}
\CONT{\CEX_{\alpha \beta}} \neq 0
 \qquad \qquad
(Z_2 \mbox{ symmetry broken}) .
\end{equation}
One can, however, obtain a condition for the continuous part of the
extrinsic curvature by considering the jump of Eq.~(\ref{genein}). We
obtain
\begin{eqnarray}
\label{jumpbrnee}
0 & = &   \frac{2}{3} \KAPPACQ
          \left(  \DISC{T_{\mu \nu}} \FFF_\alpha^\mu \FFF_\beta^\nu
                - \left(  \DISC{T_{\mu \nu} \NORMVEC^\mu \NORMVEC^\nu}
                  + \frac{1}{4} \DISC{T} \right) \FFF_{\alpha \beta}
          \right)
\nonumber \\
 &    & - \DISC{\CEX} \CONT{\CEX_{\alpha \beta}}
        + \DISC{\CEX_\alpha^\mu} \CONT{\CEX_{\mu \beta}}
        + \FFF_{\alpha \beta}
          (  \DISC{\CEX} \CONT{\CEX}
           - \DISC{\CEX^{\mu \nu}} \CONT{\CEX_{\mu \nu}})
\nonumber \\ 
 &    & - \CONT{\CEX} \DISC{\CEX_{\alpha \beta}}
        + \CONT{\CEX_\alpha^\mu} \DISC{\CEX_{\mu \beta}}
        + \DISC{\WN_{\alpha \beta}} .
\end{eqnarray}
(Notice that $\DISC{\ND{(4)}{\BRN{G}}_{\alpha \beta}} = 0$.)  This
becomes, in components,
\begin{eqnarray}
\label{grrr1}
   \BRN{\rho} \CONT{\HpN}
 & = & \frac{1}{4} \left(  \DISC{P + \rho - \PRESY}
                         + \frac{1}{\KAPPACQ} \DISC{\WEYL} \right) , \\
\label{grrr2}
   (\BRN{\rho} + 3 \BRN{P}) \CONT{\HpN} - \BRN{\rho} \CONT{\IpN}
 & = &   \frac{1}{4} \left(  \DISC{P + \rho + 3 \PRESY}
                           + \frac{1}{\KAPPACQ} \DISC{\WEYL} \right) ,
\end{eqnarray}
where, using Eq.~(\ref{weylback}) and the junction conditions
(\ref{bjcn}) and (\ref{bjca}), the jump of $\WEYL$ on the brane can be
expressed as
\begin{equation}
\label{wwweyl}
\DISC{\WEYL}
 =   \left(\pezB + 2 \CONT{\UcB} - \HcB \right) \DISC{\UcB}
   - \DISC{\ppzN} \CONT{\IpN - \HpN}
   - \KAPPACQ \CONT{\IpN} \left(2 \BRN{P} + \frac{5}{3} \BRN{\rho} \right)
   + \KAPPACQ \CONT{\HpN} \left(\BRN{P} + \frac{2}{3} \BRN{\rho} \right) .
\end{equation}
(Note that Eqns~(\ref{grrr1},\ref{grrr2}) could alternatively have
been obtained from the discontinuous part of the Einstein
equations~(\ref{backein00}--\ref{backein55}).)

Equations~(\ref{grrr1},\ref{grrr2}) allow one to fix the unknown
quantities $\CONT{\HpN}$, $\CONT{\IpN}$, {\em provided} the jumps of
both the bulk matter content and the Weyl tensor are known.  Thus the
continuous part of the extrinsic curvature depends not only on the
brane matter content but also on the discontinuity of the bulk
stress-energy and the projected Weyl tensors. If both vanish,
Eqns~(\ref{grrr1},\ref{grrr2}) allow in particular the trivial
solution $\CONT{\HpN} = \CONT{\IpN} = 0$, which holds with $Z_2$
symmetry. The jump of the Weyl tensor, Eq.~(\ref{wwweyl}), contains
first derivatives of the extrinsic curvature with respect to $\fd$,
and so it is not possible in general to determine $\CONT{\IpN}$ and
$\CONT{\HpN}$ without first solving the Einstein equations in the
bulk. Nonetheless, if the projection~(\ref{wwweyl}) of the
$5$-dimensional bulk Weyl tensor on the brane is known {\it a priori}
(as in the case of a Schwarzschild-Anti~de~Sitter bulk with a known
black hole mass on both side of the brane), then $\CONT{\IpN}$ and
$\CONT{\HpN}$ can be determined {\em directly} from
Eqns~(\ref{grrr1},\ref{grrr2}) (see~\cite{JP1,JP2,JP3} for a more
detailed discussion).

\subsection{Brane motion}

On contracting Eq.~(\ref{jumpbrnee}) with the first fundamental form
$\FFF^{\alpha \beta}$ and using the second junction
condition~(\ref{juncgen}), one obtains
\begin{equation}
\label{sail}
\BRN{T}^{\mu \nu} \CONT{\CEX_{\mu \nu}}
 = \DISC{\NORMVEC_\mu \NORMVEC_\nu T^{\mu \nu}} .
\end{equation}
This equation is known as the ``sail
equation''~\cite{JP1,JP2,JP3}. The right hand side is an external
force density on the brane due to the asymmetry of the bulk
stress-energy tensor on the two sides. In analogy with Newton's second
law (here the force is due to a pressure difference between the two
sides of the brane), $\BRN{T}^{\alpha \beta}$, $\CONT{\CEX_{\alpha
\beta}}$, and $\DISC{\NORMVEC_\mu \NORMVEC_\nu T^{\mu \nu}}$ play the
role of mass, acceleration, and force, respectively. Notice from
(\ref{contK}) that when $Z_2$ symmetry is assumed, this equation
vanishes identically.  When $Z_2$ symmetry is broken, the
``acceleration'' $\CONT{\CEX_{\mu \nu}}$ is non-zero. In this paper we
do not assume $Z_2$ symmetry, but recall that we have chosen a
coordinate system in which the brane is at rest: Eq.~(\ref{newcons2})
must therefore be understood as dictating the condition that must be
satisfied by $\CONT{\HpN}$ and $\CONT{\IpN}$ (and therefore by the
coordinate system itself) for the brane to remain at a fixed position
$\POSBRN$. Later, however, we will see that Eq.~(\ref{sail}) does
indeed give a more intuitive equation of motion for the perturbed
brane position or brane displacement (see Section~\ref{ssec_sail}).

In components the sail equation leads to
\begin{equation}
\label{newcons2}
 - \CONT{ \IpN} \BRN{\rho} + 3 \CONT{\HpN} \BRN{P} = \DISC{\PRESY} ,
\end{equation}
which can also be obtained by taking the discontinuous part of the
$\{\fin \fin\}$ component of the Einstein equation,
Eq.~(\ref{backein55}) or, of course, by a linear combination of
Eqns~(\ref{grrr1},\ref{grrr2}). This is the only combination of
Equations~(\ref{grrr1},\ref{grrr2}) which does not involve the Weyl
tensor.

\subsection{Conservation equations}

The singular part of the $5$-dimensional energy conservation
equation~(\ref{backbianchi1}) yields the stress-energy conservation
equation on the brane: we find
\begin{equation}
\pezB \BRN{\rho} + 3 \HcB (\BRN{P} + \BRN{\rho}) = - \DISC{F} .
\end{equation}
(Again the generalisation to several interacting components may be
found in Appendix D5.) Notice that the jump in the bulk energy
flux transverse to the brane enters in the conservation equation,
meaning that the brane matter content can act as a source or a
sink to the energy flux along the fifth dimension. When this
energy flux is continuous, the conservation equation on the brane
reduces to the usual one, as discussed in~\cite{SMS}. Another
conservation equation appears in brane cosmology: by considering
the singular part of Eq.~(\ref{backbianchi2}) we obtain again the
sail equation~(\ref{newcons2}).  Both equations were
first found in~\cite{Bine1} for the case a bulk cosmological
constant.

\section{Bulk perturbations}
\label{secVI}

We now turn to perturbed quantities, and begin in this section by
analysing the properties of the perturbed bulk: the perturbed brane
itself will be introduced in Section~\ref{secVII}.  We work throughout
with gauge independent perturbation variables, which are inspired from
a generalisation of the Newtonian (or longitudinal) gauge to the
$5$-dimensional case.  First we introduce the bulk metric perturbation
variables using the standard scalar, vector, tensor decomposition.  We
study their gauge transformation properties and define gauge invariant
combinations. Then, in Section~\ref{ssec_pert_se}, the perturbations
of the bulk stress-energy tensor are considered, leading, in
Section~\ref{ssec_pert_ee}, to the gauge invariant perturbed bulk
Einstein equations.  Finally we write down the perturbed conservation
equations (Bianchi identities).

\subsection{Classification of the perturbations}
\label{ssecSVTdvg}

Let us consider the perturbations of a spacetime with one timelike and
$n$ spacelike coordinates. The perturbed metric of this spacetime
possesses $\TDEMI (n + 1) (n + 2)$ different components.  Amongst
these, a coordinate transformation allows $n + 1$ of them to be fixed,
so that there are $\TDEMI n (n + 1)$ independent metric
coefficients. For example, in synchronous gauge, the $\delta g_{0
\alpha}$ are set to zero.

When solving perturbation equations about a given spacetime, one is
naturally led to classify perturbations. Two classifications are of
particular relevance. Firstly, the perturbations may be classified
according to their physical meaning, and this is done by looking at
the spin of the perturbation in a locally Minkowskian frame. The
different perturbations are density (spin $0$) modes, vorticity (spin
$1$) modes, and gravitational (spin $2$) waves. Secondly, the
perturbations may be classified more geometrically in terms of irreducible
components under the group of isometries of
 the unperturbed spacetime. This leads to scalar,
vector and tensor perturbations.  Under some circumstances, these two
classifications are identical.  In particular, this is true for a
Friedmann-Lema\^\i{}tre-Robertson-Walker (FLRW) spacetime, which can
be foliated by a set of maximally symmetric spacelike
hypersurfaces. In brane cosmology, however, the bulk is not as
symmetric as in the FLRW case, and the two classifications are
different.

Components which transform irreducibly under symmetries of the
background spacetime evolve independently (to linear order) while the
physical spin components mix.

\subsubsection{Physical splitting}

As explained above, metric perturbations can be decomposed according
to their spin with respect to a local rotation of the coordinate
system. This leads to density modes, vorticity modes, and
gravitational waves.  Gravitational (spin $2$) waves are ``true''
degrees of freedom of the gravitational field in the sense that they
can exist even in vacuum. The number of gravitational wave modes is
given by the dimension of the vector space spanned by symmetric,
transverse, traceless rank $2$ tensors in an $n$-dimensional space:
this is $\TDEMI (n - 2) (n + 1)$.  In addition, when there is a non
trivial matter content, there may be vorticity (or spin $1$) modes
arising from rotational velocity fields, which have $n - 1$
independent components.  Finally, there remain $\TDEMI n (n + 1) -
\TDEMI (n - 2) (n + 1) - (n - 1) = 2$ possible density (spin $0$)
modes, which are usually represented by the two Bardeen potentials
$\Phi$ and $\Psi$~\cite{Bar,KS}.

More schematically, let us consider the metric perturbation around a
locally inertial frame, written in synchronous gauge and in Fourier
space considering the wave vector $k^i = k \delta^i_1$~:
\begin{equation}
\label{dgphys}
\delta g_{\alpha \beta} = \left(
\begin{array}{c|c|cccc}
0 & 0 & 0 & 0 & \ldots & 0 \\
\hline 0 &  2k^2 E - 2C & i k V_2 & i k V_3 & \ldots & i k V_n \\
\hline 0 & i k V_2 & \displaystyle -2C + \sum_{i = 3}^n h^+_i
     & h^\times_{2 3} & \ldots & h^\times_{2 n} \\
0 & i k V_3 & h^\times_{2 3} & -2C - h^+_3 & & h^\times_{3 n} \\
\vdots & \vdots & \vdots & & \ddots & \vdots \\
0 & i k V_n & h^\times_{2 n} & h^\times_{3 n} & \ldots & -2 C - h^+_n
\end{array} \right) .
\end{equation}
The quantities $E$ and $C$ describe the density modes (with the
standard definition of the Bardeen potentials, one has $\Phi = - C$
and $\Psi = - \partial_t^2 E$), the $V_i$ ($i = 2, \ldots, n$)
represent the vorticity modes, and the $h^+_i$ ($i = 3, \ldots, n$)
and $h^\times_{j k}$ ($2 \leq j < k \leq n$) represent the
gravitational waves (when $n = 3$, these notations agree with the
standard definition of the $h^+$ and $h^\times$ modes).

\subsubsection{Geometrical splitting}

The three above types of perturbation generally do not evolve
independently: even at linear order, they are coupled if the
unperturbed spacetime does not possess any symmetries. However, for
most cosmological models (including the ones considered in this
paper), spacetime possesses some symmetries, being invariant under a
certain group of global transformations. We consider the symmetry
group ${\rm SO} (N)$ with $N < n$, which is of course relevant when
there exists a coordinate system in which $N$ coordinates span a
maximally symmetric space.

When this is the case, perturbations may be decomposed into components
which transform irreducibly under ${\rm SO} (N)$-rotations of the
coordinate system. This leads to what we call scalar, vector and
tensor perturbations which are perturbations whose spin with respect
to ${\rm SO} (N)$ is $0$, $1$ and $2$ respectively.  The main
advantage of this decomposition is that the three new types of
perturbation are now decoupled from each other, and hence are
convenient when studying the evolution of cosmological
perturbations. For example, consider an $n$-dimensional space with $N$
coordinates (labelled by $i$, $j$, etc) spanning an $N$-dimensional,
maximally symmetric sub-space, with metric $\gamma_{i j}$, and
associated covariant derivative $\nabla_i$. The $n - N$ remaining
coordinates will be labelled by $A$, $B$, etc.  In this case, the
metric perturbations can be decomposed as
\begin{eqnarray}
\delta g_{i j}
 & = & - 2 C \gamma_{i j} - 2 \nabla_{i j} E
       - 2 \nabla_{(i} \VV E_{j)} - 2 \TT E_{ij} , \\
\delta g_{i A} & = & \nabla_i \ea + \vea_i , \\
\delta g_{A B} & = & \eab ,
\end{eqnarray}
where barred quantities are divergenceless $N$-vectors, and double
barred quantities are divergenceless, traceless $N$-tensors of rank 2
(with respect to the covariant derivative $\nabla_i$ and metric
$\gamma_{i j}$ respectively). With our definitions, it is clear that
$C$, $E$, $\ea$, $\eab$ are scalars, $\VV{E}_i$, $\vea_i$ are vectors,
and $\TT{E}_{i j}$ are tensors under ${\rm SO} (N)$ rotations.  The
perturbed metric components can then be written as
\begin{equation}
\label{dggeom}
\delta g_{\alpha \beta} = \left(
\begin{array}{c|c|cccc|c}
0 & 0 & 0 & 0 & \ldots & 0 & 0 \\
\hline 0 & 2 k^2 E -2 C & -2 i k \VV{E}_2
     & - 2 i k \VV{E}_3 & \ldots & -2 i k \VV{E}_N & i k \ea \\
\hline 0 & - 2 i k \VV{E}_2 & \displaystyle -2 C + \sum_{i = 3}^N h^+_i
     & h^\times_{2 3} & \ldots & h^\times_{2 N} & \vea_2 \\
0 & i k \VV{E}_3 & h^\times_{2 3}
  & -2 C - h^+_3 & & h^\times_{3 N} & \vea_3 \\
\vdots & \vdots & \vdots & & \ddots & \vdots & \vdots \\
0 & i k \VV{E}_N & h^\times_{2 N}
  & h^\times_{3 N} & \ldots & -2 C - h^+_N & \vea_N \\
\hline 0 & i k \ea & \vea_2 & \vea_3
     & \ldots & \vea_N & \eab
\end{array} \right) ,
\end{equation}
with the $h^+_k, h^\times_{l m}$ describing $\TT{E}_{i j}$.
Obviously, one has
\begin{itemize}

\item $2 + (n - N) + \TDEMI (n - N) (n - N + 1)$ scalar
degrees of freedom,

\item $(N - 1) (n - N + 1)$ vector degrees of freedom and,

\item $\TDEMI (N - 2) (N + 1)$ tensor degrees of freedom.

\end{itemize}
By definition, the tensor components are spin $2$ quantities and
represent gravitational waves. It is clear that when $N \neq n$, not
all the gravitational waves are tensor perturbations (with respect to
${\rm SO} (N)$): $\TDEMI (n - N) (n + N - 1)$ of them are actually
scalar or vector perturbations. In fact, the spin of the second
decomposition can be understood as the projection of the spin of the
first decomposition on the maximally symmetric space. Therefore,
density modes are always scalar perturbations, vorticity modes can be
either scalar of vector perturbations, and gravitational waves can be
any of the three.  By comparing Eqns~(\ref{dgphys})
and~(\ref{dggeom}), it is clear that:
\begin{itemize}

\item the $2 + (n - N) + \TDEMI (n - N) (n - N + 1)$ scalars
decompose as $2$ density modes, $n - N$ vorticity modes, and $\TDEMI
(n - N) (n - N + 1)$ gravitational waves,

\item the $(N - 1) (n - N + 1)$ vectors represent $(N - 1)$
vorticity modes and $(N - 1)(n - N)$ gravitational waves,

\item the $\TDEMI (N - 2) (N + 1)$ tensors all represent
gravitational waves.

\end{itemize}
For our purpose ($n = N + 1 = 4$), this reduces to
\begin{itemize}

\item $4$ scalar degrees of freedom which split into the $2$ density
modes, $1$ vorticity mode, and $1$ gravitational wave,

\item $4$ vector degrees of freedom which go into $2$ vorticity
modes and $2$ gravitational waves,

\item $2$ tensor degrees of freedom which all represent gravitational
waves.

\end{itemize}
As expected, we have $10$ degrees of freedom $5$ of which are
gravitational waves. This decomposition ensures that even in the
vacuum, the scalar and vector parts of the Einstein equation will
allow non trivial solutions. These are usually called ``graviscalars''
and ``graviphotons''~\cite{gravitruc1,gravitruc2}. This effect
represents the most striking change to the physics of brane
cosmological perturbations as compared to that of the standard FLRW
case since it can occur at arbitrary low energy as long as the
corresponding gravitational waves exist in the bulk.

\subsubsection{The brane point of view}

The brane is, by definition, described by $N + 1$ coordinates: one
timelike and the $N$ spacelike coordinates spanning an $N$-dimensional
maximally symmetric space. For the case of one codimension, we have $N
= n - 1$. The perturbed induced metric of the maximally symmetric
space then has $\TDEMI n (n - 1)$ independent components. An important
question is how these perturbations can interact with the bulk
perturbations.  It is clear that whatever the bulk matter content,
there are at least $\TDEMI (n - 2) (n + 1) = \TDEMI n (n - 1) - 1$
degrees of freedom which arise from the gravitational waves in the
bulk.  Therefore, one can expect that $\TDEMI (n - 2) (n + 1)$ of the
brane perturbations can interact with the bulk. We will see that this
is indeed the case: the second Israel condition essentially states
that the discontinuity of some bulk perturbations which can exist even
in the vacuum describe the matter content of the brane. But this also
suggests that one additional scalar degree of freedom of the brane is
likely not to be directly related with the bulk perturbations. It
happens, indeed, that this extra degree of freedom physically
corresponds to the perturbation of the brane position in the bulk,
which is independent of the gravitational waves.  For example, if the
bulk is pure Minkowski space, one can consider a fixed coordinate
system (as, e.g., Newtonian gauge, which is unambiguously fixed).  The
position of the brane in this coordinate system is defined
independently of the metric perturbations.  This extra degree of
freedom ensures that in any situation all the brane perturbations can
be related to bulk perturbations (see also the discussion in
Ref.~\cite{Deruelle}). One of the aims of this paper is to make the
link between these two sets of perturbations.

\subsection{Geometrical perturbation variables}

We now make use of maximal symmetry on ${\cal M}$.  Due to rotational
invariance, we can split the perturbations into components which
transform irreducibly under rotations, i.e., into different ${\rm SO}
(3)$-spin components, which evolve independently to first order
perturbation theory. One could then go on and split these into
irreducible components under translations, corresponding to the
expansion in terms of eigenvectors of the Laplacian on $\cal M$ (which
is the Fourier transform in the case $\COURB = 0$)~\cite{KS}.
Following the discussion of the last paragraph, the perturbed line
element can be generally written as
\begin{equation}
\label{mpert1}
\ddd s^2 =   \IBB (1 + 2 A ) \ddd \eta^2
         + 2 \IAB B_i \ddd \eta \ddd x^i
         - \IAA (\gamma_{i j} + h_{i j}) \ddd x^i \ddd x^j
         + 2 \IBN \Bperp \ddd \eta \ddd \fd
         + 2 \INA \epi_i \ddd x^i \ddd \fd
         - \INN (1 - 2 \epp) \ddd \fd{}^2 .
\end{equation}
Here, the $\perp$ indices of $\epi_i$, $\epp$ are labels.  The
quantities $B_i$ and $\epi_i$ are vectors on ${\cal M}$ which can be
respectively decomposed into scalar (spin 0) components $B$, $\Eperp$,
and divergenceless vector (spin 1) components $\VV{B}_i$, $\Vepi_i$,
such that $\gamma^{i j} \nabla_i \VV{B}_j = \gamma^{i j} \nabla_i
\Vepi_j = 0$.  Equivalently, the tensor on ${\cal M}$, $h_{i j}$, can
be decomposed into two scalars, $C$ and $E$, a divergenceless vector,
$\VV{E}_i$, and divergenceless, traceless, tensor (spin $2$)
component, $\TT{E}_{i j}$, such that $\gamma^{i j} \nabla_i \VV{E}_j =
\gamma^{i j} \nabla_i \TT{E}_{i j} = \TT{E}{}^i_i = 0$.  This
decomposition is
\begin{eqnarray}
B_i & = & \nabla_i B + \VV{B}_i , \\
\epi_i & = & \nabla_i \Eperp + \Vepi_i , \\
h_{i j} & = & 2 C \gamma_{i j} + 2 E_{i j} , \\
E_{i j} & = & \nabla_{(i} E_{j)} + \TT{E}_{i j} , \\
E_i & = & \nabla_i E + \VV{E}_i .
\end{eqnarray}
The indices of these ${\cal M}$-quantities are raised and lowered with
the metric $\gamma_{i j}$.  The symmetries of the metric ensure that
the scalar ($A$, $B$, $C$, $E$, $\Bperp$, $\Eperp$, $\epp$), vector
($\VV{B}_i$, $\VV{E}_i$, $\Vepi_{i}$) and tensor ($\TT{E}_{i j}$)
quantities evolve independently.

\subsection{Gauge invariant metric perturbations}

Let us consider an infinitesimal coordinate transformation
\begin{equation}
\label{infi}
x^\alpha \to x^\alpha + \xi^\alpha ,
\end{equation}
with
\begin{eqnarray}
\label{cico}
\xi^\alpha & = & (T, L^i, \Lperp) , \\
L^i & = & \nabla^i L + \VV{L}^i .
\end{eqnarray}
Under this coordinate change the geometrical perturbations transform
in the following way:
\begin{eqnarray}
\label{transmetdeb}
A & \to & A + \pezB (\IB  T) + \IpN \IN \Lperp , \\
B_i & \to & B_i - \IAzB \dot L_i + \IBzA \nabla_i T , \\
C & \to & C + \HcB \IB T + \HpN \IN \Lperp , \\
\label{transE}
E_i & \to & E_i + L_i , \\
\TT{E}_{i j} & \to & \TT{E}_{i j} , \\
\Bperp & \to & \Bperp - \INzB \dot \Lperp + \IBzN T' , \\
\epi_i & \to & \epi_i - \IAzN L'_i - \INzA \nabla_i \Lperp , \\
\label{transepp}
\epp & \to &   \epp - \UcB \IB T - \ppzN (\IN \Lperp) ,  \\
\label{transABBE}
\ABBE & \to & \ABBE + T , \\
\label{transmetfin}
\ANEE & \to & \ANEE - \Lperp  .
\end{eqnarray}
(Recall that $\dot{} \equiv \partial / \partial \eta$ and that ${}'
\equiv \partial / \partial \fd$.  In this section we will use
both this notation and the $\partial_{u, n}$ defined in
Eqns~(\ref{defperpu},\ref{defperpn}): we aim to do so in such a
way as to keep the equations as simple as possible.) We can therefore
define the following four scalar and two  vector
perturbation variables, which are invariant under infinitesimal coordinate
transformations, also called gauge transformations in this context:
\begin{eqnarray}
\label{vPsi}
\Psi & = & A - \pezB \BABBE + \IpN \NANEE , \\
\label{vPhi}
\Phi & = & - C + \HcB \BABBE - \HpN \NANEE , \\
\label{vvp}
\vp & = & \Bperp - \IB \ppzN \ABBE - \IN \pezB \ANEE ,  \\
\label{vhpp}
\hpp & = & \epp + \UcB \BABBE - \ppzN \NANEE , \\
\label{vvi}
\vi_i & = & \VV{B}_i + \IAzB \dot{\VV{E}}_i , \\
\label{vhci}
\hci_i & = & \Vepi_i + \IAzN \VV{E}{}'_i .
\end{eqnarray}
The two vector variables possess two independent components (hence
four degrees of freedom). The tensor variable $\TT{E}_{i j}$ is gauge
invariant since there are no tensor type gauge transformations, and
possesses two independent components. All these quantities represent a
generalisation of the Newtonian gauge often used in FLRW cosmologies
(we can no longer call it a ``longitudinal gauge'', as $\delta g_{0
\fin} \neq 0$). It is completely fixed by setting $\ABBE$, $E^i$,
$\ANEE$ to $0$ and in this case one has
\begin{eqnarray}
\label{newtdeb}
\delta g_{0 0} & = & 2 \IBB \Psi , \\
\delta g_{0 i} & = & \IAB \vi_i , \\
\delta g_{i j} & = & 2 \IAA (\Phi \gamma_{i j} - \TT{E}_{i j} ) , \\
\delta g_{0 \fin} & = & \IBN \vp , \\
\delta g_{i \fin} & = & \INA \hci_i , \\
\label{newtfin}
\delta g_{\fin \fin} & = & 2 \INN \hpp .
\end{eqnarray}
This gauge is perfectly well-suited for describing the bulk
perturbation {\em without} a brane. In the presence of a brane,
however, things are more complicated since some of these quantities
involve first or second derivatives with respect to the fifth
dimension, and hence they are not always regular at the brane position
(see Section~\ref{ssec_reg_pert_geom}). For this reason, other gauge
choices are often preferred, but not essential\footnote{In any case,
there is no particular reason why the brane and bulk metric
perturbations should be the same as the brane perturbation depends
explicitly on the brane position, which is not a quantity that can be
defined everywhere in the bulk.}.

\subsection{Perturbed stress-energy tensor}
\label{ssec_pert_se}

We now perturb the unit vectors $U^\alpha$ and $N^\alpha$, defined in
Eqns~(\ref{Tdef}--\ref{Ndef}), which are the timelike and spacelike
eigenvectors normal to the maximally symmetric $3$-spaces.  It follows
from the normalization conditions, Eqns~(\ref{Udef},\ref{Ndef}), that
each vector has only four independent components. Furthermore, as
$U^\alpha$ and $N^\alpha$ are eigenvectors of a symmetric tensor, they
are normal to each other, $N_\mu U^\mu = 0$.  Hence at perturbed order
there are only seven independent components which we denote by
$\RVD{v}^i$, $\RVD{\dni}^i$, $\www$. They are defined by
\begin{eqnarray}
\delta U^\alpha
 & = & \left(  \IzB \GAM (\BET \www - A - \BET \Bperp),
               \IzA \RVD{v}^i,
               \IzN \GAM (\www + \BET \epp) \right) , \\
\delta N^\alpha
 & = & \left(  \IzB \GAM (- \www + \BET A + \Bperp),
               \IzA \RVD{\dni}^i,
             - \IzN \GAM (\BET \www + \epp) \right) .
\end{eqnarray}
Neglecting the metric perturbations, the quantity $\www$ represents
the perturbation of the Lorentz boost $\BET$,
\begin{equation}
\label{wperp}
\www = \frac{\delta \GAM}{\BGM}
        = \frac{\delta (\BGM)}{\GAM}
        = \GMM \delta \BET .
\end{equation}
As usual, we will decompose $\RVD{v}^i$, $\RVD{\dni}^i$, into scalar
and vector components,
\begin{eqnarray}
\label{decom1}
\RVU{v}_i & = & \nabla_i \RVD{v} + \VVRVU{v}_i , \\
\label{decom2}
\RVU{\dni}_i & = & \nabla_i \RVD{\dni} + \VVRVU{\dni}_i .
\end{eqnarray}
Finally, in order to write down the stress-energy tensor, it is also
useful to introduce the variables
\begin{eqnarray}
v^i & = & \GAM (\RVD{v}^i + \BET \RVD{\dni}^i) , \\
f^i & = & \GAM (\RVD{\dni}^i + \BET \RVD{v}^i) ,
\end{eqnarray}
which have a decomposition into scalar and vector components similar
to~(\ref{decom1},\ref{decom2}).  With these definitions, general
perturbations of the bulk stress-energy tensor, $\delta T_{\alpha
\beta}$, are
\begin{eqnarray}
\delta T_{0 0}
 & = & \IBB \left(\delta \rho + 2 \rho A \right) , \\
\delta T_{0 i}
 & = & - \IAB \left( (\rho + P) v_i - \rho B_i -  F (\dni_i + \epi_i)
              \right) , \\
\delta T_{0 \fin}
 & = & - \IBN \left(  \delta F + F (A - \epp) - \rho \Bperp \right) , \\
\delta T_{i j}
 & = & \IAA \left(  \delta P \gamma_{i j} + \Pi_{i j}
                  + 2 P (C \gamma_{i j} +  E_{i j}) \right) , \\
\delta T_{i \fin}
 & = & \INA \left( (P - \PRESY) \dni_i + F (v_i - B_i) - \PRESY \epi_i
            \right) , \\
\delta T_{\fin \fin}
 & = & \INN \left(\delta \PRESY  - 2 \PRESY \epp - 2 F \Bperp \right) .
\end{eqnarray}
Here we have defined, according
to~(\ref{cordtointdeb}--\ref{cordtointfin},\ref{wperp}):
\begin{eqnarray}
\delta \rho
 & = &   \GMM (\delta \REST{\rho} + \BTT \delta \REST{\PRESY})
       + 2 F \www , \\
\delta \PRESY
 & = &   \GMM (\delta \REST{\PRESY} + \BTT \delta \REST{\rho})
       + 2 F \www , \\
\delta P
 & = & \delta \REST{P} , \\
\delta F
 & = &   \BGG (\delta \REST{\rho} + \delta \REST{\PRESY})
       + (\rho + \PRESY) \www ,
\end{eqnarray}
and we have introduced the anisotropic stress tensor $\Pi_{i j}$,
which again may be decomposed into a scalar, (divergenceless) vector,
and (divergenceless, traceless) tensor components according to
\begin{equation}
\Pi_{i j}
 =   \left(\nabla_{i j} - \frac{1}{3} \nabla^2 \gamma_{i j} \right) \Pi
   + \nabla_{(i} \VV{\Pi}_{j)} + \TT{\Pi}_{i j} .
\end{equation}

On investigation of the behaviour of these variables under the
infinitesimal coordinate transformations~(\ref{infi}) (see
Appendix~\ref{sec_app_pert_bulk_matt}), we find the following
scalar gauge invariant variables
\begin{eqnarray}
v^\sharp
 & = & v + \IAzB \dot E , \\
\VV{v}^\sharp_i
 & = & \VV{v}_i + \IAzB \dot{\VV{E}}_i , \\
\dni^\sharp
 & = & \dni - \IAzN E' , \\
\VV{\dni}^\sharp_i
 & = & \VV{\dni}_i - \IAzN \VV{E}{}'_i , \\
\www^\sharp
 & = & \www - \frac{\dot \GAM}{\BGM} \ABBE
               + \frac{\GAM'}{\BGM} \ANEE
               - \INzB \pe \ANEE , \\
\delta \REST{X}^\sharp
 & = & \delta \REST{X} - \dot \REST{X} \ABBE + \REST{X}' \ANEE ,
\end{eqnarray}
where $\REST{X}$ is any scalar quantity (density $\REST{\rho}$,
pressure $\REST{P}$, etc).  The anisotropic stress tensor
$\Pi_{i j}$ is gauge invariant by itself due to the  Stewart-Walker
lemma~\cite{STW}. Notice that $\rho$, $\PRESY$, $F$ and $\www$
are not scalars (since they depend explicitly on the choice of the
coordinate system via the vector fields $u^\alpha$ and
$n^\alpha$), but we can, however, define the following gauge
invariant variables,
\begin{eqnarray}
\delta \rho^\sharp
 & = &   \GMM (\delta \REST{\rho}^\sharp + \BTT \delta \REST{\PRESY}^\sharp)
       + 2 F \www^\sharp , \\
\delta \PRESY^\sharp
 & = &   \GMM (\delta \REST{\PRESY}^\sharp + \BTT \delta \REST{\rho}^\sharp)
       + 2 F \www^\sharp , \\
\delta P^\sharp
 & = & \delta \REST{P}^\sharp , \\
\delta F^\sharp
 & = &   \BGG (\delta \REST{\rho}^\sharp + \delta \REST{\PRESY}^\sharp)
       + (\rho + \PRESY) \www^\sharp .
\end{eqnarray}
As an example, the perturbed stress-energy tensor for a scalar field
is given in Appendix~\ref{ssec_app_sf}.

\subsection{The perturbed Einstein equations}
\label{ssec_pert_ee}

The explicit forms of the perturbed Christoffel symbols, the
perturbed Riemann, Ricci and Einstein tensors are all given in
Appendices~\ref{ssec_app_pert_christ}, \ref{ssec_app_pert_ric},
\ref{ssec_app_pert_riem}, \ref{ssec_app_pert_ein}, where they are
expressed in terms of the gauge invariant variables introduced
above. We now write down the full perturbed bulk Einstein
equations also in terms of gauge invariant variables.  They split
into seven scalar, three vector (each with two independent
components), and one tensor (with two independent components)
equations, adding up to the required $15$ components of a
symmetric $5 \times 5$ tensor. These equations are given below,
where we indicate on the right hand side of each equation from
which component of the Einstein equations they were derived and,
when necessary, the term to which they are proportional.  The
seven scalar equations are
\begin{eqnarray}
\label{einpert00}
   \DzAA (2 \Phi + \hpp) + 6 \KzAA \Phi
\nonumber \\
 - 3   \left(2 \HcB^2 + 2 \HcB \UcB \right) \Psi
           -3 \HcB \pezB \hpp
           -3 \left(2 \HcB + \UcB \right) \pezB \Phi
\nonumber \\
 - 3 (\HpN \ppzN + 4 \HpN^2) \hpp
 - 6 \hpp \ppzN \HpN
\nonumber \\
 + 3 (\ppzN + 4 \HpN ) \ppzN \Phi
\nonumber \\
 + 3 \left(\ppzN + 3 \HpN + \IpN \right) (\HcB \vp)
 & = & \TEQREFC{\KAPPACQ \left(\delta \rho^\sharp - F \vp \right)}
               {}{00}{} , \\
\label{einpert0is}
   \DDEMI \left(\ppzN +  \HpN + 2 \IpN \right) \vp
\nonumber \\
 -  (\UcB + 2 \HcB) \Psi
         - (\pezB + \UcB - \HcB) \hpp - 2 \pezB \Phi
 & = & \TEQREFC{  \KAPPACQ \left( (P + \rho) \IA v^\sharp
               - F \IA \dni^\sharp \right)}{}{0i}{} , \\
\label{einpertijstr}
 - 2 \KzAA \Phi
\nonumber \\
 +2 \left((\UcB + 2 \HcB) (\pezB + \UcB) + 3 \HcB^2 \right) \Psi
 + 2 \Psi \pezB (\UcB + 2 \HcB)
\nonumber \\
 + 2\left( (\IpN + 2 \HpN) (\ppzN + \IpN) + 3 \HpN^2 \right) \hpp
 + 2 \hpp \ppzN (\IpN + 2 \HpN)
\nonumber \\
 + (\pezB + \UcB) \pezB (\hpp + 2 \Phi)
\nonumber \\
 - (\UcB + 2 \HcB) \pezB (\Psi - \hpp - 3 \Phi)
\nonumber \\
 + (\ppzN + \IpN) \ppzN (\Psi - 2 \Phi)
\nonumber \\
 + (\IpN + 2 \HpN) \ppzN (\Psi - \hpp - 3 \Phi)
\nonumber \\
 - \TDEMI (\ppzN \pezB + \pezB \ppzN + \IpN \pezB + \UcB \ppzN) \vp
\nonumber \\
 - \left(  (\UcB + 2 \HcB) \ppzN 
         + (\IpN + 2 \HpN) \pezB \right) \vp
\nonumber \\
 - \vp \left(  (\ppzN + \IpN) (\UcB + 2 \HcB)
             + (\pezB + \UcB) (\IpN + 2 \HpN) \right)
\nonumber \\
 - 3 \left(\UcB \pezB - \IpN \ppzN \right) \Phi
 - 6 \HcB \HpN \vp
 & = & \TEQREFC{\KAPPACQ \left(\delta P^\sharp + \frac{2}{3} \Delta \Pi
                         \right)}{}{ij}{\propto \gamma_{i j}} , \\
\label{einpertijsst}
\Phi - \Psi + \hpp
 & = & \TEQREFC{\KAPPACQ \IAA \Pi}{}{ij}{\propto \nabla_{i j}} , \\
\label{einpert05}
 - 3 \left(  (\pezB \ppzN + (\HpN - \IpN) \pezB + \HcB \ppzN) \Phi
           + \HcB \ppzN \Psi - \HpN \pezB \hpp \right)
\nonumber \\
 - 3 \vp \pezB \HcB
 - \left(  \DDEMI \DzAA
         + 3 (\HcB^2 - \HcB \UcB) \right) \vp
 & = & \TEQREFC{\KAPPACQ \left(  \delta F^\sharp + F (\Psi - \hpp) \right)}
               {}{0\fin}{} , \\
\label{einperti5s}
   \DDEMI \left(\pezB +  \HcB + 2 \UcB \right) \vp
\nonumber \\
 -  (\ppzN + \IpN - \HpN) \Psi
         - (2 \HpN + \IpN) \hpp
         + 2 \ppzN \Phi
 & = & \TEQREFC{\KAPPACQ \left(F \IA v^\sharp + (P - \PRESY) \IA \dni^\sharp
                         \right)}{}{i\fin}{} , \\
\label{einpert55}
 - \DzAA (2 \Phi - \Psi) - 6 \KzAA \Phi
\nonumber \\
 + 3 \HcB \left(\pezB + 4 \HcB \right) \Psi
 + 6 \Psi \pezB \HcB
\nonumber \\
 + 3 \left(\pezB + 4 \HcB \right) \pezB \Phi
\nonumber \\
 + 3   \left(\HpN \ppzN \right) \Psi
           + 3 \left(2 \HpN^2 + 2 \HpN \IpN \right) \hpp
           - 3 \left(2 \HpN \ppzN + \IpN \ppzN \right) \Phi
\nonumber \\
 - 3 \left(\pezB + 3 \HcB + \UcB \right) (\HpN \vp)
 & = & \TEQREFC{\KAPPACQ \left(\delta \PRESY^\sharp - F \vp \right)}
               {}{\fin\fin}{} .
\end{eqnarray}
The three vector equations are:
\begin{eqnarray}
\label{einpert0iv}
 - \DDEMI (\DzAA + 2 \KzAA) \vi_i
\nonumber \\
 - \DDEMI \left(\ppzN + 4 \HpN \right) 
          \left( (\ppzN + \IpN - \HpN) \vi_i \right)
\nonumber \\
 + \DDEMI \left(\ppzN + 4 \HpN \right)
          \left( (\pezB + \UcB - \HcB) \hci_i \right)
 & = & \TEQREFD{\KAPPACQ \left(  (P + \rho) (\VV{v}^\sharp_i - \vi_i)
                               - F (\VV{\dni}^\sharp_i + \hci_i)
                         \right)}{}{0i}{} , \\
\label{einpertijv}
   \left(\pezB + 2 \HcB + \UcB \right) \vi_i
 - \left(\ppzN + 2 \HpN + \IpN \right) \hci_i
 & = & \TEQREFD{\KAPPACQ \IA \VV{\Pi}_i}{}{ij}{} , \\
\label{einpert5iv}
   \DDEMI (\DzAA + 2 \KzAA) \hci_i
\nonumber \\
 + \DDEMI \left(\pezB + 4 \HcB \right) 
          \left( (\ppzN + \IpN - \HpN) \vi_i \right)
\nonumber \\
 - \DDEMI \left(\pezB + 4 \HcB \right)
          \left( (\pezB + \UcB - \HcB) \hci_i \right)
 & = & \TEQREFD{\KAPPACQ \left(  F (\VV{v}^\sharp_i - \vi_i)
                               + (P - \PRESY) (\VV{\dni}^\sharp_i + \hci_i )
                         \right)}{}{i\fin}{} ,
\end{eqnarray}
and the tensor equation is
\begin{eqnarray}
\label{einpertijt}
 - (\DzAA - 2 \KzAA) \TT{E}_{i j}
 + \left(\pezB + 3 \HcB + \UcB \right) \pezB \TT{E}_{i j}
 - \left(\ppzN + 3 \HpN + \IpN \right) \ppzN \TT{E}_{i j}
 & = & \TEQREFB{\KAPPACQ \TT{\Pi}_{i j}}{}{i j}{} .
\end{eqnarray}

As a small aside, it is interesting to check our analysis of
Section~\ref{ssecSVTdvg}. We shall take for simplicity an empty,
Minkowski bulk, so that the terms proportional to $\KzAA$, $\HcB$,
$\UcB$, $\HpN$ and $\IpN$ vanish.  Then the above equations reduce to
\begin{eqnarray}
\DzAA (\hpp + 2 \Phi) & = & - 3 \ppzN^2 \Phi , \\
\DzAA (\Psi - 2 \Phi) & = & - 3 \pezB^2 \Phi , \\
\DzAA \vp & = & - 6 \pezB \ppzN \Phi , \\
\left(\pezB^2 - \ppzN^2 - \DzAA \right) \Phi & = & 0 , \\
\ppzN \hci_i & = & \pezB \vi_i , \\
\left(\pezB^2 - \ppzN^2 - \DzAA \right) \vi_i & = & 0 , \\
\left(\pezB^2 - \ppzN^2 - \DzAA \right) \TT{E}_{i j} & = & 0 .
\end{eqnarray}
In the vacuum, in addition to the usual two tensor modes, there
are one scalar and two vector degrees of freedom which satisfy wave
equations and represent the graviscalar and graviphoton (for a total
of five gravitons, as expected). The remaining degrees of freedom can
only exist if matter is present. They describe either density or
vorticity modes.

\subsection{Perturbed conservation equations}

We now compute the perturbed energy momentum conservation equations.
Even though they do not contain new information, they can provide useful
evolution equations for
the matter content of the bulk.
Here we write them down just for the total bulk matter. The
generalisation to several components is straightforward and is given
in Appendix~\ref{ssec_app_pert_cons}.  Written in terms of gauge
invariant variables there are three scalar conservation equations,
\begin{eqnarray}
\label{pert_cons_0}
   (\pezB + 3 \HcB + 2 \UcB) (\delta \rho^\sharp - F \vp)
 + 3 \HcB \delta P^\sharp
 + \UcB (\delta \PRESY^\sharp - \delta \rho^\sharp)
\nonumber \\
 + (\ppzN + 3 \HpN + 2 \IpN) \left(  \delta F^\sharp + F (\Psi + \hpp) \right)
\nonumber \\
 + \DzAA \left( (P + \rho) \IA v^\sharp - F \IA \dni^\sharp \right)
\nonumber \\
 - 3 (P + \rho) \pezB \Phi - (\rho + \PRESY) \pezB \hpp - F \pezB \vp
 + F \ppzN (\Psi - \hpp - 3 \Phi)
 & = & \TEQREFA{0}{}{0}{} , \\
\label{pert_cons_i}
   \left(\pezB + 3 \HcB + \UcB \right)
        \left((P + \rho) \IA v^\sharp - F \IA \dni^\sharp \right)
\nonumber \\
 + \left(\ppzN + 3 \HpN + \IpN \right)
   \left(F \IA v^\sharp + (P - \PRESY) \IA \dni^\sharp \right)
\nonumber \\
 + \delta P^\sharp + \frac{2}{3} (\DzAA + 3 \KzAA) \IAA \Pi
 + (P + \rho) \Psi + (\PRESY - P) \hpp + F \vp
 & = & \TEQREFA{0}{}{i}{} , \\
\label{pert_cons_5}
   (\pezB + 3 \HcB + 2 \UcB)
   \left(  \delta F^\sharp - F (\Psi + \hpp) - (\rho + \PRESY) \vp \right)
\nonumber \\
 + (\ppzN + 3 \HpN + 2 \IpN)
   (\delta \PRESY^\sharp - F \vp)
 - 3 \HpN \delta P^\sharp - \IpN (\delta \PRESY^\sharp - \delta \rho^\sharp)
\nonumber \\
 + \DzAA \left( F \IA v^\sharp + (P - \PRESY) \IA \dni^\sharp \right)
\nonumber \\
 + F \pezB (\Psi - \hpp - 3 \Phi)
 + 3 (P - \PRESY) \ppzN \Phi + (\rho + \PRESY) \ppzN \Psi + F \ppzN \vp
 & = & \TEQREFA{0}{}{\fin}{} ,
\end{eqnarray}
and one vector conservation equation,
\begin{eqnarray}
   \left(\pezB + 4 \HcB + \UcB \right)
   \left(  (P + \rho) (\VV{v}^\sharp_i - \vi_i)
         - F (\VV{\dni}^\sharp_i + \hci_i) \right)
\nonumber \\
 + \left(\ppzN + 4 \HpN + \IpN \right)
   \left(  F (\VV{v}^\sharp_i - \vi_i)
         + (P - \PRESY) (\VV{\dni}^\sharp_i + \hci_i) \right)
\nonumber \\
 + \DDEMI \left(\DzAA + 2 \KzAA \right) \IA \VV{\Pi}_i
 & = & \TEQREFA{0}{}{i}{} .
\end{eqnarray}
Finally, in order to close the system, we must specify an equation of
state for $\delta P^\sharp$, $\delta \PRESY^\sharp$, $\dni_i^\sharp$
and $\Pi_{i j}$, as functions either of $\delta \rho^\sharp$ or of some
other non dynamical variables (such as the entropy). For example, all
these quantities vanish for a bulk containing non relativistic
matter. For a scalar field, most of them are also set to zero, as is
discussed in Appendix~\ref{ssec_app_sf}. We can interpret the three
scalar equations (\ref{pert_cons_0}--\ref{pert_cons_5}) as the
conservation equations for $\delta \rho^\sharp$, $v^\sharp$, and
$\delta F^\sharp$.

Notice that with these Bianchi or conservation equations, three scalar
and one vector Einstein equations are redundant and can be
dropped. Formally, the seven scalar Einstein equations can be split
into four dynamical equations for the four scalar metric perturbations
$\Phi$, $\Psi$, $\vp$ and $\hpp$, and three constraint equations. It
happens, however, that with our choice of variables, the splitting is
not completely straightforward. For example, the $\{0 \alpha\}$
Einstein equations are the constraint equations for the metric
components $g_{i j}$, $g_{i 4}$, $g_{4 4}$, the rest being the
evolution equations.  However, in terms of our gauge invariant
variables, Eq.~(\ref{einpertijsst}) is obviously a constraint
equation. This is because $\Psi$ involves first and second time
derivatives of the metric perturbation $E$.  Equivalently, the three
vector Einstein equations can be split into one constraint equation
and two dynamical equations for the variables $\vi_i$ and $\hci_i$.

\section{Bulk perturbation with a brane}
\label{secVII}

In the previous section we have considered the most general perturbed
$5$-dimensional bulk spacetime for which there is a perturbed
maximally symmetric space orthogonal to the fifth direction. We have
seen that its dynamics can be described in terms of four geometrical
scalar perturbation variables governed by four evolution equations and
three constraints, two geometrical vector perturbation variables
governed by two evolution equations and a constraint, and one
geometrical tensor variable governed by one tensor evolution
equation. In this section we add a brane to this system --- that is we
assume, as was the case for the background, that the bulk contains a
perturbed homogeneous and isotropic brane as a singular source. This
will introduce one new geometrical degree of freedom, the brane
displacement, whose dynamics has to be considered in order to fully
describe the perturbations on the brane.

\subsection{Brane position and its displacement}

The perturbed brane embedding is given by
\begin{eqnarray}
\label{embpertdeb}
X^0 & = & \sigma^0 + \ZETA^0 (\sigma^a) , \\
X^i & = & \sigma^i + \ZETA^i (\sigma^a) , \\
\label{embpertfin}
X^\fin & = & \POSBRN + \UPS (\sigma^a) .
\end{eqnarray}
Here $\UPS$ is the displacement of the brane from its background
position $X^\fin = \POSBRN$, and it is a function on the brane
worldsheet. It is a true new degree of freedom which sometimes also
called ``radion'' \cite{radion,Deruelle}. On the contrary, as we will
soon see, the perturbations in the $X^0$ and $X^i$ directions can be
set to zero without loss of generality, as they do not lead to any
physical consequences (e.g., to a physical ``deformation'' of the
brane)~\cite{Deruelle} .

It was first noticed in \cite{Tanaka} that, when studying brane
perturbations of a Randall-Sundrum background (of Ref.~\cite{RS2}),
using the transverse and traceless gauge in Gaussian normal
coordinates, the brane position is no more at constant $\fd$ in
presence of matter sources. The presence of an $\UPS \neq 0$ can be
interpreted as a bending of the brane due to the presence of matter or
gravitational waves. The bending $\UPS$ can in principle be set to
zero by choosing a convenient set of bulk coordinates such that $\fd =
\POSBRN$, since by the infinitesimal coordinate transformation defined
in the previous section, Eq.~(\ref{cico}), one has
\begin{equation}
\label{UPStrans}
\UPS \to \UPS - \Lperp .
\end{equation}
This is not, however, the most general possibility.  As was noted
in~\cite{Carsten,Anisotropicstress}, such a gauge choice will also fix
the gauge of some other perturbation variables. In fact, if we choose
$\Lperp$ such that $\UPS$ is zero, $\ANEE$ is fixed (see
Eq.~(\ref{transmetfin})). In a gauge invariant approach one must keep
$\UPS$ arbitrary.

We now can define the perturbed vector orthogonal to the brane,
$\NORMVEC^\alpha + \DNORMVEC^\alpha$. One easily obtains (See
Appendix~\ref{ssec_app_pert_brane2})
\begin{equation}
\label{ups}
\DNORMVEC_\alpha
 = \left(- \IN \dot \UPS, - \IN \nabla_i \UPS,  - \IN \epp \right) .
\end{equation}
Note that no vector perturbations enter in the above formula. This is
just a consequence of Frobenius theorem~\cite{Wald}. Also, the fact
that the perturbations $\ZETA^a$ do not enter in the above expression
illustrates that they do not correspond to any physical deformation of
the brane (see further comments below).

\subsection{Induced metric and first fundamental form}

In this subsection we calculate the perturbed first fundamental
form which will be used in the perturbed first Israel junction
condition in the following subsections.

We shall first look at the perturbation of the induced metric
$\INDMET_{a b} (\sigma^a)$. In doing so, it is important to recall
that the brane embedding in the unperturbed and perturbed bulks
(respectively given by~(\ref{embdeb}--\ref{embfin}),
(\ref{embpertdeb}--\ref{embpertfin})) are different. Therefore, each
brane variable has two contribution to its perturbation: one coming
from the perturbation of the variable and a second contribution due to
the fact that we have to evaluate it at the perturbed brane
position. We obtain
\begin{eqnarray}
\DINDMET_{0 0} (\sigma^a)
 & = & 2 \IBB
       \left(A + \dot \ZETA^0 + \IcB \IB \ZETA^0 + \IpN \IN \UPS \right) , \\
\DINDMET_{0 i} (\sigma^a)
 & = & \IAB
       \left(B_i - \IAzB \dot \ZETA_i + \IBzA \nabla_i \ZETA^0 \right) , \\
\DINDMET_{i j} (\sigma^a)
 & = & - 2 \IAA
         \left(  \left(C + \HcB \IB \ZETA^0 + \HpN \IN \UPS \right) 
                 \gamma_{i j}
               + E_{i j} + \nabla_{(i} \ZETA_{j)} \right) ,
\end{eqnarray}
where, to first order in perturbation theory, the right hand side of
these equations are evaluated at $x^\alpha = X^\alpha$ (corresponding
to the unperturbed embedding). As usual, we can decompose the
perturbation $\ZETA^i$ into its scalar and vector parts using the
metric $\gamma_{i j}$ evaluated at the unperturbed brane position:
$\ZETA_i = \nabla_i \ZETA + \VV{\ZETA}_i$. This is possible since, for
perturbations, the time derivative $\pe$ and spatial derivatives
$\partial_i$ are equivalent to derivation with respect to $\sigma^0$
and $\sigma^i$ respectively.

A word of caution is in order here. Recall that the quantities $A$,
$C$, $\IpN$, $\HpN$, etc., are defined in the bulk.  However, due to
the presence of the brane, they may be (and in fact, are --- see
below) discontinuous at the brane position. For this reason, they may
only be evaluated on each side of the brane.  Therefore we must check
that the above expressions are consistent in the sense that they have
the same value when evaluated {\em on both sides} of the brane ---
only if that is the case, they can be considered well defined {\em on}
the brane. As we shall soon see, this consistency in fact results from
the first Israel condition.  Anticipating this result, the above
equations allow us to define in the standard way the brane
perturbations $\BRN{A}$, $\BRN{B}_i$, etc.,
\begin{eqnarray}
\BRN{A} & \equiv & A + \dot \ZETA^0 + \IcB \IB \ZETA^0 + \IpN \IN \UPS , \\
\BRN{B} & \equiv & B - \IAzB \dot \ZETA + \IBzA \ZETA^0 , \\
\BRN{\VV{B}}_i & \equiv & \VV{B}_i - \IAzB \dot{\VV{\ZETA}}_i , \\
\BRN{C} & \equiv & C + \HcB \IB \ZETA^0 + \HpN \IN \UPS , \\
\BRN{E} & \equiv & E + \ZETA , \\
\BRN{\VV{E}}_i & \equiv & \VV{E}_i + \VV{\ZETA}_i , \\
\BRN{\TT{E}}_{i j} & \equiv & \TT{E}_{i j} .
\end{eqnarray}
Using the standard $4$-dimensional perturbation theory, we construct
the two Bardeen potentials, as well as the brane vector and tensor
metric perturbations,
\begin{eqnarray}
\label{corina}
\BRNPsi
 & \equiv & \BRN{A} - \pezB \left(\IA \BRN{B} + \IAAzB \BRN{\dot E} \right)
   =   \Psi + \IpN \left(\IN \UPS - \NANEE \right) , \\
\BRNPhi
 & \equiv & - \BRN{C} + \HcB \left(\IA \BRN{B} + \IAAzB \BRN{\dot E} \right)
   =   \Phi - \HpN \left(\IN \UPS - \NANEE \right) , \\
\BRN{\vi}_i
 & \equiv & \BRN{\VV{B}}_i + \IAzB \dot{\BRN{\VV{E}}}_i
   =   \vi_i , \\
\label{corina2}
\BRN{\TT{E}}_{i j}
 & \equiv & \TT{E}_{i j} .
\end{eqnarray}
(The two first equations are equivalent to Eqns~(5.21) of
Ref.~\cite{bp8}.) Finally, using Eq.~(\ref{def_q}), the perturbed
first fundamental form is
\begin{eqnarray}
\label{gina}
\DFFF_{0 0} & = & 2 \IBB (A + \IpN \IN \UPS) , \\
\DFFF_{0 i} & = & \IAB B_i , \\
\DFFF_{i j} & = & - 2 \IAA (C + \HpN \IN \UPS) \gamma_{i j}
                  - 2 \IAA E_{i j} , \\
\label{rona}
\DFFF_{0 \fin}
 & = & - \IBN \left(\Bperp - \INzB \dot \UPS \right) , \\
\label{brigitta}
\DFFF_{i \fin} & = & \INA \left(\epi_i - \INzA \nabla_i \UPS \right) , \\
\label{corine} 
\DFFF_{\fin \fin} & = & 0 .
\end{eqnarray}
Notice that the $\ZETA^i$ do not appear in
equations~(\ref{corina}--\ref{corine}): this is again related to the
fact that they do not represent physical degrees of freedom. These
above expressions can also be obtained by starting from the definition
\begin{equation}
\FFF_{\alpha \beta} = g_{\alpha \beta} + \NORMVEC_\alpha \NORMVEC_\beta ,
\end{equation}
paying attention that in the perturbed and unperturbed cases, the bulk
metric is not evaluated at the same position ($\fd = \POSBRN + \UPS$
and $\fd = \POSBRN$ respectively)~\cite{Deruelle}.

\subsection{First Israel condition for the standard 4-dimensional
perturbation variables}
\label{ssec_reg_pert_geom}

Using Eqns~(\ref{transmetdeb}--\ref{transmetfin},\ref{UPStrans}),
the coordinate transformations of the  following variables are obviously
continuous as they do
not involve derivatives with respect to $\fd$:
\begin{eqnarray}
A + \IpN \IN \UPS & \to & A + \IpN \IN \UPS + \pezB (\IB T) , \\
B_i & \to & B_i - \IAzB \dot L_i + \IBzA \nabla_i T , \\
C + \HpN \IN \UPS & \to & C + \HpN \IN \UPS + \HcB \IB T , \\
E_i & \to & E_i + L_i , \\
\TT{E}_{i j} & \to & \TT{E}_{i j} , \\
\ABBE & \to & \ABBE + T .
\end{eqnarray}
Given the first fundamental form~(\ref{gina}--\ref{corine}), the first
Israel conditions imply that these quantities, which are linear
combinations of the components (\ref{gina}) to (\ref{rona}) of the
perturbed first fundamental form $\DFFF_{\alpha \beta}$, are
continuous,
\begin{eqnarray}
\DISC{A} + \DISC{\IpN \IN} \UPS & = & 0 , \\
\DISC{B_i} & = & 0 , \\
\DISC{C} + \DISC{\HpN \IN} \UPS & = & 0 , \\
\DISC{E_i} & = & 0 , \\
\DISC{\TT{E}_{i j}} & = & 0 , \\
\DISC{\ABBE} & = & 0 ,
\end{eqnarray}
or, equivalently
\begin{eqnarray}\
\label{Bardeeepsi}
\DISC{\Psi} & = & - \DISC{\IpN \IN \UPS - \IpN \NANEE } , \\
\label{Bardeeephi}
\DISC{\Phi} & = & \DISC{\HpN \IN \UPS - \HpN \NANEE} , \\
\DISC{\vi_i} & = & 0 , \\
\DISC{\TT{E}_{i j}} & = & 0 .
\end{eqnarray}
Hence tensor perturbations are continuous and only the two scalar
quantities $\vp$, $\hpp$ and the vector quantity $\hci_i$ may
jump. The first two equations are equivalent to
\begin{equation}
\DISC{\BRNPhi} = \DISC{\BRNPsi} = 0 ,
\end{equation}
which ensures that the brane Bardeen potentials $\BRNPhi$ and
$\BRNPsi$ are well defined.  We also see that the bulk vector and
tensor perturbation $\vi_i$ and $\TT{E}_{i j}$ may be defined on the
brane where they reduce to the standard vector and tensor metric
perturbations of a $4$-dimensional spacetime with maximally symmetric
spacelike hypersurfaces. Equivalently, the corresponding first
fundamental form can be rewritten as
\begin{eqnarray}
\DFFF_{0 0}
& = & 2 \IBB \BRNPsi + (\dot{\BRN q}_{0 0} + 2\BRN q_{0 0} \pe) \ABBE , \\
\DFFF_{0 i}
& = & - \IBzA \FFF_{i j} \vi^j
      + \FFF_{0 0} \nabla_i \ABBE +\BRN q_{i j} \dot E^j , \\
\DFFF_{i j}
& = &   2\BRN q_{k (i} \left(E_{j)}^k - \delta^k_{j)} \BRNPhi
                                \right) ,
\end{eqnarray}
which indeed reduce to $2 \IBB \BRNPsi$, $- \IBzA \FFF_{i j} \vi^j$
and $2 \IAA (\BRNPhi \gamma_{i j} - \TT{E}_{i j})$, in longitudinal
gauge ($B = E = 0$) for scalar perturbations and in the gauge
$\VV{E}_i=0$ for vector perturbations.

\subsection{Regularity conditions for coordinate transformations and
non-standard {\protect $4$}-dimensional perturbation variables}

So far we have given the relationship between the intrinsic brane
metric perturbations, the brane displacement $\UPS$, and some of the
bulk metric perturbations. Things become a little bit more involved
when we consider the other bulk metric perturbations that appear in
$\delta g_{\alpha\fin}$.

As we already noticed, the first Israel condition does not constrain
$\epp$ (see Eq.~(\ref{corine})). Nevertheless, since the
transformation law for $\epp$ can be written as
\begin{equation}
\epp \to \epp - \UcB \IB T - \ppzN (\IN \Lperp) ,
\end{equation}
and since all the metric coefficients must remain finite, it follows
that $\DISC{\IN \Lperp} = 0$ (see also
Appendix~\ref{sec_app_pert_bulk_geom}) . Furthermore, as the
coordinate transformation $x^\alpha \to x^\alpha + \xi^\alpha$ must be
invertible, we also require $\DISC{\Lperp} = 0$. Thus if $\IN$ is not
continuous, then $\Lperp (\POSBRN) = 0$:
\begin{equation}
\DISC{\IN} \neq 0
 \qquad \Rightarrow \qquad
\Lperp (\fd = \POSBRN) = 0 .
\end{equation}
This is the only additional requirement that one must impose on the
coordinate system in the vicinity of the brane. Note that if $Z_2$
symmetry is assumed, $\Lperp (\fd) = 0$ must be imposed even though
$\IN$ is continuous~\cite{Carsten}.

As mentioned in Section~\ref{ssec_Israel_and}, the only place where
discontinuities or singularities are allowed is the brane
position. When $\UPS = 0$, the brane is at $\fd = \POSBRN$.  However,
if the brane position is perturbed, $\UPS \neq 0$, the above
requirement implies
\begin{equation}
\DISC{\IN} \UPS = 0 ,
\end{equation}
and hence $\IN$ is not allowed to jump if the brane position is
perturbed. If the unperturbed metric has a zeroth order discontinuity
in the coefficient $\IN$, the brane position must remain at $\UPS = 0$
to first order perturbation theory (see Figure~\ref{fig4}). This
statement is in fact valid even in the absence of metric perturbations
(see Appendix~\ref{sec_app_pert_bulk_geom}).
\begin{figure}
\centerline{\psfig{file=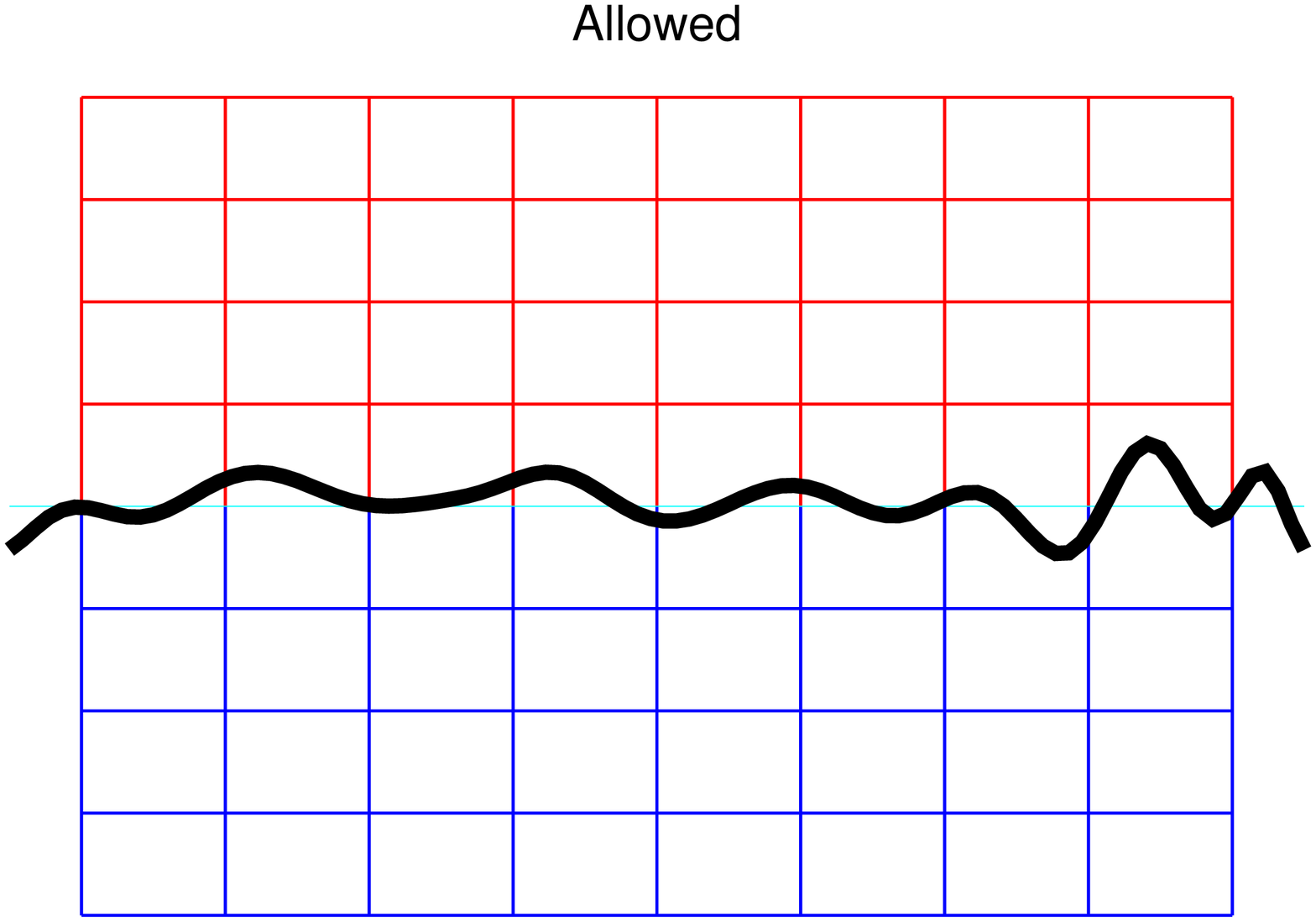,angle=270,width=3.5in}
            \psfig{file=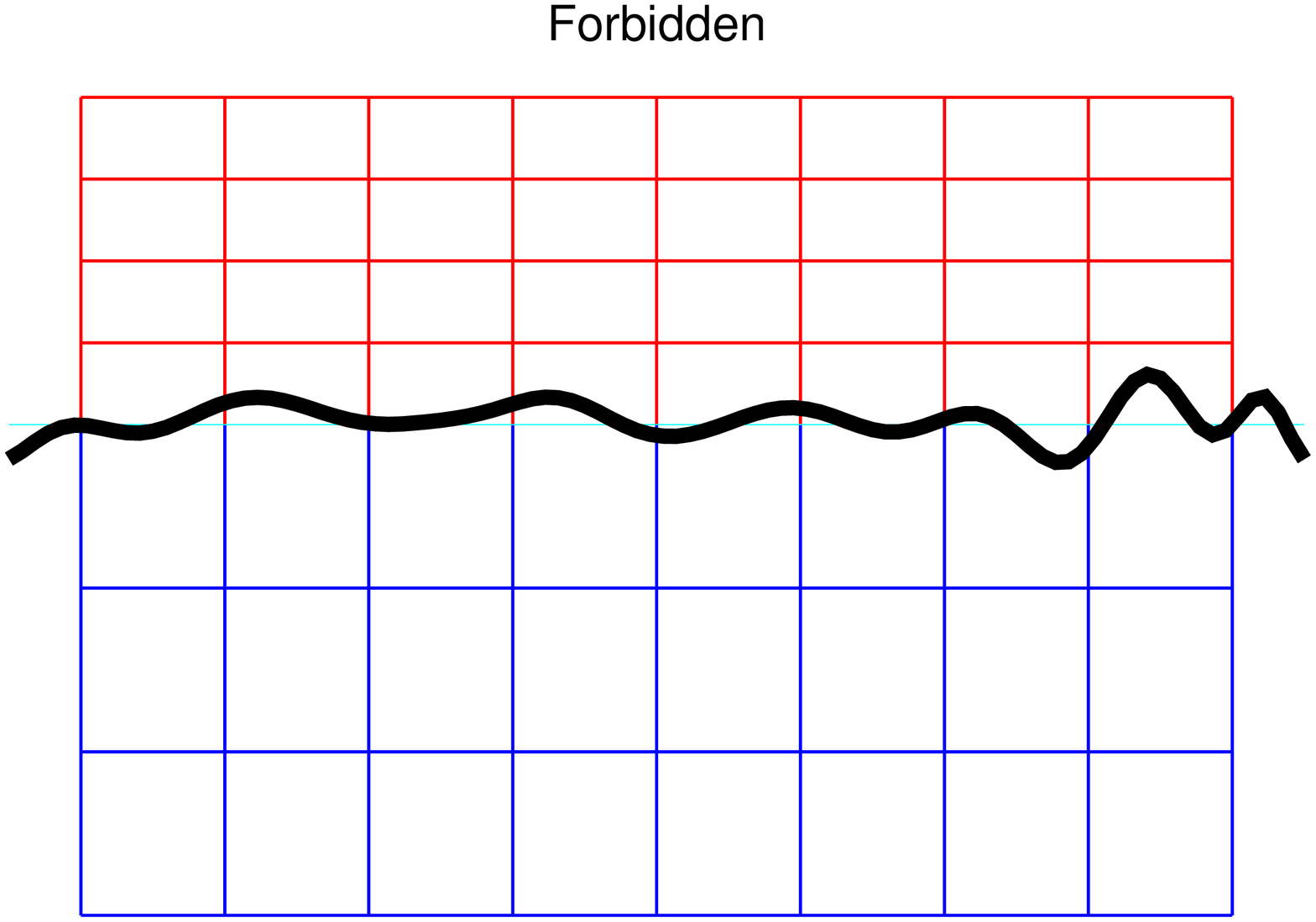,angle=270,width=3.5in}}
\caption{Illustration of the constraints on the brane position
$\UPS$ implied by the unperturbed ``scale factor'' $\IN$. When
$\DISC{\IN} = 0$ at zeroth order, then the brane position $\UPS$ can
be non zero at first order (left panel).  On the contrary, if
$\DISC{\IN} \neq 0$ at zeroth order, then the brane position must be
zero at first order, in order the perturbation theory to be valid, so
that right panel represents a forbidden situation. This is due to the
fact that in the situation of the right panel, an infinitesimal, first
order coordinate change could change by a large (zeroth order) amount
some perturbed metric coefficients in the vicinity of the brane.
(Here we have represented a situation corresponding to an unperturbed
bulk, but of course it also holds when it is perturbed. In fact the
perturbation of the brane position necessarily induces some metric
perturbations in the bulk.)} 
\label{fig4}
\end{figure}

Let us now consider the coordinate transformations of the variables
$\DFFF_{0 \fin}$ and $\DFFF_{i \fin}$ given in
Eqns~(\ref{rona},\ref{brigitta}),
\begin{eqnarray} 
\label{t0perp}
\IN \Bperp - \INNzB \dot \UPS
 & \to & \IN \Bperp - \INNzB \dot \UPS + \IB T' , \\
\IN \epi_i - \INNzA \nabla_i \UPS   
\label{tiperp}
 & \to & \IN \epi_i - \INNzA \nabla_i \UPS - \IA L'_i .
\end{eqnarray}
The first Israel condition also states that
\begin{eqnarray}
\DISC{\IN \Bperp} - \DISC{\INNzB} \dot \UPS
   =   \DISC{\IN \Bperp}
 & = & 0 , \\
\DISC{\IN \epi_i} - \DISC{\INNzA} \nabla_i \UPS
   =   \DISC{\IN \epi_i}
 & = & 0 .
\end{eqnarray}
Therefore, in order for the
transformations~(\ref{t0perp},\ref{tiperp}) to be continuous we have
to imply that a valid coordinate change satisfies
\begin{eqnarray}
\DISC{T'} & = & 0 , \\
\DISC{L^i{}'} & = & 0 .
\end{eqnarray}
These conditions ensure that $T$ and $L^i$ admit second
derivatives. This standard requirement for any valid coordinate
transformation is therefore preserved even in the presence of a brane.
In particular (see Eqns~\ref{transABBE},\ref{transE})), this means
that {\em if} $\ABBE'$ or $E^i{}'$ are discontinuous (which is
allowed), then $\ABBE$ and $E^i$ {\em cannot} be transformed
identically to zero by coordinate changes.  This does not prevent the
quantities $\Phi$, $\Psi$, $\vp$, $\hpp$ from being well defined.
However, it does mean that there may not be a coordinate system in
which Eqns~(\ref{newtdeb}--\ref{newtfin}) are valid (which we never
needed to suppose).

The first Israel condition does not require the continuity of $\vp$
and $\hci_i$, see Eqns~(\ref{isrvp},\ref{isrhi}) below. Finally, from
\begin{equation}
\NANEE \to \NANEE - \IN \Lperp ,
\end{equation}
and using the fact that $\DISC{\IN \Lperp} = 0$, the jump
\begin{equation}
\DISC{\NANEE} \equiv \BRNXI
\end{equation}
is gauge invariant. Therefore the gauge invariant quantity $\hpp$
given in Eqn~(\ref{vhpp}) may contain a singular part,
\begin{equation}
\label{singhpp}
\hpp =   \epp + \UcB \BABBE - \ppzN \CONT{\NANEE}
       - \DIR \BRNXI ,
\end{equation}
which again shows that $\hpp$ cannot always be a component of the
perturbed metric tensor in the vicinity of the brane. Of course, this
does not invalidate the results found previously, but simply suggests
that other variables may be more suitable to describe the metric
perturbations in the vicinity of the brane.

After these remarks on the regularity requirements of gauge
transformations in a bulk-brane system, we can now define some further
gauge invariant scalar variables in terms of which we will express the
second Israel condition. They will also be used to write the Einstein
equations for an observer on the brane when we want to compare our
results with the usual $4$-dimensional cosmological perturbation
theory.

Let us first define the gauge invariant combination
\begin{equation}
\label{defups}
\UPS^\sharp \equiv \UPS - \ANEE .
\end{equation}
Since $\ANEE$ can be discontinuous, $\UPS^\sharp$ is defined on each
side of the brane.  Note that we have
\begin{equation}
\DISC{\IN \UPS^\sharp} = - \BRNXI .
\end{equation}
Furthermore, we set
\begin{equation}
\CONT{\IN \UPS^\sharp} \equiv \BRUPS .
\end{equation}
When $\IN$ is continuous, $\BRUPS / \IN$ can be interpreted as the
``gauge invariant brane position'' --- that is, the position of the
brane unambiguously defined when $\CONT{\ANEE}$ is set to zero by a
suitable coordinate change (which always exists {\em if} $\IN$ is
continuous).

In principle, derivatives normal to the brane are not defined for
brane variables. But in what follows we will also use $\ppzN (\IN
\UPS^\sharp)$ which we simply define as
\begin{equation}
\label{UPnor1}
\ppzN (\IN \UPS^\sharp)
 \equiv \UpN \UPS - \ppzN \NANEE .
\end{equation}
In other words, the operator $\ppzN$ acts on every metric perturbation
defined in the bulk but not on $\UPS$ (i.e., we define $\ppzN \UPS
\equiv 0$).  Along similar lines, one can also define $\ppzN^2 (\IN
\UPS^\sharp)$ and $\ppzN^2 \BRUPS$.  The quantity $\ppzN (\IN
\UPS^\sharp)$ can contain a singular term because of the discontinuous
part of $\NANEE$.  Therefore, it is useful to define $\ppzN \BRUPS$ by
\begin{equation}
\label{UPnor2}
\ppzN \BRUPS
 \equiv \ppzN (\IN \UPS^\sharp) + \DIR \BRNXI .
\end{equation}
This new quantity can take different values on each side of the brane,
so that we can define $\DISC{\ppzN \BRUPS}$ and $\CONT{\ppzN \BRUPS}$
following Eqns~(\ref{discbrn},\ref{contbrn}).

We may not simply continue $\UPS$ into the bulk as a variable $\UPS$
is independent of $\fd$. This is a gauge dependent continuation and
the definitions for $\ppzN (\IN \UPS^\sharp)$ and $\ppzN^2 (\IN
\UPS^\sharp)$ given above would be valid only in the gauge where
$\UPS$ is independent of $\fd$. From the above expressions it is also
clear that the variables $\ppzN (\IN \UPS^\sharp)$ and $\ppzN^2 (\IN
\UPS^\sharp)$ and all brane perturbation variables which contain these
derivatives, like e.g. $\BRNhpp$ below, are gauge invariant only with
respect to gauge transformations parallel to the brane. Therefore, it
is important to keep in mind that these quantities are defined only on
(each side of) the brane, although the notation $\ppzN (\IN
\UPS^\sharp)$, $\ppzN \BRUPS$ may be slightly misleading. They simply
refer to Eqns~(\ref{UPnor1},\ref{UPnor2}).

Using equation~(\ref{defups}), one can define several gauge invariant
quantities which can also only be evaluated on either side of the
brane (that is either at $\fd = \POSBRN + \UPS{}^+$ or at $\fd =
\POSBRN + \UPS{}^-$)~:
\begin{eqnarray}
\BRNvp & \equiv & \Bperp - \IBzN{\ABBE}' - \INzB \dot{\UPS}
         = \vp - (\pezB - \UcB) (\IN \UPS^\sharp) , \\
\BRNhpp & \equiv & \epp +\UcB \BABBE - \UpN \IN \UPS
          =        \hpp - \ppzN (\IN \UPS^\sharp) .
\end{eqnarray}
By comparing the last equation to Eq.~(\ref{singhpp}), it appears that
$\BRNhpp$ does not contain a singular term (and hence it is a quantity
that has a meaning on each side of the brane).

Using Eqns~(\ref{Bardeeepsi},\ref{Bardeeephi}) we have
\begin{eqnarray}
\BRNPsi & = & \Psi + \IpN \IN \UPS^\sharp , \\
\BRNPhi & = & \Phi - \HpN \IN \UPS^\sharp .
\end{eqnarray}
The derivatives $\ppzN \BRNPsi$ and $\ppzN \BRNPhi$ are defined via
Eq.~(\ref{UPnor1},\ref{UPnor2}) above.  The above equations will
become very useful when considering the second Israel conditions and
writing down the perturbed Weyl tensor.

The first Israel condition states that $\BRNPhi$ and $\BRNPsi$ are defined
on the brane, and that $\ppzN \BRNPhi$, $\ppzN \BRNPsi$, $\BRNvp$, and
$\BRNhpp$ are well-defined on both sides of the brane, but it does not
imply their continuity. In fact, it is their discontinuity which will
enter into the perturbed second Israel condition. Using the above
definitions we have the following relations for the discontinuous and
the continuous parts of the gauge invariant scalar perturbation
variables:
\begin{eqnarray}
\DISC{\Psi}
 & = & - \DISC{\IpN} \BRUPS
       + \CONT{\IpN} \BRNXI , \\
\DISC{\Phi}
 & = &   \DISC{\HpN} \BRUPS
       - \CONT{\HpN} \BRNXI , \\
\DISC{\vp}
 & = &   \DISC{\BRNvp}
       - \left(\pezB - \CONT{\UcB} \right) \BRNXI
       - \DISC{\UcB} \BRUPS , \\
\DISC{\hpp}
 & = &   \DISC{\BRNhpp}
       + \DISC{\ppzN \BRUPS} ,
\end{eqnarray}
and
\begin{eqnarray}
\CONT{\Psi}
 & = &   \BRNPsi
       - \CONT{\IpN} \BRUPS
       + \frac{1}{4} \DISC{\IpN} \BRNXI , \\
\CONT{\Phi}
 & = &   \BRNPhi
       + \CONT{\HpN} \BRUPS
       - \frac{1}{4} \DISC{\HpN} \BRNXI , \\
\CONT{\vp}
 & = &   \CONT{\BRNvp}
       + \left(\pezB - \CONT{\UcB} \right) \BRUPS
       + \frac{1}{4} \DISC{\UcB} \BRNXI , \\
\CONT{\hpp}
 & = &   \CONT{\BRNhpp}
       + \CONT{\ppzN \BRUPS} .
\end{eqnarray}
If $Z_2$ symmetry is assumed, the first of these relations reduce to
\begin{eqnarray}
\DISC{\Psi} & = & - \DISC{\IpN} \BRUPS , \\
\DISC{\Phi} & = &   \DISC{\HpN} \BRUPS ,
\end{eqnarray}
and
\begin{eqnarray}
\CONT{\Psi} & = & \BRNPsi + \frac{1}{4} \DISC{\IpN} \BRNXI , \\
\CONT{\Phi} & = & \BRNPhi - \frac{1}{4} \DISC{\HpN} \BRNXI .
\end{eqnarray}
These relations are very important. They allow us to move freely
between the bulk (non underlined) perturbation variables, which are
defined everywhere, and the brane-related (underlined) variables,
which are well defined only on the brane, i.e. they are either defined
{\em on} the brane, like $\BRNPhi$ and $\BRNPsi$, or {\em on both
sides} of the brane, like $\IN \UPS^\sharp$, $\ppzN \BRNPhi$, $\ppzN
\BRNPsi$, $\BRNvp$, or $\BRNhpp$, etc.

The main difference between the brane and bulk perturbation variables
is $\UPS^\sharp$ which appears in the former.  The brane displacement,
however, is not unrelated to the bulk metric perturbations: a
displacement of the brane induces metric perturbation in the bulk (as
one could have guessed by making an analogy with a charged surface in
electromagnetism).

\subsection{Extrinsic curvature and second Israel condition}
\label{secVIII}

We define the perturbed stress-energy tensor on the brane as
\begin{equation}
\label{tensorbranebeg}
\BRN{\delta T}_{\alpha \beta} =
   (\BRN{\delta \rho} + \BRN{\delta P})\BRN{u}_\alpha \BRN{u}_\beta
 + 2 (\BRN{\rho} + \BRN{P}) \BRN{u}_{(\alpha} \BRN{\delta u}_{\beta)}
 - \BRN{\delta P} \FFF_{\alpha \beta}
 - \BRN{P} \DFFF_{\alpha \beta}
 + a^2 \BRN{\Pi}_{\alpha \beta} ,
\end{equation}
where $\BRN{\delta u}_\alpha$ is the perturbation of the energy
velocity on the brane which is given by
\begin{equation}
\BRN{\delta u}^\alpha
 = \left(- \IzB A, \IzA \BRN{v}^i, \INNzB \dot \UPS \right) .
\end{equation}
(The $\BRN{\delta u}^\fin$ component is determined by the condition
$(\NORMVEC_\mu + \DNORMVEC_\mu) (\BRN{u}^\mu + \BRN{\delta u}^\mu) =
0$.) The variable $\BRN{\Pi}_{\alpha \beta}$ is the anisotropic stress
tensor and it is gauge invariant by itself.

As discussed in Appendix~\ref{ssec_app_pert_stt}, we define the gauge
invariant perturbations for the energy density and the pressure on the
brane by
\begin{eqnarray}
\BRN{\delta \rho}^\sharp
 & = & \BRN{\delta \rho} - \dot {\BRN{\rho}} \ABBE , \\
\BRN{\delta P}^\sharp
 & = & \BRN{\delta P} - \dot {\BRN{P}} \ABBE .
\end{eqnarray}
Similarly we define the gauge invariant perturbation variables for
the velocity on the brane,
\begin{eqnarray}
\BRN{v}^\sharp & = & \BRN{v} + \IAzB \dot E , \\
\label{tensorbraneend}
\BRN{\VV{v}}^\sharp_i
 & = & \BRN{\VV{v}}_i + \IAzB \dot{\VV{E}}_i . 
\end{eqnarray}

To impose the second Israel junction condition, we need to compute
$\DCEX_{\alpha \beta}$ which is the difference between the perturbed
value of $\CEX_{\alpha \beta}$ at the brane position $\fd = \POSBRN +
\UPS$, and the background value of $\CEX_{\alpha \beta}$. The results
are given in Appendix~\ref{ssec_app_pert_cex}.  The extrinsic
curvature has to be compared to the perturbation of the surface stress
tensor,
\begin{equation}
\DSTT_{\alpha \beta}
 =   \BRN{\delta T}_{\alpha \beta}
   - \frac{1}{3} \BRN{\delta T} \FFF_{\alpha \beta}
   - \frac{1}{3} \BRN{T} \DFFF_{\alpha \beta} ,
\end{equation}
whose components are given in Appendix~\ref{ssec_app_pert_stt}.

The perturbation of the second Israel condition,
\begin{equation}
\label{pertks}
\DISC{\DCEX_{\alpha \beta}} = - \KAPPACQ \DSTT_{\alpha \beta} ,
\end{equation}
yields four discontinuity conditions for the scalar perturbation
variables $\BRNvp$, $\BRNhpp$, $\BRNXI$, and the first derivatives
$\ppzN \BRN{\Psi}$ and $\ppzN \BRN{\Phi}$,
\begin{eqnarray}
\label{isrrho}
 - \DzAA \BRNXI
 + 3 \DISC{\ppzN  \BRNPhi - \HpN \BRNhpp \HcB \DISC{\BRNvp}}
 & = & \KAPPACQ \BRN{\delta \rho}^\sharp , \\
\label{isrvp}
\DDEMI \DISC{\BRNvp - (\pezB + \UcB - 2 \HcB) (\IN \UPS^\sharp)}
 & = & \KAPPACQ (\BRN{P} + \BRN{\rho}) \IA \BRN{v}^\sharp , \\
\label{isrP}
   \DISC{\ppzN \BRNPsi + \IpN \BRNhpp - \left(\pezB + \UcB \right) \BRNvp}
 & = & \KAPPACQ \left(  \BRN{\delta P}^\sharp
                    + \frac{2}{3} \BRN{\delta \rho}^\sharp \right) , \\
\label{isrpi}
 - \BRNXI
 & = & \KAPPACQ \IAA \BRN{\Pi} .
\end{eqnarray}
In terms of the gauge invariant bulk variables these conditions
read
\begin{eqnarray}
\label{isrrho2}
 - \DzAA \BRNXI
 + 3 \DISC{\ppzN  \Phi - \HpN \hpp + \HcB \vp}
\nonumber \\
 + 3 \BRUPS \left(\HcB \DISC{\UcB} - \DISC{\ppzN} \CONT{\HpN} \right)
\nonumber \\ 
 - 3 \BRNXI \left(\HcB \CONT{\UcB} - \CONT{\ppzN} \CONT{\HpN} \right)
 + 3 \pezB \BRNXI
 & = & \KAPPACQ \BRN{\delta \rho}^\sharp , \\
\label{isrvp2}
\DDEMI \DISC{\vp} + (\pezB -  \HcB) \BRNXI
 & = & \KAPPACQ (\BRN{P} + \BRN{\rho}) \IA \BRN{v}^\sharp , \\
\label{isrP2}
   \DISC{\ppzN \Psi + \IpN \hpp - \left(\pezB + \UcB \right) \vp}
\nonumber \\
 + \BRUPS \left(\DISC{\ppzN} \CONT{\IpN} - \DISC{(\pezB + \UcB) \UcB} \right)
\nonumber \\ 
 - \BRNXI \left(\CONT{\ppzN} \CONT{\IpN} - \CONT{(\pezB + \UcB) \UcB} \right)
 - \pezB^2 \BRNXI
 & = & \KAPPACQ \left(  \BRN{\delta P}^\sharp
                    + \frac{2}{3} \BRN{\delta \rho}^\sharp \right) , \\
\label{isrpi2}
 - \BRNXI
 & = & \KAPPACQ \IAA \BRN{\Pi} .
\end{eqnarray}
Notice that when $Z_2$ symmetry is imposed $\BRUPS$ never appears in
these equations.

For the vector perturbation variables we obtain two discontinuity
conditions for $\hci_i$ and the first derivative $\ppzN \vi_i$,
\begin{eqnarray}
 - \DDEMI
   \DISC{\ppzN + \IpN - \HpN} \vi_i
 + \DDEMI \DISC{(\pezB + \UcB - \HcB) \hci_i}
 & = & \KAPPACQ
       (\BRN{P} + \BRN{\rho}) (\BRN{v}^\sharp_i - \vi_i) , \\
\label{isrhi}
 - \DISC{\hci_i}
 & = & \KAPPACQ \IA \BRN{\VV{\Pi}}{}_i .
\end{eqnarray}
Finally, there is also a discontinuity condition for the normal
derivative of the tensor perturbation variable $\TT{E}_{i j}$:
\begin{eqnarray}
 - \DISC{\ppzN \TT{E}_{i j}}
 & = & \KAPPACQ \BRN{\TT{\Pi}}{}_{i j} .
\end{eqnarray}
Note that the Israel conditions do not give any constraint on the
$\{\alpha \fin\}$ components of Eq.~(\ref{pertks}). The above
constraints can also be found directly from the singular part of
Einstein's equations~(\ref{einpert00}--\ref{einpertijt}). This is
relatively straightforward for the vector and tensor modes, but much
more involved for the scalar part, as one must rewrite the equations
using the underlined quantities defined above, and also because one
has to consider the perturbation of the covariant Dirac function
$\DIR$. However, for completeness, this has been undertaken in
Appendix~\ref{ssec_app_pert_newee}, and we have checked that both
approaches lead to the same result.

\subsection{Sail equation}
\label{ssec_sail}

As we have seen, the junction conditions are conveniently written
using the underlined variables $\BRNPhi$, etc. In order to use them,
we must know $\IN \UPS^\sharp$. The jump of this quantity, $\BRNXI$,
is given by Eq.~(\ref{isrpi}). As its continuous part $\BRUPS$
represents the brane displacement, it is natural to seek an equation
describing the brane motion. As for the unperturbed case, such an
equation is found by taking the discontinuous part of the $\{\fin
\fin\}$ component of Einstein's equations.  This yields
\begin{eqnarray}
\label{sailpert}
   3 \CONT{\HpN} \BRN{\delta P}^\sharp
 - \CONT{\IpN} \BRN{\delta \rho}^\sharp
\nonumber \\
 - 3 \BRN{P}
   \CONT{  \ppzN \BRNPhi - \HpN \BRNhpp + \HcB \BRNvp
         + \frac{1}{3} \DzAA \BRUPS }
\nonumber \\ 
 - \BRN{\rho}
   \CONT{  \ppzN \BRNPsi + \IpN \BRNhpp - \left(\pezB + \UcB \right) \BRNvp }
 & = & \DISC{\UDT{\delta \PRESY}^\sharp} , 
\end{eqnarray}
where $\UDT{\delta \PRESY}^\sharp$ corresponds to the pressure
perturbation along the extra dimension as measured by an observer at
rest with respect to the brane. The relationship between $\delta
\PRESY^\sharp$ and $\UDT{\delta \PRESY}^\sharp$ is given in
Eq.~(\ref{udtpresy}) below. Equation~(\ref{sailpert}) is the typical
equation for the displacement of a membrane (it involves the Laplacian
of the displacement $\BRUPS$).  When going back to the bulk (non
underlined) perturbations, Eq.~(\ref{sailpert}) becomes, as expected,
a wave equation for $\BRUPS$:
\begin{eqnarray}
\label{sail3}
 - \pezB^2 (\BRN{\rho} \BRUPS)
 - 3 \HcB \pezB \left(2 \BRN{\rho} \BRUPS + \BRN{P} \BRUPS \right)
 - \BRN{P}\DzAA \BRUPS
 + 2 \KzAA \BRN{\rho} \BRUPS
\nonumber \\
 - 3 \left(\BRN{P} + \frac{2}{3} \BRN{\rho} \right)
   \BRUPS \left(2 \pezB \HcB + 4 \HcB^2 \right)
\nonumber \\
 - (\BRN{P} + \BRN{\rho}) \BRUPS
   \left(  3 \CONT{\HpN} \CONT{\HpN - \IpN}
         + \frac{\KAPPACQ}{4} \BRN{\rho} (\BRN{P} + \BRN{\rho}) \right)
\nonumber \\
 + \BRUPS \left(  3 \CONT{\HpN} \DISC{\PRESY - P}
                  + \CONT{\IpN} \DISC{\PRESY + \rho} \right)
 & = &   \DISC{  \delta \PRESY^\sharp}
       - 3 \CONT{\HpN}\BRN{\delta P}^\sharp
       + \CONT{\IpN} \BRN{\delta \rho}^\sharp
\nonumber \\ & &
       + 3 \BRN{P} \CONT{  \ppzN \Phi - \HpN \hpp + \HcB \vp }
\nonumber \\ & &
       + \BRN{\rho}
         \CONT{  \ppzN \Psi + \IpN \hpp - \left(\pezB + \UcB \right) \vp }
\nonumber \\ & &
       + 2 \pezB \CONT{F} \BRNXI
       - \BRNXI \left(\pezB - N \HcB \right)  \CONT{F}
\nonumber \\ & &
       + 3 \CONT{\HpN} \BRNXI
         \left(  \frac{\KAPPACQ}{4} (\BRN{P} + \BRN{\rho})^2
               + \CONT{\PRESY - P} \right)
\nonumber \\ & & 
       + \CONT{\IpN} \BRNXI
         \left(- \frac{\KAPPACQ}{4} (\BRN{P} + \BRN{\rho}) \BRN{\rho}
               + \CONT{\PRESY + \rho} \right) .
\end{eqnarray}
Note that there is nothing which guarantees {\it a priori} that the
motion of the brane is stable. Even in the simplest case ($Z_2$
symmetry, $\COURB = 0$, no bulk perturbation contributing to the right
hand side of the above equation and brane stress-energy tensor
dominated by a constant tension term), this equation becomes
\begin{equation}
\left(  \partial_\eta^2 + 2 \frac{\dot \IA}{\IA} \partial_\eta
      - \nabla^2 - 2 \frac{\ddot \IA}{\IA} \right) \BRUPS = 0 ,
\end{equation}
and the mass term becomes negative for sufficiently fast expansion
rate of the brane!

\section{The brane point of view}
\label{secIX}

In the previous sections we have derived the bulk perturbation
equations and their boundary conditions on the brane. This allows
us in principle to solve the full system of perturbation equations
in the bulk for given initial conditions. From these one can
determine also the perturbed Weyl tensor and the second
fundamental form.

In order to make contact with $4$-dimensional cosmology in this
section, we want to write the perturbed version of the $4$-dimensional
Einstein equations on the brane. As for the background, this can
either be done directly from the perturbed bulk Einstein
equations~(\ref{einpert00}--\ref{einpertijt}), or using the
Gauss-Codacci equation.

\subsection{Projected Weyl tensor on the brane}

The full expression of the perturbed Weyl tensor $\delta C_{\alpha
\beta \gamma \delta}$ is given in
Appendix~\ref{ssec_app_pert_weyl}. Here we write only the components
of the perturbed projected Weyl tensor, $\DWN_{\alpha \beta}$, on the
spacelike direction $\NORMVEC^\alpha + \DNORMVEC^\alpha$, written in
terms of the underlined gauge invariant variables. We have
\begin{eqnarray}
\DWN_{0 0} & = &
   \frac{1}{2} \IBB \BRN{\delta \WEYL}^\sharp + 2 \WN_{0 0} \BRNPsi
\nonumber \\ & &
 + (\dot{\WN}_{0 0} + 2 \WN_{0 0} \pe) \ABBE , \\
\DWN_{0 i} & = &
 - \IB \nabla_i \DWN^v - \IAB \VV{\DWN}{}^v_i
\nonumber \\ & &
 + \WN_{0 0} \nabla_i \ABBE
 + \WN_{i j} \dot E^j , \\
\DWN_{i j} & = &
   \frac{1}{6} \IAA \gamma_{i j} \BRN{\delta \WEYL}^\sharp
\nonumber \\ & &
 + \left(\nabla_{i j} - \frac{1}{3} \nabla^2 \gamma_{i j} \right) \DWN^\Pi
 + \IA \nabla_{(i} \VV{\DWN}{}^\Pi_{j)}
 + \IAA \TT{\DWN}{}^\Pi_{i j}
\nonumber \\ & &
 + \dot{\WN}_{i j} \ABBE
 + 2 \WN_{k (i} (E^k_{j)} - \delta^k_{j)} \BRNPhi) , \\
\DWN_{0 \fin} & = &
   \WN^0_0 \DFFF_{0 \fin} , \\
\DWN_{i \fin} & = &
   \WN^j_i \DFFF_{j \fin} , \\
\DWN_{\fin \fin} & = & 0 ,
\end{eqnarray}
where we have set
\begin{eqnarray}
\BRN{\delta \WEYL}^\sharp
 & \equiv &   \frac{2}{3} (\DzAA + 3 \KzAA) \BRNPhi
            + \frac{1}{3} \DzAA (\BRNPsi - \BRNhpp)
\nonumber \\ & &
            + \left(\pezB + \UcB \right)
              \left(\pezB \BRNPhi + \HcB \BRNPsi \right)
            - \left(\pezB + 2 \UcB - \HcB \right)
              \left(\pezB \BRNhpp + \UcB \BRNPsi \right)
\nonumber \\ & &
            - \left(\ppzN + \IpN \right)
              \left(  \ppzN \BRNPhi - \HpN \BRNhpp + \HcB \BRNvp
                    + \frac{1}{3} \DzAA (\IN \UPS^\sharp) \right)
\nonumber \\ & &
            - \left(\ppzN + 2 \IpN - \HpN \right)
              \left(  \ppzN \BRNPsi + \IpN \BRNhpp
                    - \left(\pezB + \UcB \right) \BRNvp \right)
\nonumber \\ & &
            - \BRNPsi \pezB \left(\UcB - \HcB \right)
            - \left(\BRNhpp \ppzN - \BRNvp \pezB \right)
              \left(\IpN - \HpN \right) , \\
\DWN^v
 & \equiv &   \frac{2}{3}
              \left(- \left(\HcB \BRNPsi + \pezB \BRNPhi \right)
                    + \left(\pezB \BRNhpp + \UcB \BRNPsi \right)
                    + \left(\UcB - \HcB \right) \BRNhpp
                    + \left(\HpN - \IpN \right) \BRNvp
\right. \nonumber \\ & & \left. \qquad \qquad
                    - \TDEMI \ppzN
                      \left(  \BRNvp
                            - \left(\pezB + \UcB - 2 \HcB \right)
                              (\IN \UPS^\sharp) \right) \right) , \\
\DWN^\Pi
 & \equiv & \frac{1}{3}
            \left(  \BRNPhi - \BRNPsi - 2 \BRNhpp
                  + (\HpN + \IpN - 2 \ppzN) (\IN \UPS^\sharp)
            \right) .
\end{eqnarray}
Since, by solving the bulk equations, we in principle obtain the
non underlined variables, it is useful to express the above
components also in terms of these,
\begin{eqnarray}
\BRN{\delta \WEYL}^\sharp
 & = &   \frac{2}{3} (\DzAA + 3 \KzAA) \Phi
       + \frac{1}{3} \DzAA (\Psi - \hpp)
\nonumber \\ & &
       + \left(\pezB + \UcB \right)
         \left(\pezB \Phi + \HcB \Psi \right)
       - \left(\pezB + 2 \UcB - \HcB \right)
         \left(\pezB \hpp + \UcB \Psi \right)
\nonumber \\ & &
       - \left(\ppzN + \IpN \right)
         \left(  \ppzN \Phi - \HpN \hpp + \HcB \vp \right)
\nonumber \\ & &
       - \left(\ppzN + 2 \IpN - \HpN \right)
         \left(  \ppzN \Psi + \IpN \hpp
               - \left(\pezB + \UcB \right) \vp \right)
\nonumber \\ & &
       - \Psi \pezB \left(\UcB - \HcB \right)
       - \left(\hpp \ppzN - \vp \pezB \right)
         \left(\IpN - \HpN \right)
\nonumber \\ & &
       + \WEYL' \UPS^\sharp , \\
\DWN^v
 & = &   \frac{2}{3}
         \left(- \left(\HcB \Psi + \pezB \Phi \right)
               + \left(\pezB \hpp + \UcB \Psi \right)
               + \left(\UcB - \HcB \right) \hpp
               - \left(\TDEMI \ppzN + \IpN - \HpN \right) \vp \right) , \\
\DWN^\Pi
 & = & \frac{1}{3} \left(  \Phi - \Psi - 2 \hpp \right) .
\end{eqnarray}
For the vector and tensor part of the projected Weyl tensor we have defined
\begin{eqnarray}
\VV{\DWN}{}^v_i
 & \equiv & - \frac{1}{6} (\DzAA + 2 \KzAA) \vi_i
            + \frac{1}{3}
              \left( \ppzN + \HpN \right)
              \left(  \left(\ppzN + \IpN - \HpN \right) \vi_i
                    - \left(\pezB + \UcB - \HcB \right) \hci_i \right) , \\
\VV{\DWN}{}^\Pi_i
 & \equiv & \frac{1}{3}
            \left(  \left(\pezB + 2 \left(\HcB - \UcB \right)
                    \right) \vi_i
                  + \left(2 \ppzN + \left(\HpN - \IpN \right)
                    \right) \hci_i \right) ,
\end{eqnarray}
\begin{eqnarray}
\TT{\DWN}{}^\Pi_{i j}
 & \equiv & \frac{1}{3}
            \left(  \left(  \pezB + 3 \HcB - 2 \UcB \right) \pezB
                  - \left(\DzAA - 2 \KzAA \right)
                  + \left(  2 \ppzN + 3 \HpN - \IpN \right) \ppzN
            \right) \TT{E}_{i j} .
\end{eqnarray}

\subsection{Perturbed Einstein equations on the brane}

With these definitions, we can now write the
projected perturbed Einstein equations on the brane. They split into
four scalar equations,
\begin{eqnarray}
   2 \left(\DzAA + 3 \KzAA \right) \BRNPhi
\nonumber \\
 - 6 \HcB
   \left(\HcB \BRNPsi + \pezB \BRNPhi \right)
 & = &   \frac{1}{6} \KAPPACQ^2
         \left( \sum_\SBRN \BRN{\rho}_\SBRN \right)
         \sum_\SBRN \BRN{\delta \rho}_\SBRN^\sharp
\nonumber \\ & &
       - 2 \CONT{\HpN}
         \CONT{  3 \ppzN \BRNPhi- 3 \HpN \BRNhpp + 3 \HcB \BRNvp
               + \DzAA \IN \UPS^\sharp  }
\nonumber \\ & &
       + \frac{1}{2} \KAPPACQ
         \sum_\SBLK \CONT{  \UDT{\delta P}^\sharp_\SBLK
                          + \UDT{\delta \rho}^\sharp_\SBLK
                          - \UDT{\delta \PRESY}^\sharp_\SBLK }
       + \frac{1}{2} \CONT{\BRN{\delta \WEYL}^\sharp} , \\
 - 2 \left(\HcB \BRNPsi + \pezB \BRNPhi \right)
 & = &   \frac{1}{6} \KAPPACQ^2
         \left( \sum_\SBRN \BRN{\rho}_\SBRN \right)
         \sum_\SBRN \left((\BRN{P}_\SBRN + \BRN{\rho}_\SBRN)
                          \IA v_\SBRN^\sharp \right)
\nonumber \\ & &
       - \frac{3}{2} \CONT{\HpN}
         \CONT{\BRNvp - (\pezB + \UcB - 2 \HcB) (\IN \UPS^\sharp)}
\nonumber \\ & &
       + \frac{2}{3} \KAPPACQ
         \sum_\SBLK \CONT{  (P_\SBLK + \rho_\SBLK) v_\SBLK^\sharp
                          - F_\SBLK \UDT{\dni}^\sharp_\SBLK }
       + \CONT{\DWN^v} , \\
 + \frac{2}{3} \left(\DzAA \BRNPsi - (\DzAA + 3 \KzAA) \BRNPhi \right)
\nonumber \\
 + 2 \left(\pezB + 3 \HcB \right)
   \left(\HcB \BRNPsi + \pezB \BRNPhi \right)
\nonumber \\
 + 2 \BRNPsi \pezB \HcB
 & = &   \frac{1}{6} \KAPPACQ^2
         \left( \sum_\SBRN \BRN{\rho}_\SBRN \right)
         \sum_\SBRN \BRN{\delta P}_\SBRN^\sharp
\nonumber \\ & &
       - 2 \CONT{\HpN}
         \CONT{  \ppzN \BRNPsi + \IpN \BRNhpp - (\pezB + \UcB) \BRNvp }
\nonumber \\ & &
       + 2 \CONT{ \HpN + \IpN}
         \CONT{  \ppzN \BRNPhi - \HpN \BRNhpp + \HcB \BRNvp
               + \frac{1}{3} \DzAA \IN \UPS^\sharp }
\nonumber \\ & &
       + \frac{1}{6} \KAPPACQ
         \sum_\SBLK \CONT{  \UDT{\delta P}^\sharp_\SBLK
                          + \UDT{\delta \rho}^\sharp_\SBLK
                          + 3 \UDT{\delta \PRESY}^\sharp_\SBLK }
\nonumber \\ & &
       + \frac{1}{6} \KAPPACQ^2
         \left( \sum_\SBRN (\BRN{P}_\SBRN + \BRN{\rho}_\SBRN) \right)
         \sum_\SBRN \BRN{\delta \rho}_\SBRN^\sharp
       + \frac{1}{6} \CONT{\BRN{\delta \WEYL}^\sharp} , \\
 \BRNPhi - \BRNPsi
 & = &   \frac{1}{6} \KAPPACQ^2
         \left(\sum_\SBRN \BRN{\rho}_\SBRN \right)
         \IAA \sum_\SBRN \BRN{\Pi}_\SBRN
\nonumber \\ & &
       - \CONT{ \HpN + \IpN}
         \BRUPS
\nonumber \\ & &
       - \frac{1}{4} \KAPPACQ^2
         \left(\sum_\SBRN (\BRN{P}_\SBRN + \BRN{\rho}_\SBRN)
         \right) \IAA \sum_\SBRN \BRN{\Pi}_\SBRN
\nonumber \\ & &
       + \frac{2}{3} \KAPPACQ \IAA \sum_\SBLK \CONT{\Pi_\SBLK}
       + \CONT{\DWN^\Pi} ,
\end{eqnarray}
two vector equations,
\begin{eqnarray}
 - \DDEMI \left(\DzAA + 2 \KzAA \right) \vi_i
 & = &   \frac{1}{6} \KAPPACQ^2
         \left( \sum_\SBRN \BRN{\rho}_\SBRN \right)
         \sum_\SBRN \left((\BRN{P}_\SBRN + \BRN{\rho}_\SBRN)
                          (\VV{v}_i^{\SBRN\;\sharp} - \vi_i) \right)
\nonumber \\ & &
       + \frac{2}{2} \CONT{\HpN}
         \CONT{  (\ppzN + \IpN - \HpN) \vi_i
               - (\pezB + \UcB - \HcB) \hci_i}
\nonumber \\ & &
       + \frac{2}{3} \KAPPACQ
         \sum_\SBLK \CONT{  (P_\SBLK + \rho_\SBLK)
                            (\VV{v}^{\SBLK\;\sharp}_i - \vi_i)
                          - F_\SBLK (\VV{\dni}^{\SBLK\;\sharp}_i + \hci_i) }
       + \CONT{\VV{\DWN}{}^v_i} , \\
\left(\pezB + 2 \HcB \right) \vi_i
 & = &   \frac{1}{6} \KAPPACQ^2
         \left( \sum_\SBRN \BRN{\rho}_\SBRN \right)
         \IA \sum_\SBRN \BRN{\VV{\Pi}}^\SBRN_i
\nonumber \\ & &
       + \CONT{ \HpN + \IpN}
         \CONT{\hci_i}
\nonumber \\ & &
       - \frac{1}{4} \KAPPACQ^2
         \left( \sum_\SBRN (\BRN{P}_\SBRN + \BRN{\rho}_\SBRN) \right)
         \IA \sum_\SBRN \BRN{\VV{\Pi}}^\SBRN_i
\nonumber \\ & &
       + \frac{2}{3} \KAPPACQ \IA \sum_\SBLK \CONT{\VV{\Pi}^\SBLK_i}
       + \CONT{\VV{\DWN}{}^\Pi_i} ,
\end{eqnarray}
and one tensor equation
\begin{eqnarray}
   \left(\pezB + 3 \HcB \right)
   \pezB \TT{E}_{i j}
 - \left(\DzAA - 2 \KzAA \right) \TT{E}_{i j}
 & = &   \frac{1}{6} \KAPPACQ^2
         \left( \sum_\SBRN \BRN{\rho}_\SBRN \right)
         \sum_\SBRN \BRN{\TT{\Pi}}{}^\SBRN_{i j}
\nonumber \\ & &
       + \CONT{ \HpN + \IpN}
         \CONT{\ppzN \TT{E}_{i j}}
\nonumber \\ & &
       - \frac{1}{4} \KAPPACQ^2
         \left( \sum_\SBRN (\BRN{P}_\SBRN + \BRN{\rho}_\SBRN) \right)
         \sum_\SBRN \BRN{\TT{\Pi}}{}^\SBRN_{i j}
\nonumber \\ & &
       + \frac{2}{3} \KAPPACQ \sum_\SBLK \CONT{\TT{\Pi}{}^\SBLK_{i j}}
       + \CONT{\TT{\DWN}{}^\Pi_{i j}} ,
\end{eqnarray}
where we have set, for the bulk matter quantities evaluated at the
brane position,
\begin{eqnarray}
\UDT{\delta \rho}^\sharp
 & =  &   \delta \rho^\sharp
        + \rho' \UPS^\sharp
        - 2 F \INzB \dot \UPS^\sharp , \\
\UDT{\delta P}^\sharp
 & = &   \delta P^\sharp
       + P' \UPS^\sharp , \\
\UDT{\delta F}^\sharp
 & = &   \delta F^\sharp
       + F' \UPS^\sharp
       - (\rho + \PRESY) \INzB \dot \UPS^\sharp , \\
\label{udtpresy}
\UDT{\delta \PRESY}^\sharp
 & =  &   \delta \PRESY^\sharp 
        + \PRESY' \UPS^\sharp
        - 2 F \INzB \dot \UPS^\sharp , \\
\IA \UDT{\dni}^\sharp
 & = &   \IA \dni^\sharp
       - \IN \UPS^\sharp .
\end{eqnarray}
These corrections follow from the fact that we have to go in a
coordinate system which follows the brane.  The terms proportional to
$X' \UPS^\sharp$ are here because we consider the bulk matter content
at $\fd = \POSBRN + \UPS$ rather than at $\fd = \POSBRN$, the terms
proportional to $\dot \UPS^\sharp$ come from the fact that we also
perform a Lorentz boost in order to follow the brane motion, and the
term $\IN \UPS^\sharp$ in the last equation comes from the fact that
the brane is bent.

As for the unperturbed case, the continuous parts of the bulk
stress-energy tensor and of the projected Weyl tensor appear on
the right hand side of these equations, as well as the components
of the continuous part of the perturbed extrinsic curvature. These
are related through the discontinuous part of the Einstein
equations to the discontinuity of the perturbed Weyl tensor and of
the bulk perturbed matter content. The corresponding equations can
be found in Appendix~\ref{ssec_app_pffffff}.

\subsection{Perturbed conservation equation}

The brane matter conservation equations follow from the singular part
of $D_\mu T^{\mu \alpha} = 0$ or from the Bianchi identities. One
obtains (see Appendix~\ref{ssec_app_pert_cons})
\begin{eqnarray}
   \pezB \BRN{\delta \rho}^\sharp
 + 3 \HcB (  \BRN{\delta \rho}^\sharp
                        + \BRN{\delta P}^\sharp)
\nonumber \\
 + (\BRN{P} + \BRN{\rho})
   \DzAA \IA \BRN{v}^\sharp
 - 3 (\BRN{P} + \BRN{\rho}) \pezB \BRNPhi
 & = & - \DISC{\UDT{\delta F}^\sharp + F \BRNPsi} , \\
   \left(  \pezB + 3 \HcB \right)
   \left((\BRN{P} + \BRN{\rho}) \IA \BRN{v}^\sharp \right)
 + \BRN{\delta P}^\sharp
\nonumber \\
 + \frac{2}{3} \left(\DzAA + 3 \KzAA \right) \IAA \BRN{\Pi}
 + (\BRN{P} + \BRN{\rho}) \BRNPsi
 & = & - \DISC{  F \IA v^\sharp
               + (P - \PRESY)
                 \IA \UDT{\dni}^\sharp} , \\
  \left(\pezB + 4 \HcB \right)
  \left(  (\BRN{P} + \BRN{\rho})
          (\VV{\BRN{v}}^\sharp_i - \vi_i)  \right)
\nonumber \\
 + \DDEMI \left(\DzAA + 2 \KzAA \right) \IA \VV{\BRN{\Pi}}_i
 & = & - \DISC{  F (\VV{v}^\sharp_i - \vi_i)
               + (P - \PRESY)
                 (\VV{\dni}^\sharp_i + \hci_i) } .
\end{eqnarray}
Again, when there are no discontinuities in the bulk matter
perturbations, one obtains the usual conservation equations.

\section{Conclusion}
\label{conclusions}

In this paper we have derived gauge invariant cosmological
perturbation theory in braneworld scenarios with one codimension. The
unperturbed background system we considered
(Sections~\ref{secII}--\ref{secIV}) consists of a $5$-dimensional bulk
spacetime with a maximally symmetric $3$-dimensional subspace of
curvature $\COURB$, containing arbitrary (possibly interacting) matter
with energy-momentum tensor $T_{\alpha \beta}$, and a homogeneous and
isotropic $3$-brane again with arbitrary stress energy tensor
$\BRN{T}_{\alpha \beta}$. We have not assumed $Z_2$ symmetry across
the brane. As such, our work generalises that of previous authors who
have considered perturbation theory mainly in the $Z_2$-symmetric
case, and with specific bulk (and brane) matter (e.g., a bulk
cosmological constant~\cite{bp8} or scalar field~\cite{fields}). We
believe that the general setup considered here is a necessary
component of any serious attempt which may be made to tackle such
important questions as the cosmic microwave background anisotropies in
braneworlds.

The only coordinate choice we have made is to fix the unperturbed
brane to be at a given position $\POSBRN$ in the extra dimension. The
bulk metric is explicitly time-dependent. When the bulk contains only
a cosmological constant, this is not the most natural coordinate
system: there one would work with (static) Schwarzschild-AdS$_5$ and a
dynamical brane~\cite{Christos,Kraus,Ida,Muko}.  However, in the case
of arbitrary bulk matter and especially for the study of
perturbations, we have found it more convenient to work in a
coordinate system in which the brane is at rest.

In Sections~\ref{secII}--\ref{secIV} we derived all the relevant
background equations, ending with the brane Friedmann
equation~(\ref{bf1}--\ref{bf2}). As discussed in Section~\ref{secIV},
when $Z_2$ symmetry is not assumed, one has additional contributions
to the $4$-dimensional Einstein tensor on the brane. In order to study
these terms one has to include equations for the extrinsic curvature.

In the remainder of the paper we studied perturbations of this system,
setting up a completely gauge invariant formalism.
Section~\ref{secVI} contains a general discussion of the
classification of perturbations in an $n+1$-dimensional space time, as
well as the interplay between bulk and brane perturbations.  An
important point which we note there is the existence of one extra
scalar degree of freedom on the brane which is not a metric
perturbation (although it interacts with some bulk metric
perturbations): this is the brane displacement. In
Section~\ref{secVII} we have derived an equation of motion for the
gauge invariant perturbation variable describing this quantity.

In Section~\ref{secVI} we introduced the perturbed $5$-dimensional
bulk spacetime. This led to the definition of four scalar, two vector
and one tensor gauge invariant bulk perturbation variables given in
equations~(\ref{vPsi}--\ref{vhci}).  Following the definition of gauge
invariant variables for the perturbations of the bulk matter, we were
able to write down the perturbed bulk Einstein equations in a gauge
invariant manner.  The perturbed brane was then introduced in
Section~\ref{secVII}.  In analogy with usual $4$-dimensional
cosmological perturbation theory, our aim was to introduce two scalar
gauge invariant brane perturbation variables (the Bardeen potentials),
one vector and one tensor metric perturbation.  The correct definition
of these variables can only be given once the perturbed brane metric
and Israel junction conditions are used determine the brane variables
in terms of the continuous part and the jump of the bulk
perturbations. The brane variables are defined in
equations~(\ref{corina}--\ref{corina2}). The gauge invariant brane
displacement also enters in these definitions.  Finally the perturbed
Einstein equations on the brane were derived in
Section~\ref{secIX}. As in the unperturbed case, they contain a
contribution from the projection of the perturbed bulk Weyl tensor
which in general have to be determined by solving the bulk equations.

Despite the fact that we have tried to present our results as clearly
as possible, the formalism presented in this paper is technically
rather complicated.  This reflects the fact that we have considered a
very general scenario. The corollary is, however, that our results
should be applicable to a whole variety of different (and possibly
simpler) situations of interest in braneworld scenarios.  In a
forthcoming paper, we plan to apply this formalism to a specific model
and solve some of the perturbation equations presented here.

\acknowledgments

We would like to thank Francis Bernardeau, Timon Boehm, Philippe
Brax, David Langlois, Roy Maartens and Jean-Philippe Uzan for
enlightening discussions.  A.R.\ thanks Chlo\'e Riazuelo for
pointing out an error in the perturbed Weyl tensor~(\ref{cloclo}).
R.D.\ and F.V.\ thank the Intitute for Advanced Study for
hospitality. F.V.\ and D.S.\ thank the Swiss Foundation Ernst et
Lucie Schmidheiny for travel support. D.S.\ is grateful to the
Rockefeller Foundation for wonderful hospitality in Bellagio,
Italy, during the final stages of this work. This work is
supported by the Swiss National Science Foundation.

\appendix

\newAPP

\section*{Appendix}

In this Appendix, we give all the necessary formulae that were
used to obtain the results presented in the text.

Here we will consider an $N+1+1$-dimensional bulk with,
$N$-dimensional maximally symmetric, spacelike hypersurfaces of
constant curvature $\COURB$. Therefore, here $\alpha = 0$, $1$,
$\ldots$, $\SSfinN$, where $\SSfinN = N + 1$, and $i = 1$, $2$,
$\ldots$, $N$.  We will consider both the unperturbed
(Section~\ref{sec_app_bulk_geom}--\ref{sec_app_brane_pov}) and
perturbed cases
(Section~\ref{sec_app_pert_bulk_geom}--\ref{sec_app_pert_brane_pov}).
Both the bulk and the brane matter content are arbitrary as well as
the global geometry of the bulk. Furthermore, we do not assume $Z_2$
symmetry.

\APPsection{Some useful formulae}

\subAPPsection{Some ensor definitions and sign conventions}

Following the definitions of~\cite{MTW}, we use the sign convention
$(-++)$, that is the signature of the metric is $(+-\ldots-)$, and the
Riemann and Ricci tensors are respectively defined by
\begin{eqnarray}
R^\alpha{}_{\beta \gamma \delta}
 & = &   \partial_\gamma \Gamma^{\;\alpha}_{\beta \delta}
       - \partial_\delta \Gamma^{\;\alpha}_{\beta \gamma}
       + \Gamma^{\;\alpha}_{\gamma \mu} \Gamma^{\;\mu}_{\beta \delta}
       - \Gamma^{\;\alpha}_{\delta \nu} \Gamma^{\;\nu}_{\beta \gamma} , \\
R_{\alpha \beta}
 & = & R^\mu{}_{\alpha \mu \beta} .
\end{eqnarray}
The Weyl tensor is defined by
\begin{eqnarray}
C_{\alpha \beta \gamma \delta}
 & = &   R_{\alpha \beta \gamma \delta}
       - \frac{1}{N} \left(  R_{\alpha \gamma} g_{\beta \delta}
                           - R_{\alpha \delta} g_{\beta \gamma}
                           + R_{\beta \delta} g_{\alpha \gamma}
                           - R_{\beta \gamma} g_{\alpha \delta} \right) 
\nonumber \\ & & 
       + \frac{1}{N (N + 1)} R 
         (  g_{\alpha \gamma} g_{\beta \delta}
          - g_{\alpha \delta} g_{\beta \gamma} ) .
\end{eqnarray}

\subAPPsection{Brane-related metric quantities}

The induced metric is defined by
\begin{equation}
\INDMET_{a b} \equiv g_{\mu \nu} \; \partial_a X^\mu \; \partial_b X^\nu .
\end{equation}
The metric can be projected back in the bulk to give the first
fundamental form
\begin{equation}
\FFF^{\alpha \beta}
 \equiv \INDMET^{p q} \partial_p X^\alpha \; \partial_q X^\beta .
\end{equation}
More generally, any tensor $\BRN{X}_{a_1 \ldots a_n}$ defined for the
brane can be projected back in the bulk using
\begin{equation}
\BRN{X}^{\alpha_1 \ldots \alpha_n} 
 = \BRN{X}^{p_1 \ldots p_n} \;   
   \partial_{p_1} X^{\alpha_1} \; \ldots \;
   \partial_{p_n} X^{\alpha_n} .
\end{equation}
In particular, for the brane Riemann tensor, one has
\begin{equation}
\BRN{R}^{\alpha \beta \gamma \delta}
 = \partial_p X^\alpha \; \partial_q X^\beta \;
   \partial_r X^\gamma \; \partial_s X^\delta \;
   \BRN{R}_{p q r s} .
\end{equation}
One can also define the normal spacelike unit vector $\NORMVEC_\alpha$
to the brane according to
\begin{equation}
\label{app_norm_vec_def}
\NORMVEC_\mu \partial_a X^\mu = 0 
\quad , \qquad
\NORMVEC_\mu \NORMVEC^\mu = - 1 ,
\end{equation}
and the bulk metric evaluated at the brane position can be split into
\begin{equation}
g_{\alpha \beta} = \FFF_{\alpha \beta} - \NORMVEC_{\alpha \beta} , 
\end{equation}
with
\begin{equation}
\NORMVEC_{\alpha \beta} \equiv \NORMVEC_\alpha \NORMVEC_\beta . 
\end{equation}
Note that Eq.~(\ref{app_norm_vec_def}) implies
\begin{equation}
\NORMVEC_\sigma \partial^2_{a b} X^\sigma
 = - \partial_a X^\mu \; \partial_b X^\nu
     (\CEX_{\mu \nu} + \Gamma^{\;\rho}_{\mu \nu} \NORMVEC_\rho) .
\end{equation} 
One can also define the extrinsic curvature according to
\begin{equation}
\CEX_{\alpha \beta} \equiv \FFF^\mu_{(\alpha} D_\mu \NORMVEC_{\beta)} ,
\end{equation}
which obeys the following relations
\begin{eqnarray}
\FFF^\mu_\alpha \CEX _{\mu \beta} & = & \CEX_{\alpha \beta} , \\
\NORMVEC^\mu_\alpha \CEX _{\mu \beta} & = & 0 .
\end{eqnarray}

With these definition, the brane Riemann and Ricci tensors, the brane
scalar curvature and the brane Einstein tensor can be rewritten
\begin{eqnarray}
\BRN{R}_{a b c d}
 & = & \partial_a X^\mu \; \partial_b X^\nu \;
       \partial_c X^\rho \; \partial_d X^\sigma \;
       \left(  R_{\mu \nu \rho \sigma}
             - D_{(\mu} \NORMVEC_{\rho)} D_{(\nu} \NORMVEC_{\sigma)}
             - D_{(\mu} \NORMVEC_{\sigma)} D_{(\nu} \NORMVEC_{\rho)} 
       \right) , \\
\BRN{R}_{\alpha \beta \gamma \delta}
 & = &   \FFF^\mu_\alpha \FFF^\nu_\beta \FFF^\rho_\gamma \FFF^\sigma_\delta
         R_{\mu \nu \rho \sigma}
       - \CEX_{\alpha \gamma} \CEX_{\beta \delta}
       + \CEX_{\alpha \delta} \CEX_{\beta \gamma} , \\
\BRN{R}_{\alpha \beta} 
 & = &   \frac{N - 1}{N} 
         \left(  \FFF_\alpha^\mu \FFF_\beta^\nu R_{\mu \nu}
               - \frac{1}{N - 1} \FFF_{\alpha \beta} 
                 \left(  G_{\mu \nu} \NORMVEC^{\mu \nu} 
                       + \frac{1}{N + 1} G \right) \right)
\nonumber \\ & & 
       - \CEX \CEX_{\alpha \beta} + \CEX_\alpha^\mu \CEX_{\mu \beta}
       + \WN_{\alpha \beta} , \\
\BRN{R} & = &   R + R_{\mu \nu} \NORMVEC^{\mu \nu}
              + \CEX_{\mu \nu} \CEX^{\mu \nu} - \CEX^2 , \\
\BRN{G}_{\alpha \beta}
 & = &   \frac{N - 1}{N} 
         \left(  \FFF_\alpha^\mu \FFF_\beta^\nu G_{\mu \nu}
               - \FFF_{\alpha \beta} \left(  G_{\mu \nu} \NORMVEC^{\mu \nu} 
                                           + \frac{1}{N + 1} G \right) \right)
\nonumber \\ & & 
       - \CEX \CEX_{\alpha \beta} + \CEX_\alpha^\mu \CEX_{\mu \beta}
       + \DDEMI \FFF_{\alpha \beta} 
         \left(\CEX^2 - \CEX_{\mu \nu} \CEX^{\mu \nu} \right)
       + \WN_{\alpha \beta} .
\end{eqnarray}

\APPsection{Background geometric quantities}
\label{sec_app_bulk_geom}

\subAPPsection{Metric}

\begin{eqnarray}
g_{0 0} = \IBB
 \quad & , & \quad
g^{0 0} = \IzBB , \\
g_{i j} = - \IAA \gamma_{i j}
 \quad & , & \quad
g^{i j} = - \IzAA \gamma^{i j} , \\
g_{\finN \finN} = - \INN
 \quad & , & \quad
g^{\finN \finN} = - \IzNN .
\end{eqnarray}

\subAPPsection{Notations}

\begin{eqnarray}
\Hconf = \frac{\dot \IA}{\IA}
 \quad , \qquad
\Iconf & = & \frac{\dot \IB}{\IB}
 \quad , \qquad
\Uconf = \frac{\dot \IN}{\IN} , \\
\Hperp = \frac{\IA'}{\IA}
 \quad , \qquad
\Iperp & = & \frac{\IB'}{\IB}
 \quad , \qquad
\Uperp = \frac{\IN'}{\IN} .
\end{eqnarray}

\subAPPsection{Christoffel symbols}
\label{ssec_app_christ}

\begin{eqnarray}
\Gamma^{\;0}_{i j} & = & \IAAzBB \Hconf \gamma_{i j}
 \quad, \qquad
\Gamma^{\;0}_{\finN \finN} = \INNzBB \Uconf , \\
\Gamma^{\;0}_{0 0} = \Iconf
 \quad, \qquad
\Gamma^{\;i}_{j 0} & = & \Hconf \delta^i_j
 \quad, \qquad
\Gamma^{\;\finN}_{\finN 0} = \Uconf , \\
\Gamma^{\;0}_{0 \finN} = \Iperp
 \quad, \qquad
\Gamma^{\;i}_{j \finN} & = & \Hperp \delta^i_j
 \quad, \qquad
\Gamma^{\;\finN}_{\finN \finN} = \Uperp , \\
\Gamma^{\;\finN}_{0 0} = \IBBzNN \Iperp
 \quad, \qquad
\Gamma^{\;\finN}_{i j} & = & - \IAAzNN \Hperp \gamma_{i j} , \\
\Gamma^{k}_{i j} & = & \ND{(N)}{\Gamma}^{k}_{i j} , \\
\Gamma^{\;\mu}_{\mu 0} = \Iconf + N \Hconf + \Uconf
 \quad & , & \qquad
\Gamma^{\;\mu}_{\mu \finN} = \Iperp + N \Hperp + \Uperp .
\end{eqnarray}
The superscript $N$ means that the corresponding quantity is evaluated
using the metric $\gamma_{i j}$.

\subAPPsection{Ricci tensor}
\label{ssec_app_ric}

\begin{eqnarray}
R_{0 0} & = &
 - (\pe + \Uconf - \Iconf) \Uconf
 - N (\pe + \Hconf - \Iconf) \Hconf
 + \IBBzNN (\pp + \Iperp - \Uperp) \Iperp
 + N \IBBzNN \Hperp \Iperp , \\
R_{i j} & = &
   (N - 1) \gamma_{i j} \COURB
 + \IAAzBB \gamma_{i j} (\pe + N \Hconf + \Uconf - \Iconf) \Hconf
 - \IAAzNN \gamma_{i j} (\pp + N \Hperp + \Iperp - \Uperp) \Hperp , \\
R_{0 \finN} & = &
 N (- \dot {\Hperp} - \Hconf \Hperp + \Hconf \Iperp + \Uconf \Hperp) , \\
R_{\finN \finN} & = &
 - (\pp + \Iperp - \Uperp) \Iperp
 - N (\pp + \Hperp - \Uperp) \Hperp
 + \INNzBB (\pe + \Uconf - \Iconf) \Uconf
 + N \INNzBB \Hconf \Uconf .
\end{eqnarray}

\subAPPsection{Scalar curvature}

\begin{equation}
R = - N (N + 1) \left(  \frac{\Hconf^2}{\IBB} + \frac{\COURB}{\IAA}
                      - \frac{\Hperp}{\INN} \right)
    + 2 N \frac{\COURB}{\IAA}
    - \frac{2}{\IBB} (\pe + \Uconf - \Iconf) (\Uconf + N \Hconf)
    + \frac{2}{\INN} (\pp + \Iperp - \Uperp) (\Iperp + N \Hperp) .
\end{equation}

\subAPPsection{Einstein tensor}
\label{ssec_app_ein}

\begin{eqnarray}
G_{0 0} & = &
   \frac{N (N - 1)}{2}\left(\Hconf^2+\IBBzAA\COURB - \IBBzNN \Hperp^2 \right)
 + N \Hconf \Uconf - N \IBBzNN (\Hperp' + \Hperp^2 - \Hperp \Uperp) , \\
G_{i j} & = &
 - \frac{N (N - 1)}{2}\left(\IAAzBB\Hconf^2 + \COURB - \IAAzNN\Hperp^2 \right)
   \gamma_{i j}
 + (N - 1) \COURB \gamma_{i j}
\nonumber \\ & &
 - \IAAzBB \gamma_{i j} (\pe + \Uconf - \Iconf)
                        \left(\Uconf + (N - 1) \Hconf \right)
 + \IAAzNN \gamma_{i j} (\pp + \Iperp - \Uperp)
                        \left(\Iperp + (N - 1) \Hperp \right) , \\
\label{back_ee_05}
G_{0 \finN} & = &
 N (- \dot{ \Hperp} - \Hconf \Hperp + \Hconf \Iperp + \Uconf \Hperp) , \\
G_{\finN \finN} & = &
 - \frac{N (N - 1)}{2}\left(\INNzBB\Hconf^2 + \INNzAA\COURB - \Hperp^2 \right)
 - N \INNzBB (\dot {\Hconf} + \Hconf^2 - \Hconf \Iconf) + N \Hperp \Iperp .
\end{eqnarray}

\subAPPsection{Riemann tensor}
\label{ssec_app_riem}

\begin{eqnarray}
R_{0 \finN 0 \finN} & = &
   \INN (\pe + \Uconf - \Iconf) \Uconf
 - \IBB (\pp + \Iperp - \Uperp) \Iperp , \\
R_{0 i 0 j} & = &
   \IAA \gamma_{i j} (\pe + \Hconf - \Iconf) \Hconf
 - \IAABBzNN \gamma_{i j} \Hperp \Iperp , \\
R_{0 i \finN j} & = &
   \IAA \gamma_{i j}
   (\dot {\Hperp} + \Hconf \Hperp - \Hconf \Iperp - \Uconf \Hperp) , \\
R_{\finN i \finN j} & = &
   \IAA \gamma_{i j} (\pp + \Hperp - \Uperp) \Hperp
 - \INNAAzBB \gamma_{i j} \Hconf \Uconf , \\
R_{i j k l} & = &
 - \IAAAA (\gamma_{i k} \gamma_{j l} - \gamma_{i l} \gamma_{j k})
   \left(\frac{\Hconf^2}{\IBB} + \frac{\COURB}{\IAA} -
        \frac{\Hperp^2}{\INN} \right) .
\end{eqnarray}

\subAPPsection{Weyl tensor}
\label{ssec_app_weyl}

\begin{eqnarray}
C_{0 \finN 0 \finN} & = &
   \frac{N - 1}{N + 1} \IBB \INN \WEYL , \\
C_{0 i 0 j} & = &
 - \frac{1}{N} \frac{N - 1}{N + 1} \IAA \IBB \WEYL \gamma_{i j} , \\
C_{\finN i \finN j} & = &
   \frac{1}{N} \frac{N - 1}{N + 1} \INN \IAA \WEYL \gamma_{i j} , \\
C_{i j k l} & = &
 - \frac{2}{N (N + 1)} (\gamma_{i k} \gamma_{j l} - \gamma_{i l} \gamma_{j k})
   \IAAAA \WEYL ,
\end{eqnarray}
with
\begin{equation}
\WEYL =   \frac{\COURB}{\IAA}
        + \IzBB (\pe + \Uconf - \Iconf) (\Uconf - \Hconf)
        - \IzNN (\pp + \Iperp - \Uperp) (\Iperp - \Hperp) .
\end{equation}

\APPsection{Background matter content}
\label{sec_app_bulk_matt}

\subAPPsection{Unit vectors}

From the fields $\XXE = \eta$ and $\XXP = \fd$, one can build two unit
vectors $u^\alpha$ and $n^\alpha$:
\begin{eqnarray}
u_\alpha & = & \frac{D_\alpha \XXE}
                    {\sqrt{D_\mu \XXE D^\mu \XXE}} , \\
u_\alpha & = & (\IB, {\bf 0}, 0) , \\
u^\alpha & = & \left(\IzB, {\bf 0}, 0 \right) , \\
u_\mu u^\mu & = & 1 , \\
n_\alpha & = & \frac{D_\alpha \XXP}
                    {\sqrt{- D_\mu \XXP D^\mu \XXP}} , \\
n_\alpha & = & (0, {\bf 0}, \IN) , \\
n^\alpha & = & \left(0, {\bf 0}, - \IzN \right) , \\
n_\mu n^\mu & = & -1 .
\end{eqnarray}
One can define the following operators:
\begin{eqnarray}
\pezB & \equiv & u^\mu \partial_\mu = \displaystyle \IzB \pe , \\
\ppzN & \equiv & - n^\mu \partial_\mu = \displaystyle \IzN \pp .
\end{eqnarray}
As in the main text, we also use
\begin{eqnarray}
\HcB = \frac{\pezB \IA}{\IA}
 \quad , \qquad
\IcB & = & \frac{\pezB \IB}{\IB}
 \quad , \qquad
\UcB = \frac{\pezB \IN}{\IN} , \\
\HpN = \frac{\ppzN \IA}{\IA}
 \quad , \qquad
\IpN & = & \frac{\ppzN \IB}{\IB}
 \quad , \qquad
\UpN = \frac{\ppzN \IN}{\IN} .
\end{eqnarray}

\subAPPsection{Stress-energy tensor}
\label{ssec_app_set}

For the bulk matter, it is more convenient to introduce the unit
vector $U_\alpha$ which represents the bulk $N+2$-velocity of the
fluid:
\begin{eqnarray}
U_\alpha & = & (\IB \GAM, {\bf 0} , - \IN \BGM) , \\
U^\alpha & = & \left(\IzB \GAM, {\bf 0}, \IzN \BGM \right) , \\
U_\mu U^\mu & = & 1 , \\
\GMM (1 - \BTT) & = & 1 .
\end{eqnarray}
Here $\BET$ represents the Lorentz boost which must be performed along
the $\fd$ axis in order to be in the rest frame of the bulk matter.
As usual $\GAM = 1 / \sqrt{1 - \BET^2}$.  Due to the symmetries of
spacetime, the stress energy tensor of any component possesses $N$
identical eigenvalues $\REST{P}$. The other eigenvalues are
$\REST{\rho}$ (associated to the timelike eigenvector $U^\alpha$) and
$\REST{\PRESY}$ (associated to the spacelike eigenvector
$N^\alpha$). One has
\begin{eqnarray}
N^\alpha & = & \left(- \IzB \BGM, {\bf 0}, - \IzN \GAM \right) , \\
N_\alpha & = & (- \IB \BGM, {\bf 0}, \IN \GAM ) .
\end{eqnarray}
\begin{eqnarray}
T_{\alpha \beta}
 & = &   (\REST{P} + \REST{\rho}) U_\alpha U_\beta
       - (\REST{P} - \REST{\PRESY}) N_\alpha N_\beta
       - \REST{P} g_{\alpha \beta} , \\
 & = &   (P + \rho) u_\alpha u_\beta
       - (P - \PRESY) n_\alpha n_\beta
       - P g_{\alpha \beta}
       - 2 F u_{(\alpha} n_{\beta)} .
\end{eqnarray}

\begin{eqnarray}
T_{0 0} = \IBB \GMM (\REST{\rho} + \BTT \REST{\PRESY}) \equiv \IBB \rho
 \quad & , & \qquad
T^{0 0} = \IzBB \rho , \\
T_{i j} = \IAA \REST{P} \gamma_{i j} \equiv \IAA P \gamma_{i j}
 \quad & , & \qquad
T^{i j} = \IzAA P \gamma^{i j} , \\
T_{0 \finN} = - \IBN \BGG (\REST{\PRESY} + \REST{\rho}) \equiv - \IBN F
 \quad & , & \qquad
T^{0 \finN} = \IzBN F , \\
T_{\finN \finN} =      \INN \GMM (\REST{\PRESY} + \BTT \REST{\rho})
                \equiv \INN \PRESY
 \quad & , & \qquad
T^{\finN \finN} = \IzNN \PRESY .
\end{eqnarray}

\begin{eqnarray}
\rho & = & \GMM (\REST{\rho} + \BTT \REST{\PRESY}) , \\
\PRESY & = & \GMM (\REST{\PRESY} + \BTT \REST{\rho}) , \\
P & = & \REST{P} , \\
F & = & \BGG (\REST{\rho} + \REST{\PRESY}) ,
\end{eqnarray}

\begin{eqnarray}
\frac{\BET}{1 + \BTT} & = & \frac{F}{\rho + \PRESY} , \\
\BET & = & \frac{\rho + \PRESY}{2 F}
           \left(1 - \sqrt{1 - \left(\frac{2 F}{\rho + \PRESY}\right)^2}
           \right) , \\
\REST{\rho} & = &   \frac{\rho - \BTT \PRESY}{1 + \BTT}
\nonumber \\
            & = &   \frac{\rho - \PRESY}{2}
                  + \sqrt{\frac{(\rho + \PRESY)^2}{4} - F^2} , \\
\REST{\PRESY} & = &   \frac{\PRESY - \BTT \rho}{1 + \BTT}
\nonumber \\
              & = &   \frac{\PRESY - \rho}{2}
                    + \sqrt{\frac{(\rho + \PRESY)^2}{4} - F^2} , \\
\REST{P} & = & P .
\end{eqnarray}

\subAPPsection{Einstein equations}

\begin{eqnarray}
   \frac{N (N - 1)}{2}
   \left(\HcB^2 + \KzAA - \HpN^2 \right)
 + N \HcB \UcB
 - N (\ppzN \HpN + \HpN^2)
 & = & \KAPPAN \rho , \\
 - \frac{N (N - 1)}{2}
   \left(\HcB^2 + \KzAA - \HpN^2 \right)
 + (N - 1) \KzAA
 & & 
\nonumber \\
 - (\pezB + \UcB) \left(\UcB + (N - 1) \HcB \right)
 + (\ppzN + \IpN) \left(\IpN + (N - 1) \HpN \right)
 & = & \KAPPAN P , \\
   N (\pezB \HpN + \HcB \HpN - \HcB \IpN)
 = N (\ppzN \HcB + \HcB \HpN - \HpN \UcB)
 & = & \KAPPAN F , \\
 - \frac{N (N - 1)}{2}
   \left(\HcB^2 + \KzAA - \HpN^2 \right)
 - N (\pezB \HcB + \HcB^2)
 + N \HpN \IpN
 & = & \KAPPAN \PRESY .
\end{eqnarray}

\subAPPsection{Conservation equations}
\label{ssec_app_cons}

For any species,
\begin{eqnarray}
D_\mu T^{\mu \alpha}_f & = & Q^\alpha_f , \\
Q_f^\alpha & = & \REST{\Gamma}^f U_f^\mu - \REST{D}^f N_f^\mu , \\
Q^f_\alpha & = & (  \IB \GAM_f (\REST{\Gamma}^f + \beta_f \REST{D}^f), {\bf 0},
                  - \IN \GAM_f (\REST{D}^f + \beta_f \REST{\Gamma}^f))
             \equiv (\IB \Gamma_f, {\bf 0}, - \IN D_f) , \\
\Gamma_f & = & \GAM (\REST{\Gamma}^f + \BET \REST{D}^f) , \\
D_f & = & \GAM (\REST{D}^f + \BET \REST{\Gamma}^f) , \\
\REST{\Gamma}^f & = & \GAM (\Gamma_f - \BET D_f) , \\
\REST{D}^f & = & \GAM ( D_f - \BET \Gamma_f) , \\
Q_f^\alpha & = & \left(\IzB \Gamma_f, {\bf 0}, \IzN D_f \right) , \\
\sum_f \Gamma_f = \sum_f D_f & = & 0 ,
\end{eqnarray}
\begin{eqnarray}
\label{back_cons_bulk}
   \pezB \rho_f + N \HcB (P_f + \rho_f) + \UcB (\PRESY_f + \rho_f)
 + (\ppzN + N \HpN + 2 \IpN) F_f
 & = & \Gamma_f , \\
   (\pezB + N \HcB + 2 \UcB) F_f
 + \ppzN \PRESY_f + \IpN (\PRESY_f + \rho_f) + N \HpN (\PRESY_f - P_f)
 & = &  D_f .
\end{eqnarray}

\APPsection{Background brane-related quantities}
\label{sec_app_brane}

\subAPPsection{Brane position}

In general, one has the brane position $X^\alpha$ as a function of $N
+ 1$ variables $\sigma^a$. We choose
\begin{eqnarray}
X^0 & = & \sigma^0 , \\
X^i & = & \sigma^i , \\
X^\finN & = & \POSBRN .
\end{eqnarray}

\subAPPsection{Induced metric}

One first builds the unit vector orthogonal to the brane:
\begin{eqnarray}
\NORMVEC_\mu \frac{\partial X^\mu}{\partial \sigma^a} & = & 0 , \\
\NORMVEC_\alpha & = & (0, {\bf 0}, \IN) , \\
\NORMVEC^\alpha & = & \left(0, {\bf 0}, - \IzN \right) .
\end{eqnarray}
The components of this vector have the same functional form but
possibly different numerical values when evaluated at $\fd =
\POSBRN^+$ and $\fd = \POSBRN^-$. Then the induced metric
is given by:
\begin{eqnarray}
\FFF_{\alpha \beta} & = & g_{\alpha \beta}
+ \NORMVEC_\alpha \NORMVEC_\beta , \\
\FFF_{\alpha \mu} \NORMVEC^\mu & = & 0 .
\end{eqnarray}
\begin{eqnarray}
\FFF_{0 0} & = & \IBB , \\
\FFF_{i j} & = & - \IAA \gamma_{i j} .
\end{eqnarray}

\subAPPsection{First Israel conditions}
\label{ssec_app_lperp}

For any quantity $f$, we define
\begin{equation}
f =   \DISC{f} \left(\theta (\fd - \POSBRN) - \TDEMI \right)
    + \CONT{f} ,
\end{equation}
where $\DISC{f}$ is the discontinuity of $f$, $\CONT{f}$ is the
continuous part of $f$ and $\theta$ is the Heaviside function.  The
first Israel condition states that $\FFF_{\alpha \beta} (\POSBRN^+) =
\FFF_{\alpha \beta} (\POSBRN^-)$~:
\begin{eqnarray}
\DISC{\IA} & = & 0 , \\
\DISC{\IB} & = & 0 .
\end{eqnarray}
In particular, this means that $\IN$ is allowed to be
discontinuous at the brane position. Also, the derivatives of
$\IA$ and $\IB$ with respect to $\fd$ can be discontinuous.

\subAPPsection{Extrinsic curvature}

\begin{eqnarray}
\CEX_{\alpha \beta}
 & = & \FFF_{(\alpha}^\mu D_\mu \NORMVEC_{\beta)} , \\
\CEX_{\alpha \mu} \NORMVEC^\mu & = & 0 .
\end{eqnarray}
\begin{eqnarray}
\CEX_{0 0} & = & - \IBBzN \Iperp , \\
\CEX_{i j} & = & \IAAzN \Hperp \gamma_{i j} , \\
\CEX & \equiv &   g^{\mu \nu} \CEX_{\mu \nu}
       =          \FFF^{\mu \nu} \CEX_{\mu \nu}
       =        - \IzN (\Iperp + N \Hperp) , \\
\CEX^{\mu \nu} \CEX_{\mu \nu}
 & = & \IzNN (\Iperp^2 + N \Hperp^2) , \\
\CEX^2 - \CEX^{\mu \nu} \CEX_{\mu \nu}
 & = & \frac{N}{\INN} ((N - 1) \Hperp^2 + 2 \Hperp \Iperp) .
\end{eqnarray}

\subAPPsection{Stress-energy tensor}

Formally, one can take a stress-energy tensor of the above form to
describe the brane content, provided that we have
\begin{eqnarray}
\REST{\rho} & \equiv & \DIR \BRN{\rho} , \\
\REST{P} & \equiv & \DIR \BRN{P} , \\
\REST{\PRESY} & \equiv & 0 , \\
\GAM & = & 1 , \\
\BET & = & 0 ,
\end{eqnarray}
with
\begin{equation}
\label{def_beta}
\DIR = \frac{\sqrt{|\FFF|}}{\sqrt{|g|}}
       \delta(\fd - \POSBRN) .
\end{equation}
The condition $\GAM = 1$ and $\BET = 0$ has not necessarily to be
satisfied but is a consequence of the coordinate choice to put the
brane at rest with respect to the coordinate system. Because of the
Dirac term~(\ref{def_beta}), $\BRN{P}$ and $\BRN{\rho}$ depend on
$\eta, x^i$ only. Since the stress energy tensor of the brane is
strictly zero elsewhere, its eigenvectors are not defined outside this
hypersurface.  The vector $\NORMVEC^\alpha$ appears as the analog of
the vector $N^\alpha$ as an eigenvector associated to the eigenvalue
$\REST{\PRESY} = 0$. Equivalently one can define an $N+2$-velocity
$\BRN{u}_\alpha$ which corresponds to the eigenvector associated with
the eigenvalue $\REST{\rho}$.  Therefore,
\begin{eqnarray}
T_{\alpha \beta}^\SBRN & = & \DIR \BRN{T}_{\alpha \beta} , \\
\BRN{T}_{\alpha \mu} \NVEC^\mu & = & 0 , \\
\BRN{T}_{\alpha \beta}
 & = &   (\BRN{P} + \BRN{\rho}) \BRN{u}_\alpha \BRN{u}_\beta
       - \BRN{P} \FFF_{\alpha \beta} , \\
\BRN{u}_\alpha & \equiv & (\IB, {\bf 0}, 0) .
\end{eqnarray}

\subAPPsection{{\protect $\STT_{\alpha \beta}$} tensor}

\begin{equation}
\STT_{\alpha \beta}
 = \BRN{T}_{\alpha \beta} - \frac{1}{N} \BRN{T} \FFF_{\alpha \beta} .
\end{equation}
\begin{eqnarray}
\STT_{0 0}
 & = & \IBB \left(\frac{N - 1}{N} \BRN{\rho} + \BRN{P} \right) , \\
\STT_{i j}
 & = & \frac{1}{N} \IAA \BRN{\rho} \gamma_{i j} .
\end{eqnarray}

\subAPPsection{Second Israel condition}
\label{ssec_app_isr2}

\begin{equation}
\DISC{\CEX_{\alpha \beta}} = - \KAPPAN \STT_{\alpha \beta} .
\end{equation}
\begin{eqnarray}
\DISC{\frac{\Iperp}{\IN}}
 & = & \KAPPAN \left(\frac{N - 1}{N} \BRN{\rho} + \BRN{P} \right) , \\
\DISC{\frac{\Hperp}{\IN}} & = & - \KAPPAN \frac{1}{N} \BRN{\rho} , \\
- N \DISC{\frac{\Hperp}{\IN}} & = & \KAPPAN \BRN{\rho} , \\
\DISC{\frac{\Iperp}{\IN}} + (N - 1) \DISC{\frac{\Hperp}{\IN}}
 & = & \KAPPAN \BRN{P} , \\
\DISC{\frac{\Iperp}{\IN}} - \DISC{\frac{\Hperp}{\IN}}
 & = & \KAPPAN (\BRN{P} + \BRN{\rho}) .
\end{eqnarray}

\subAPPsection{Projected Weyl tensor}

\begin{eqnarray}
\WN_{\alpha \beta} & = & C_{\alpha \mu \beta \nu} \NORMVEC^\mu
\NORMVEC^\nu , \\
g^{\mu \nu} \WN_{\mu \nu} = \FFF^{\mu \nu} \WN_{\mu \nu}
                          = \NORMVEC^\mu \WN_{\mu \alpha} & = & 0 .
\end{eqnarray}
\begin{eqnarray}
\WN_{0 0} & = & \frac{N - 1}{N + 1} \IBB \WEYL , \\
\WN_{i j} & = & \frac{N - 1}{N (N + 1)} \IAA \WEYL \gamma_{i j} .
\end{eqnarray}

\APPsection{Brane point of view, unperturbed case}
\label{sec_app_brane_pov}

Unless otherwise noted, all the quantities are evaluated at the brane
position. The quantity $\CONT{\ppzN} X$ stands for the continuous part
of $\ppzN X$ at the brane position.

\subAPPsection{Friedmann equation}

Taking the continuous part of the Einstein equation at the brane
position, we get
\begin{eqnarray}
   \frac{N (N - 1)}{2} \left(\HcB^2 + \KzAA \right)
 & = &   \frac{\KAPPAN^2}{8} \frac{N + 1}{N} \BRN{\rho}^2
       + \KAPPAN \CONT{\rho_\SBLK}
\nonumber \\ & &
       - N \HcB \CONT{\UcB}
       + N \CONT{\ppzN} \CONT{\HpN}
       + \frac{1}{2} N (N + 1) \CONT{\HpN}^2 , \\
 - \DDEMI (N - 1) \left(\HcB^{- 1} \pezB + N \right)
   \left(\HcB^2 + \KzAA \right) 
 & = & - \frac{\KAPPAN^2}{8} 
         \left(  2 \BRN{P}^2 
               + 2 \frac{N - 1}{N} \BRN{P} \BRN{\rho}
               + \frac{N - 1}{N} \BRN{\rho}^2 \right)
       + \KAPPAN \CONT{P_\SBLK}
\nonumber \\ & &
       + \CONT{\left(  \pezB + \UcB + (N - 1) \HcB \right) \UcB }
\nonumber \\ & &
       - \CONT{\ppzN + \IpN} \CONT{\IpN + (N - 1) \HpN }
       - \DDEMI N (N - 1) \CONT{\HpN}^2 .
\end{eqnarray}
As such, these equations are not yet very useful because they
involve many terms which are not explicit `brane variables'.

\subAPPsection{New Friedmann equation}
\label{ssec_app_newfried}

Consider the combination $\CONT{\HcB \{\SfinN \SfinN\} + \HpN
\{0 \SfinN\}}$ of the Einstein equations.  It yields
\begin{equation}
\left(\pezB + (N + 1) \HcB \right)
\left(  \frac{N (N - 1)}{2}
        \left( \HcB^2 + \KzAA - \CONT{\HpN}^2 \right)
      - \frac{N - 1}{8 N} \KAPPAN^2 \BRN{\rho}^2 \right)
 = - (N - 1) \KAPPAN \CONT{\HcB \PRESY + \HpN F} .
\end{equation}
In the case $\CONT{\HpN F} = \pe \CONT{\PRESY} = 0$, they can be
integrated exactly and we find
\begin{eqnarray}
\frac{N (N - 1)}{2} \left(\HcB^2 + \KzAA \right)
 & = &   \frac{N - 1}{8 N} \KAPPAN^2 \BRN{\rho}^2
       - \frac{N - 1}{N + 1} \KAPPAN \CONT{\PRESY}
       + \frac{\cal C}{\IA^{N + 1}} .
\end{eqnarray}

\subAPPsection{Friedmann equations using the Weyl tensor}

In general, the Friedmann equation can conveniently be rewritten using the
Weyl tensor. One has
\begin{eqnarray}
\frac{N (N - 1)}{2} \left(\HcB^2 + \KzAA \right)
 & = &   \frac{N - 1}{8 N} \KAPPAN^2 \left(\sum_\SBRN \BRN{\rho}_\SBRN\right)^2
\nonumber \\ & &
       + \frac{N (N - 1)}{2} \CONT{\HpN}^2
\nonumber \\ & &
       + \frac{N - 1}{N + 1} \KAPPAN
\sum_\SBLK \CONT{P_\SBLK + \rho_\SBLK -\PRESY_\SBLK}
       + \frac{N - 1}{N + 1} \CONT{\WEYL} , \\
 - \frac{N - 1}{2} \left(N + \HcB^{- 1} \pezB \right)
   \left(\HcB^2 + \KzAA \right)
 & = &   \frac{N - 1}{8 N} \KAPPAN^2
         \left(\sum_\SBRN \BRN{\rho}_\SBRN \right)
         \left(\sum_\SBRN (\BRN{\rho}_\SBRN + 2 \BRN{P}_\SBRN) \right)
\nonumber \\ & &
       - \frac{N (N - 1)}{2} \CONT{\HpN}^2
       - (N - 1) \CONT{\HpN} \CONT{\IpN - \HpN}
\nonumber \\ & &
       + \frac{N - 1}{N (N + 1)} \KAPPAN
         \sum_\SBLK \CONT{P_\SBLK + \rho_\SBLK + N \PRESY_\SBLK}
       + \frac{N - 1}{N (N + 1)} \CONT{\WEYL} .
\end{eqnarray}

\subAPPsection{Relationship between {\protect $\CONT{\CEX_{\alpha
\beta}}$} and {\protect $\DISC{\WN_{\alpha \beta}}$}}

\begin{eqnarray}
\BRN{\rho} \CONT{\HpN}
 & = & \frac{1}{N + 1} \left(  \DISC{P + \rho - \PRESY}
                             + \frac{1}{\KAPPAN} \DISC{\WEYL} \right) , \\
(N \BRN{P} + \BRN{\rho}) \CONT{\HpN} - \BRN{\rho} \CONT{\IpN}
 & = & \frac{1}{N + 1} \left(  \DISC{P + \rho + N \PRESY}
                             + \frac{1}{\KAPPAN} \DISC{\WEYL} \right) .
\end{eqnarray}

\subAPPsection{Conservation equation}

These can be found either by taking the singular part
of~(\ref{back_cons_bulk}), or by considering the discontinuity
of~(\ref{back_ee_05}). A bulk energy exchange term $\Gamma_\SBLK$ can
have a singular component $\Gamma_\SBLK^{(\DIR)}$ so that
$\Gamma_\SBLK = \DIR \Gamma_\SBLK^{(\DIR)} + \DISC{\Gamma_\SBLK}
\left(\theta (\fd - \POSBRN) - \TDEMI \right) + \CONT{\Gamma_\SBLK}$.
\begin{eqnarray}
\pezB \BRN{\rho}_\SBRN + N \HcB (\BRN{P}_\SBRN + \BRN{\rho}_\SBRN)
 & = & \BRN{\Gamma}_\SBRN , \\
\DISC{F_\SBLK} & = & \Gamma_\SBLK^{(\DIR)} , \\
\sum_\SBRN \BRN{\Gamma}_\SBRN & = & - \sum_\SBLK \DISC{F_\SBLK} .
\end{eqnarray}
One also has the sail equation
\begin{eqnarray}
 - \CONT{\IpN} \sum_\SBRN \BRN{\rho}_\SBRN
 + N \CONT{\HpN} \sum_\SBRN \BRN{P}_\SBRN
 & = & \sum_\SBLK \DISC{\PRESY_\SBLK} .
\end{eqnarray}

\APPsection{Perturbed geometric variables}
\label{sec_app_pert_bulk_geom}

\subAPPsection{Metric}

\begin{eqnarray}
\delta g_{0 0} = 2 \IBB A
 \quad & , & \qquad
\delta g^{0 0} = - \IzBB 2 A , \\
\delta g_{0 i} = \IAB B_i
 \quad & , & \qquad
\delta g^{0 i} = \IzAB B^i , \\
\delta g_{i j} = - \IAA h_{i j}
 \quad & , & \qquad
\delta g^{i j} = \IzAA h^{i j} , \\
\delta g_{0 \finN} = \IBN \Bperp
 \quad & , & \qquad
\delta g^{0 \finN} = \IzBN \Bperp , \\
\delta g_{i \finN} = \INA \epi_i
 \quad & , & \qquad
\delta g^{i \finN} = - \IzNA \eip^i , \\
\delta g_{\finN \finN} = 2 \INN \epp
 \quad & , & \qquad
\delta g^{\finN \finN} = - \IzNN 2 \epp .
\end{eqnarray}
\begin{eqnarray}
B_i & = & \nabla_i B + \VV{B}_i , \\
h_{i j} & = & 2 C \gamma_{i j} + 2 E_{i j} , \\
E_{i j} & = & \nabla_{(i} E_{j)} + \TT{E}_{i j} , \\
E_i & = & \nabla_i E + \VV{E}_i , \\
\epi_i & = & \nabla_i \Eperp + \Vepi_i .
\end{eqnarray}
All $3$-vectors indices are raised and lowered using metric $\gamma_{i
j}$. $\nabla_i$ represents its associated covariant derivative and
$\nabla^2 = \nabla_i \nabla^i$. Barred vectors are divergenceless,
double barred tensors are divergenceless and traceless with respect to
$\gamma_{i j}$ and $\nabla_i$.

\subAPPsection{Infinitesimal coordinate transformation}

Under an infinitesimal coordinate transformation $x^\alpha \to
x^\alpha + \xi^\alpha$, the perturbed part of a tensor transforms as
\begin{equation}
\delta T^{\alpha_1 ... \alpha_u}_{\beta_1 ... \beta_d}
 \to   \delta T^{\alpha_1 ... \alpha_u}_{\beta_1 ... \beta_d}
     + \xi^\mu \partial_\mu T^{\alpha_1 ... \alpha_u}_{\beta_1 ... \beta_d}
     - \partial_{\mu_i} \xi^{\alpha_i}
       T^{\alpha_1 ... \mu_i ... \alpha_u}_{\beta_1 ... \beta_d}
     + \partial_{\beta_j} \xi^{\nu_j}
       T^{\alpha_1 ... \alpha_u}_{\beta_1 ... \nu_j ... \beta_d} .
\end{equation}
Setting
\begin{eqnarray}
\xi^\alpha & = & (T, L^i, \Lperp) , \\
L^i & = & \nabla^i L + \VV{L}^i ,
\end{eqnarray}
the metric perturbations transform into
\begin{eqnarray}
A & \to & A + \dot T + \Iconf T + \Iperp \Lperp , \\
B_i & \to & B_i - \IAzB \dot L_i + \IBzA \nabla_i T , \\
C & \to & C + \Hconf T + \Hperp \Lperp , \\
E_i & \to & E_i + L_i , \\
\TT{E}_{i j} & \to & \TT{E}_{i j} , \\
\Bperp & \to & \Bperp - \INzB \dot \Lperp + \IBzN T' , \\
\epi_i & \to & \epi_i - \IAzN L'_i - \INzA \nabla_i \Lperp , \\
\epp & \to &   \epp - \Uconf T
                      - \Lperp{}' - \Uperp \Lperp , \\
\ABBE & \to & \ABBE + T , \\
\ANEE & \to & \ANEE - \Lperp .
\end{eqnarray}

There is one subtlety due to the fact that $\IN$ may be
discontinuous. We consider the above infinitesimal coordinate
transformation. For the $\{\SfinN \SfinN\}$ component, we have
\begin{equation}
g_{\finN \finN} \to g_{\finN \finN} - T \pe(\INN) - 2 \IN \pp(\IN \Lperp) .
\end{equation}
For the coordinate change to be valid, the metric components must
remain finite, therefore one {\em must} have
\begin{equation}
[\IN \Lperp] = 0 .
\end{equation}
If $\IN$ is continuous, then $\Lperp$ can be an arbitrary (continuous)
coordinate transformation, but if $\IN$
is discontinuous, then $\Lperp (\eta, x^i, \POSBRN) =
0$. Geometrically, this is related to the fact that the coordinate
system is allowed to exhibit some pathologies only at the brane
position, but not in its vicinity.

\subAPPsection{Gauge invariant metric perturbations}

Using the transformation laws for $\ABBE$, $E^i$, $- \ANEE$, it is
possible to construct the following gauge invariant quantities:
\begin{eqnarray}
\Psi & = & A - (\pe + \Iconf) \ABBE + \Iperp \ANEE , \\
\Phi & = & - C + \Hconf \ABBE - \Hperp \ANEE , \\
\vp & = & \Bperp - \IBzN \pp \ABBE - \INzB \pe \ANEE , \\
\hpp & = & \epp +\Uconf \ABBE - (\pp + \Uperp) \ANEE , \\
\vi_i & = & \VV{B}_i + \IAzB \dot{\VV{E}}_i , \\
\hci_i & = & \Vepi_i + \IAzN \VV{E}{}'_i .
\end{eqnarray}
\begin{eqnarray}
\delta g_{0 0}
 & = &   2 \IBB \Psi
       + (2 g_{0 0} \pe + \dot g_{0 0}) \ABBE - g_{0 0}' \ANEE , \\
\delta g_{0 i}
 & = & - \IBzA g_{i j} \vi^j
       + g_{0 0} \nabla_i \ABBE + g_{i j} \dot E^j , \\
\delta g_{i j}
 & = &   \dot g_{i j} \ABBE - g_{i j}' \ANEE
       + 2 g_{k (i} (E^k_{j)} - \delta^k_{j)} \Phi) , \\
\delta g_{0 \finN}
 & = &   \IBN \vp
       + g_{0 0} \pp \ABBE - g_{\finN \finN} \pe \ANEE , \\
\delta g_{i \finN}
 & = & - \INzA g_{i j} \hic^j
       - g_{\finN \finN} \nabla_i \ABBE + g_{i j} E^j{}' , \\
\delta g_{\finN \finN}
 & = &   2 \INN \hpp
       + \dot g_{\finN \finN} \ABBE
       - (2 g_{\finN \finN} \pp + g_{\finN \finN}') \ANEE .
\end{eqnarray}

\subAPPsection{Christoffel symbols}
\label{ssec_app_pert_christ}

\begin{eqnarray}
\delta \Gamma^{\;0}_{0 0} & = &
\dot A - \IBzN \Iperp \Bperp , \\
\delta \Gamma^{\;0}_{0 i} & = &
\nabla_i A - \IAzB \Hconf B_i , \\
\delta \Gamma^{\;0}_{i j} & = &
   \IAAzBB \left(- 2 A \Hconf \gamma_{i j}
                + \Hconf h_{i j} + \DDEMI \dot h_{i j} \right)
 + \IAzB \nabla_{(i} B_{j)}
 + \IAAzBN \Hperp \Bperp \gamma_{i j} , \\
\delta \Gamma^{\;0}_{0 \finN} & = &
A' - \INzB \Uconf \Bperp , \\
\delta \Gamma^{\;0}_{i \finN} & = &
   \DDEMI \IAzB (\pp + \Iperp - \Hperp) B_i
 - \DDEMI \INAzBB (\pe + \Hconf + \Uconf) \epi_i
 + \DDEMI \INzB \nabla_i \Bperp , \\
\delta \Gamma^{\;0}_{\finN \finN} & = &
   \INzB (\pp + \Iperp) \Bperp
 - \INNzBB (\pe  + 2 \Uconf) \epp - 2 \INNzBB \Uconf A ,
\end{eqnarray}
\begin{eqnarray}
\delta \Gamma^{\;i}_{0 0} & = &
   \IBBzAA \nabla^i A
 - \IBzA (\pe + \Hconf) B^i
 + \IBBzNA \Iperp \eip^i , \\
\delta \Gamma^{\;i}_{0 j} & = &
   \DDEMI \dot h^i_j
 + \DDEMI \IBzA (\nabla^i B_j - \nabla_j B^j) , \\
\delta \Gamma^{\;i}_{j k} & = &
   \DDEMI (\nabla_j h^i_k + \nabla_k h^i_j - \nabla^k h_{i j})
 + \gamma_{j k} \left(\IAzB \Hconf B^i - \IAzN \Hperp \eip^i \right) , \\
\delta \Gamma^{\;i}_{0 \finN} & = &
 - \DDEMI \INzA (\pe + \Hconf - \Uconf) \eip^i
 - \DDEMI \IBzA (\pp + \Hperp - \Iperp) B^i
 + \DDEMI \IBNzAA \nabla^i \Bperp , \\
\delta \Gamma^{\;i}_{j \finN} & = &
   \DDEMI h^i_j{}'
 + \DDEMI \INzA (\nabla^i \epi_j - \nabla_j \eip^i) , \\
\delta \Gamma^{\;i}_{\finN \finN} & = &
 - \INzA (\pp + \Hperp) \eip^i
 + \INNzAB \Uconf B^i
 + \INNzAA \nabla^i \epp ,
\end{eqnarray}
\begin{eqnarray}
\delta \Gamma^{\;\finN}_{0 0} & = &
 - \IBzN (\pe + \Uconf) \Bperp
 + \IBBzNN (\pp + 2 \Iperp) A + 2 \IBBzNN \Iperp \epp , \\
\delta \Gamma^{\;\finN}_{0 i} & = &
   \DDEMI \IABzNN (\pp + \Iperp + \Hperp) B_i
 - \DDEMI \IAzN (\pe + \Uconf - \Hconf) \epi_i
 - \DDEMI \IBzN \nabla_i \Bperp , \\
\delta \Gamma^{\;\finN}_{i j} & = &
 - \IAAzNN \left(  2 \Hperp \epp \gamma_{i j}
                 + \Hperp h_{i j} + \DDEMI h'_{i j}\right)
 - \IAzN \nabla_{(i} \epi_{j)}
 + \IAAzBN \Hconf \Bperp \gamma_{i j} , \\
\delta \Gamma^{\;\finN}_{0 \finN} & = &
 - \dot \epp
 + \IBzN \Iperp \Bperp , \\
\delta \Gamma^{\;\finN}_{i \finN} & = &
 - \nabla_i \epp
 + \IAzN \Hperp \epi_i , \\
\delta \Gamma^{\;\finN}_{\finN \finN} & = &
 - \epp{}' + \INzB \Uconf \Bperp ,
\end{eqnarray}
\begin{eqnarray}
\delta \Gamma^{\;\mu}_{\mu 0} & = &
 \pe (A - \epp + N C + \nabla^2 E) , \\
\delta \Gamma^{\;\mu}_{\mu i} & = &
 \nabla_i (A - \epp + N C + \nabla^2 E) , \\
\delta \Gamma^{\;\mu}_{\mu \finN} & = &
 \pp (A - \epp + N C + \nabla^2 E) .
\end{eqnarray}

\subAPPsection{Ricci tensor}
\label{ssec_app_pert_ric}

From now on, we shall write any perturbation variables mostly with
the gauge invariant quantities defined above. The non-gauge
invariant terms will be put under the form of some factors
involving components of the corresponding unperturbed tensor
multiplied by $\ABBE$, $E^i$ and $\ANEE$.

\begin{eqnarray}
\delta R_{0 0} & = &
   \IBBzAA \nabla^2 \Psi
\nonumber \\ & &
 + (N \Hconf + \Uconf) \pe \Psi
 + (\pe^2 + (2 \Uconf - \Iconf) \pe) \hpp
 + N (\pe^2 + (2 \Hconf - \Iconf) \pe) \Phi
\nonumber \\ & &
 + \IBBzNN (\pp + N \Hperp + \Iperp - \Uperp)
          (\pp \Psi + 2 \Iperp \Psi + 2 \Iperp \hpp)
 - \IBBzNN \Iperp \pp (\Psi + \hpp + N \Phi)
\nonumber \\ & &
 - \IBzN (  \pe \pp + \Uconf \pp + (N \Hperp + \Iperp) \pe
         + N \Hconf \Iperp + N \Uconf \Hperp + \dot {\Iperp} + \Uconf') \vp
\nonumber \\ & &
 + (\dot R_{0 0} + 2 R_{0 0} \pe) \ABBE
 - (R'_{0 0} + 2 R_{0 \finN} \pe) \ANEE , \\
\delta R_{0 i} & = &
   ((N - 1) \Hconf + \Uconf) \nabla_i \Psi
 + (\pe - \Hconf + \Uconf) \nabla_i \hpp
 + (N - 1) \pe \nabla_i \Phi
\nonumber \\ & &
 - \DDEMI \IBzN (\pp + (N - 2) \Hperp + 2 \Iperp) \vp
\nonumber \\ & &
 + \DDEMI \IBzA (\nabla^2 + K (N - 1)) \vi_i - \IBzA R_{i j} \vi^j
\nonumber \\ & &
 + \DDEMI \IABzNN
   (\pp + (N + 1) \Hperp - \Uperp)
   \left((\pp + \Iperp - \Hperp) \vi_i \right)
 - \DDEMI \IAzN
   (\pp + (N + 1) \Hperp - \Iperp)
   \left( (\pe + \Uconf - \Hconf) \hci_i \right)
\nonumber \\ & &
 + R_{0 0} \nabla_i \ABBE
 - R_{0 \finN} \nabla_i \ANEE
 + R_{i j} \dot E^j , \\
\delta R_{i j} & = &
   (\gamma_{i j} \nabla^2 + N \nabla_{i j} + 2 K (N - 1) \gamma_{i j}) \Phi
 - \nabla_{i j} (\Psi + 2 \Phi - \hpp)
\nonumber \\ & &
 - \IAAzBB \gamma_{i j}
   \left(  2 \dot {\Hconf} + 2 \Hconf (N \Hconf + \Uconf - \Iconf) \Psi
         + \Hconf \pe (\Psi + \hpp + N \Phi) \right)
\nonumber \\ & &
 - \IAAzNN \gamma_{i j}
   \left(  2 \Hperp' + 2 \Hperp (N \Hperp + \Iperp - \Uperp) \hpp
         + \Hperp \pp (\Psi + \hpp + N \Phi) \right)
\nonumber \\ & &
 + \IAAzBN \gamma_{i j}
   \left(  \Hconf \pp + \Hperp \pe
         + 2 N \Hconf \Hperp + \Hconf' + \dot{\Hperp} \right) \vp
\nonumber \\ & &
 + \IAzB (\pe + (N - 1) \Hconf + \Uconf) \nabla_{(i} \vi_{j)}
 - \IAzN (\pp + (N - 1) \Hperp + \Iperp) \nabla_{(i} \hci_{j)}
\nonumber \\ & &
 + \left(  \IAAzBB (\pe^2 + (N \Hconf + \Uconf - \Iconf) \pe)
         - \IAAzNN (\pp^2 + (N \Hperp + \Iperp - \Uperp) \pp) \right)
   (\TT{E}_{i j} - \gamma_{i j} \Phi)
 - (\nabla^2 - 2 K) \TT{E}_{i j}
\nonumber \\ & &
 + \dot R_{i j} \ABBE
 - R'_{i j} \ANEE
 + 2 R_{k (i} (E^k_{j)} - \delta^k_{j)} \Phi) , \\
\delta R_{0 \finN} & = &
   N \Hconf \pp \Psi - N \Hperp \pe \hpp
 + N (  \pe \pp
      + (\Hperp - \Iperp) \pe
      + (\Hconf - \Uconf) \pp) \Phi
\nonumber \\ & &
 + \DDEMI \IBNzAA \nabla^2 \vp
 + \IBzN (\Iperp' + \Iperp (N \Hperp + \Iperp - \Uperp)) \vp
 - \INzB (\dot {\Uconf} + \Uconf (N \Hconf + \Uconf - \Iconf)) \vp
\nonumber \\ & &
 + (\dot R_{0 \finN} + R_{0 \finN} \pe + R_{0 0} \pp) \ABBE
 - (   R'_{0 \finN} + R_{0 \finN} \pp
    + R_{\finN \finN} \pe) \ANEE , \\
\delta R_{i \finN} & = &
 - ((N - 1) \Hperp + \Iperp) \nabla_i \hpp
 - (\pp - \Hperp + \Iperp) \nabla_i \Psi
 + (N - 1) \pp \nabla_i \Phi
\nonumber \\ & &
 + \DDEMI \INzB (\pe + (N - 2) \Hconf + 2 \Uconf) \nabla_i \vp
\nonumber \\ & &
 + \DDEMI \INzA (\nabla^2 + K (N - 1)) \hci_i - \INzA R_{i j} \hic^j
\nonumber \\ & &
 + \DDEMI \IAzB
   (\pe + (N + 1) \Hconf - \Uconf)
   \left( (\pp + \Iperp - \Hperp) \vi_i \right)
 - \DDEMI \INAzBB
   (\pe + (N + 1) \Hconf - \Iconf)
   \left( (\pe + \Uconf - \Hconf) \hci_i \right)
\nonumber \\ & &
 + R_{0 \finN} \nabla_i \ABBE
 - R_{\finN \finN} \nabla_i \ANEE
 + R_{i j} E^j{}' , \\
\delta R_{\finN \finN} & = &
   \INNzAA \nabla^2 \hpp
\nonumber \\ & &
 - \INNzBB (\pe + N \Hconf + \Uconf - \Iconf)
          (\pe \hpp + 2 \Uconf \hpp + 2 \Uconf \Psi)
 + \INNzBB \Uconf \pe (\Psi + \hpp - N \Phi)
\nonumber \\ & &
 - (\pp^2 + (2 \Iperp - \Uperp) \pp) \Psi
 - (N \Hperp + \Iperp) \pp \hpp
 + N (\pp^2 + (2 \Hperp - \Uperp) \pp) \Phi
\nonumber \\ & &
 + \INzB (  \pe \pp + (N \Hconf + \Uconf) \pp + \Iperp \pe
         + N \Hconf \Iperp + N \Uconf \Hperp + \dot{\Iperp} + \Uconf') \vp
\nonumber \\ & &
 + (\dot R_{\finN \finN} + 2 R_{\finN \finN} \pp) \ABBE
 - (R'_{\finN \finN} + 2 R_{0 \finN} \pp) \ANEE .
\end{eqnarray}

\subAPPsection{Scalar curvature}

\begin{eqnarray}
\delta R & = &
   \frac{2}{\IAA}
   \left(  \nabla^2 (\Psi - \hpp - (N - 1) \Phi - K N (N - 1) \Phi \right)
\nonumber \\ & &
 + \frac{2}{\IBB}
   \left(  2 (\Uconf + N \Hconf) (\pe + \Uconf - \Iconf) \Psi
         + 2 (\dot{\Uconf} + N \dot{\Hconf}) \Psi
\right. \nonumber \\ & & \qquad \left.
         + (\pe + \Uconf - \Iconf) (\pe \hpp + N \pe \Phi)
\right. \nonumber \\ & & \qquad \left.
         + (N \Hconf + \Uconf) \pe (\hpp - \Psi)
         + N (N + 1) (\Hconf^2 \Psi + \Hconf \pe \Phi)
   \right)
\nonumber \\ & &
 + \frac{2}{\INN}
   \left(  2 (\Iperp + N \Hperp) (\pp + \Iperp - \Uperp) \hpp
         + 2 (\Iperp{}' + N \Hperp{}') \hpp
\right. \nonumber \\ & & \qquad \left.
         + (\pp + \Iperp - \Uperp) (\pp \hpp + N \pp \Phi) )
\right. \nonumber \\ & & \qquad \left.
         + (N \Hperp + \Iperp) \pp (\Psi - \hpp)
         + N (N + 1) (\Hperp^2 \hpp - \Hperp \pp \Phi)
   \right)
\nonumber \\ & &
 - \frac{2}{\IBN}
   \left(  \pp \pe
         + (N \Hconf + \Uconf) \pp
         + (N \Hperp + \Iperp) \pe
\right. \nonumber \\ & & \qquad \left.
         + N \Hconf{}' + \Uconf{}' + N \dot{\Hperp} + \dot{\Iperp}
         + N (N + 1) \Hconf \Hperp
   \right) \vp
\nonumber \\ & &
 + \dot R \ABBE - R' \ANEE ,
\end{eqnarray}

\subAPPsection{Einstein tensor}
\label{ssec_app_pert_ein}

\begin{eqnarray}
\delta G_{0 0} & = &
   \IBBzAA (\nabla^2 - K N (N - 1)) (\hpp - \Phi)
\nonumber \\ & &
 + N \left(  \IBBzAA \nabla^2 - (N \Hconf + \Uconf) \pe
           + \IBBzNN (\pp^2 + (N \Hperp - \Uperp) \pp) \right) \Phi
\nonumber \\ & &
 - N \left(  \IBBzNN \Hperp \pp (\hpp - \Phi) + \Hconf \pe (\hpp - \Phi)
           + ((N - 1) \Hconf^2 + 2 \Hconf \Uconf) (\Psi + \hpp) \right)
\nonumber \\ & &
 + N \IBzN \left(\Hconf \pp + N \Hconf \Hperp + \Hconf'\right) \vp
\nonumber \\ & &
 + 2 G_{0 0} (\Psi + \hpp) - \IBzN G_{0 \finN} \vp
\nonumber \\ & &
 + (\dot G_{0 0} + 2 G_{0 0} \pe) \ABBE
 - (G'_{0 0} + 2 G_{0 \finN} \pe) \ANEE , \\
\delta G_{0 i} & = &
   \nabla_i (  (\Uconf - \Hconf) (\Psi + \hpp) + \pe (\hpp - \Phi)
             + N (\Hconf \Psi + \dot \Phi) )
\nonumber \\ & &
 - \DDEMI \IBzN (\pp + (N - 2) \Hperp + 2 \Iperp) \nabla_i \vp
\nonumber \\ & &
 + \DDEMI \IABzNN (\pp + (N + 1) \Hperp - \Uperp) 
                  \left( (\pp + \Iperp - \Hperp) \vi_i \right)
 - \DDEMI \IAzN (\pp + (N + 1) \Hperp - \Iperp) 
                \left( (\pe + \Uconf - \Hconf) \hci_i \right)
\nonumber \\ & &
 + \DDEMI \IBzA (\nabla^2 + K (N - 1)) \vi_i - \IBzA G_{i j} \vi^j
\nonumber \\ & &
 + G_{0 0} \nabla_i \ABBE
 - G_{0 \finN} \nabla_i \ANEE
 + G_{i j} \dot E^j , \\
\delta G_{i j} & = &
   (\gamma_{i j} \nabla^2 - \nabla_{i j})
   (\Psi + \Phi - (\hpp - \Phi) - N \Phi)
\nonumber \\ & &
  + \IAAzBB \gamma_{i j}
    \left(  2 (\Uconf + (N - 1) \Hconf) (\pe + \Uconf - \Iconf)
          + N (N - 1) \Hconf^2 \right) (\Phi + \Psi)
\nonumber \\ & &
 + \IAAzBB \gamma_{i j}
   \left(  (\pe^2 + (\Uconf - \Iconf) \pe) (\hpp - (N - 1) \Phi)
         - ((N - 1) \Hconf + \Uconf) \pe (\Psi + \hpp - N \Phi) \right)
\nonumber \\ & &
  + \IAAzNN \gamma_{i j}
    \left(  2 (\Iperp + (N - 1) \Hperp) (\pp + \Iperp - \Uperp)
          + N (N - 1) \Hperp^2 \right) (\hpp - \Phi)
\nonumber \\ & &
 + \IAAzNN \gamma_{i j}
   \left(  (\pp^2 + (\Iperp - \Uperp) \pp) (\Psi - (N - 1) \Phi)
         - ((N - 1) \Hperp + \Iperp) \pp (\Psi + \hpp + N \Phi) \right)
\nonumber \\ & &
 - \IAAzBN \gamma_{i j}
   \left(  \pp \pe
         + ((N - 1) \Hconf + \Uconf) \pp
         + ((N - 1) \Hperp + \Iperp) \pe
         + N (N - 1) \Hconf \Hperp \right) \vp
\nonumber \\ & &
 + 2 \IAA \gamma_{i j}
   \left(  \IzBB (\Phi + \Psi) \pe (\Uconf + (N - 1) \Hconf)
         + \IzNN (\hpp - \Phi) \pp (\Iperp + (N - 1) \Hperp) \right)
\nonumber \\ & &
 - \IAAzBN \gamma_{i j} \vp \left(  (N - 1) \Hconf{}' + \Uconf{}'
                                  + (N - 1) \dot{\Hperp} + \dot{\Iperp} \right)
\nonumber \\ & &
 - N \left(\IAAzBB \Uconf \pe - \IAAzNN \Iperp \pp \right) \Phi
 - 2 G_{i j} \Phi
\nonumber \\ & &
 + \IAzB \left(\pe + (N - 1) \Hconf + \Uconf \right) \nabla_{(i} \vi_{j)}
 - \IAzN \left(\pp + (N - 1) \Hperp + \Iperp \right) \nabla_{(i} \hci_{j)}
\nonumber \\ & &
 + \left(  \IAAzBB (\pe^2 + (N \Hconf + \Uconf - \Iconf) \pe)
         - (\nabla^2 - 2 K)
         - \IAAzNN (\pp^2 + (N \Hperp + \Iperp - \Uperp) \pp)
   \right) \TT{E}_{i j}
\nonumber \\ & &
 + \dot G_{i j} \ABBE
 - G'_{i j} \ANEE
 + 2 G_{k (i} (E^k_{j)} - \delta^k_{j)} \Phi) , \\
\delta G_{0 \finN} & = &
   \DDEMI \left(  \IBNzAA \nabla^2
                + N \INzB (  \dot{ \Hconf} + \Hconf^2
                          - \Hconf \Iconf - \Hconf \Uconf)
                - N \IBzN (  \Hperp' + \Hperp^2
                          - \Hperp \Iperp - \Hperp \Uperp)
          \right) \vp
\nonumber \\ & &
 + N (  \Hconf \pp (\Phi + \Psi) - \Hperp \pe (\hpp - \Phi)
      + (\pe \pp - \Uconf \pp - \Iperp \pe) \Phi)
\nonumber \\ & &
 + \DDEMI \left(\INzB G_{0 0} - \IBzN G_{\finN \finN}\right) \vp
\nonumber \\ & &
 + (\dot G_{0 \finN} + G_{0 \finN} \pe + G_{0 0} \pp ) \ABBE
 - (   G'_{0 \finN} + G_{0 \finN} \pp + G_{\finN \finN} \pe) \ANEE , \\
\delta G_{i \finN} & = &
 - \nabla_i (  (\Iperp - \Hperp) (\Psi + \hpp) + \pp (\Phi + \Psi)
             + N (\Hperp \hpp + \Phi') )
\nonumber \\ & &
 + \DDEMI \INzB (\pe + (N - 2) \Hconf + 2 \Uconf) \nabla_i \vp
\nonumber \\ & &
 + \DDEMI \IAzB (\pe + (N + 1) \Hconf - \Uconf)
                \left( (\pp + \Iperp - \Hperp) \vi_i \right)
 - \DDEMI \INAzBB (\pe + (N + 1) \Hconf - \Iconf)
                  \left( (\pe + \Uconf - \Hconf) \hci_i \right)
\nonumber \\ & &
 + \DDEMI \INzA (\nabla^2 + K (N - 1)) \hci_i - \INzA G_{i j} \hic^j
\nonumber \\ & &
 + G_{0 \finN} \nabla_i \ABBE
 - G_{\finN \finN} \nabla_i \ANEE
 + G_{i j} E^j{}' , \\
\delta G_{\finN \finN} & = &
   \INNzAA (\nabla^2 - K N (N - 1)) (\Psi + \Phi)
\nonumber \\ & &
 - N \left(  \INNzAA \nabla^2 - \INNzBB (N \Hconf - \Iconf) \pe
           + (N \Hperp + \Iperp) \pp \right) \Phi
\nonumber \\ & &
 + N \left(  \INNzBB \Hconf \pe (\Psi + \Phi)
           + \Hperp \pp (\Phi + \Psi)
           + ((N - 1) \Hperp^2 + 2 \Hperp \Iperp) (\Phi + \Psi) \right)
\nonumber \\ & &
 - N \INzB \left(\Hperp \pe + N \Hconf \Hperp + \dot{ \Hperp} \right) \vp
\nonumber \\ & &
 - 2 G_{\finN \finN} (\Psi + \hpp) + \INzB G_{0 \finN} \vp
\nonumber \\ & &
 + (\dot G_{\finN \finN} + 2 G_{0 \finN} \pp) \ABBE
 - (G'_{\finN \finN} + 2 G_{\finN \finN} \pp) \ANEE .
\end{eqnarray}

\subAPPsection{Riemann tensor}
\label{ssec_app_pert_riem}

\begin{eqnarray}
\delta R_{0 i 0 j} & = &
 - \left(  \IBB \nabla_{i j}
         + \IAA \Hconf \gamma_{i j} \pe
         + \IAABBzNN \Hperp \gamma_{i j} \pp \right) \Psi
 - 2 \IAABBzNN \Hperp \Iperp \gamma_{i j} (\Psi + \hpp)
\nonumber \\ & &
 + \IAABzN \gamma_{i j} (\Hperp \pe + \Hconf \Iperp + \Uconf \Hperp) \vp
 + \IAB (\pe + \Hconf) \nabla_{(i} \vi_{j)}
 - \IABBzN \Iperp \nabla_{(i} \hci_{j)}
\nonumber \\ & &
 + \left(\IAA (\pe^2 + (2 \Hconf - \Iconf) \pe) - \IAABBzNN \Iperp \pp \right)
   (\TT{E}_{i j} - \gamma_{i j} \Phi)
\nonumber \\ & &
 + (\dot R_{0 i 0 j} + 2 R_{0 i 0 j} \pe) \ABBE
 - (R'_{0 i 0 j} + R_{\finN i 0 j} \pe + R_{0 i \finN j} \pe) \ANEE
\nonumber \\ & &
 + R_{0 k 0 j} (E^k_i - \delta^k_i \Phi)
 + R_{0 i 0 k} (E^k_j - \delta^k_j \Phi) , \\
\delta R_{0 i j k} & = &
   \IAA \nabla_j \pe (\TT{E}_{i k} - \gamma_{i k} \Phi)
 - \IAA \Hconf \gamma_{i k} \nabla_j \Psi
 + \IAB \nabla_j \nabla_{(i} \vi_{k)}
\nonumber \\ & &
 + \DDEMI \IAA \Hperp \gamma_{i k}
   \left(  \IBzN \nabla_j \vp
         - \IABzNN (\pp + \Iperp - \Hperp) \vi_j
         + \IAzN (\pe + \Uconf - \Hconf) \hci_j \right)
\nonumber \\ & &
 - [j \leftrightarrow k]
\nonumber \\ & &
 + (R_{0 i 0 k} \nabla_j + R_{0 i j 0} \nabla_k) \ABBE
 - (R_{0 i \finN k} \nabla_j + R_{0 i j \finN} \nabla_k) \ANEE
\nonumber \\ & &
 + R_{l i j k} \left(\nabla^l \dot E - \IBzA \vi^l \right) , \\
\delta R_{0 i 0 \finN} & = &
 - \IBB (\pp + \Iperp - \Hperp) \nabla_i \Psi - \IBB \Iperp \nabla_i \hpp
 + \DDEMI \IBN (\pe + 2 \Uconf - \Hconf) \nabla_i \vp
\nonumber \\ & &
 + \DDEMI \IAB (\pe + 2 \Hconf - \Uconf) 
               \left( (\pp + \Uperp - \Hperp) \vi_i \right)
 - \DDEMI \INA (\pe + 2 \Hconf - \Iconf)
               \left( (\pe + \Iconf - \Hconf) \hci_i \right)
\nonumber \\ & &
 - R_{0 \finN 0 \finN} \nabla_i \ANEE
 + R_{0 i j \finN} \left(\dot E^j - \IBzA \vi^j \right)
 + R_{0 i 0 j} \left(E^j{}' - \INzA \hic^j \right) , \\
\delta R_{0 i \finN j} & = &
   \IAA \gamma_{i j} (\Hperp \dot \hpp - \Hconf \Psi')
 - \DDEMI \left(  \IBN \nabla_{i j} - \INAAzB \gamma_{i j} \Hconf \Uconf
                + \IAABzN \gamma_{i j} \Hperp \Iperp \right) \vp
\nonumber \\ & &
 + \IAA (\pe \pp + (\Hconf - \Uconf) \pp + (\Hperp - \Iperp) \pe )
   (\TT{E}_{i j} - \gamma_{i j} \Phi)
\nonumber \\ & &
 + \DDEMI \IAB (\pp + \Hperp - \Iperp) \nabla_{(i} \vi_{j)}
 + \DDEMI \INA (\pe + \Hconf - \Uconf) \nabla_{(i} \hci_{j)}
\nonumber \\ & &
 + (\dot R_{0 i \finN j} + R_{0 i \finN j} \pe + R_{0 i 0 j} \pp) \ABBE
\nonumber \\ & &
 - (R'_{0 i \finN j} + R_{0 i \finN j} \pp + R_{\finN i \finN j} \pe) \ANEE
\nonumber \\ & &
 + R_{0 k \finN j} (E^k_i - \delta^k_i \Phi)
 + R_{0 i \finN k} (E^k_j - \delta^k_j \Phi) , \\
\delta R_{i j k l} & = &
   \left(  \IAAAAzBB \Hconf^2 \Psi + \IAAAAzNN \Hperp^2 \hpp
         - \IAAAAzBN \Hconf \Hperp \vp \right) \gamma_{i k} \gamma_{j l}
 - \gamma_{k i} \left(  \IAAAzB \Hconf \nabla_{(j} \vi_{l)}
                      - \IAAAzN \Hperp \nabla_{(j} \hci_{l)} \right)
\nonumber \\ & &
 + \left(  \IAA (\nabla_{k i} + K \gamma_{k i})
         - \IAAAAzBB \gamma_{k i} \Hconf \pe
         + \IAAAAzNN \gamma_{k i} \Hperp \pp \right)
   (\TT{E}_{j l} - \gamma_{j l} \Phi)
\nonumber \\ & &
 - [i \leftrightarrow j]
 + [ik \leftrightarrow jl]
 - [k \leftrightarrow l]
\nonumber \\ & &
 + R_{m j k l} (E^m_i - \delta^m_i \Phi)
 + R_{i m k l} (E^m_j - \delta^m_j \Phi)
 + R_{i j m l} (E^m_k - \delta^m_k \Phi)
 + R_{i j k m} (E^m_l - \delta^m_l \Phi)
\nonumber \\ & &
 + \dot R_{i j k l} \ABBE - R'_{i j k l} \ANEE , \\
\delta R_{i j 0 \finN} & = &
   \nabla_i \left(  \IAB (\pp + \Iperp - \Hperp) \vi_j
                  - \INA (\pe + \Uconf - \Hconf) \hci_j \right)
 - [i \leftrightarrow j] , \\
\delta R_{\finN i j k} & = &
   \IAA \nabla_j \pp (\TT{E}_{i j} - \gamma_{i j} \Phi)
 + \IAA \Hperp \gamma_{i k} \nabla_j \hpp
 + \INA \nabla_j \nabla_{(i} \hci_{k)}
\nonumber \\ & &
 - \DDEMI \IAA \Hconf \gamma_{i k}
   \left(  \INzB \nabla_j \vp
         + \IAzB (\pp + \Iperp - \Hperp) \vi_j
         - \INAzBB (\pe + \Uconf - \Hconf) \hci_j \right)
\nonumber \\ & &
 - [j \leftrightarrow k]
\nonumber \\ & &
 + (R_{\finN i 0 k} \nabla_j + R_{\finN i j 0} \nabla_k) \ABBE
 - (R_{\finN i \finN k} \nabla_j + R_{\finN i j \finN} \nabla_k) \ANEE
\nonumber \\ & &
 + R_{l i j k} \left(\nabla^l E' - \INzA \hic^l \right) , \\
\delta R_{0 \finN 0 \finN} & = &
 - \IBB (\pp - \Uperp) \pp \Psi - 2 \IBB (\pp + \Iperp - \Uperp) (\Iperp \Psi)
 - \IBB \Iperp \pp \hpp
\nonumber \\ & &
 - \INN (\pe - \Iconf) \pe \hpp - 2 \INN (\pe + \Uconf - \Iconf) (\Uconf \hpp)
 - \INN \Uconf \pe \Psi
\nonumber \\ & &
 + \IBN \left(\pe (\TDEMI \pp + \Iperp) + \pp (\TDEMI \pe + \Uconf) \right) \vp
\nonumber \\ & &
 + (\dot R_{0 \finN 0 \finN} + 2 R_{0 \finN 0 \finN} \pe) \ABBE
 - (R'_{0 \finN 0 \finN} + 2 R_{0 \finN 0 \finN} \pp) \ANEE , \\
\delta R_{\finN 0 \finN i} & = &
 - \INN (\pe + \Uconf - \Hconf) \nabla_i \hpp - \INN \Uconf \nabla_i \Psi
 + \DDEMI \IBN (\pp + 2 \Iperp - \Hperp) \nabla_i \vp
\nonumber \\ & &
 - \DDEMI \IAB (\pp + 2 \Hperp - \Uperp)
               \left( (\pp + \Iperp - \Hperp) \vi_i \right)
 + \DDEMI \INA (\pp + 2 \Hperp - \Iperp)
               \left( (\pe + \Uconf - \Hconf) \hci_i \right)
\nonumber \\ & &
 + R_{\finN 0 \finN 0} \nabla_i \ABBE
 + R_{j 0 \finN i} \left(E^j{}' - \INzA \hic^j \right)
 + R_{\finN j \finN i} \left(\dot E^j - \IBzA \vi^j \right) , \\
\delta R_{\finN i \finN j} & = &
   \left(- \INN \nabla_{i j}
         + \IAA \Hperp \gamma_{i j} \pp
         + \INNAAzBB \Hconf \gamma_{i j} \pe \right) \hpp
 + 2 \INNAAzBB \Hconf \Uconf \gamma_{i j} (\Psi + \hpp)
\nonumber \\ & &
 - \INAAzB \gamma_{i j} (\Hconf \pp + \Hconf \Iperp + \Uconf \Hperp) \vp
 - \INNAzB \Uconf \nabla_{(i} \vi_{j)}
 + \INA (\pp + \Hperp) \nabla_{(i} \hci_{j)}
\nonumber \\ & &
 + \left(\IAA (\pp^2 + (2 \Hperp - \Uperp) \pp) - \INNAAzBB \Uconf \pe \right)
   (\TT{E}_{i j} - \gamma_{i j} \Phi)
\nonumber \\ & &
   (  \dot R_{\finN i \finN j}
    + 2 R_{0 i \finN j} \pe + 2 R_{\finN i 0 j} \pe) \ABBE
\nonumber \\ & &
 - (R'_{\finN i \finN j} + 2 R_{\finN i \finN j} \pp) \ANEE
\nonumber \\ & &
 + R_{\finN k \finN j} (E^k_i - \delta^k_i)
 + R_{\finN i \finN k} (E^k_j - \delta^k_j) .
\end{eqnarray}

\subAPPsection{Weyl tensor}
\label{ssec_app_pert_weyl}

Defining
\begin{eqnarray}
\EZA_{i j} & = &
   (\pe + \Uconf - \Iconf) \pe \TT{E}_{i j}
 + \IBzA (\pe + \Uconf - \Hconf) \nabla_{(i} \vi_{j)} , \\
\EZB_{i j} & = &
   (\Hconf - \Uconf) \pe \TT{E}_{i j}
 + \IBzA (\Hconf - \Uconf) \nabla_{(i} \vi_{j)} , \\
\EPA_{i j} & = &
   (\pp + \Iperp - \Uperp) \pp \TT{E}_{i j}
 + \INzA (\pp + \Iperp - \Hperp) \nabla_{(i} \hci_{j)} , \\
\EPB_{i j} & = &
   (\Hperp - \Iperp) \pp \TT{E}_{i j}
 + \INzA (\Hperp - \Iperp) \nabla_{(i} \hci_{j)} , \\
\XZ & = &
(\pe + \Uconf - \Iconf)
\left((\Hconf - \Uconf) (\Psi + \hpp) - \pe(\hpp - \Phi) \right) , \\
\XP & = &
(\pp + \Iperp - \Uperp)
\left((\Hperp - \Iperp) (\Psi + \hpp) - \pp(\Phi + \Psi) \right) , \\
\VP & = &
(  \pe \pp
 + (\Iperp - \Hperp) \pe + (\Uconf - \Hconf) \pp 
 + \dot{\Iperp} - \dot{\Hperp} + \Uconf{}' - \Hconf{}' ) \vp .
\end{eqnarray}
\begin{eqnarray}
\delta C_{0 i 0 j} & = &
 - \IBB \left(\nabla_{i j} - \frac{1}{N} \nabla^2 \gamma_{i j} \right)
   \left(\frac{N - 1}{N} (\Phi + \Psi) + \frac{1}{N} (\hpp - \Phi) \right)
\nonumber \\ & &
 - \IBB \frac{N - 1}{N^2 (N + 1)} \gamma_{i j}
   \left(\nabla^2 + K N \right) (2 \Phi + \Psi - \hpp)
\nonumber \\ & &
 + C_{0 i 0 j} (\Psi + \hpp) + \IBB \frac{1}{N} (\nabla^2 - 2 K) \TT{E}_{i j}
\nonumber \\ & &
 - \frac{N - 1}{N (N + 1)} \gamma_{i j}
   \left(\IAA \XZ + \IAABBzNN \XP + \IAABzN \VP \right)
\nonumber \\ & &
 + \IAA \left(\frac{N - 1}{N} \EZA_{i j} + \EZB_{i j} \right)
 + \IAABBzNN \left(\frac{1}{N} \EPA_{i j} + \EPB_{i j} \right)
\nonumber \\ & &
   (\dot C_{0 i 0 j} + 2 C_{0 i 0 j} \pe) \ABBE
 - (C'_{0 i 0 j} + C_{\finN i 0 j} \pe + C_{0 i \finN j} \pe) \ANEE
\nonumber \\ & &
 + C_{0 k 0 j} (E^k_i - \delta^k_i \Phi)
 + C_{0 i 0 k} (E^k_j - \delta^k_j \Phi) , \\
\delta C_{0 i j k} & = &
   \frac{1}{N} \gamma_{i k} \nabla_j
   \left(  \IAA \pe (\hpp - \Phi)
         + \IAA (\Uconf - \Hconf) (\Psi + \hpp)
         - \IAABzN (\TDEMI \pp + \Iperp - \Hperp) \vp
   \right)
\nonumber \\ & &
 + \DDEMI \frac{1}{N} \gamma_{i k}
   \left(  \IAAABzNN (\pp + \Hperp - \Uperp)
                     \left( (\pp + \Iperp - \Hperp) \vi_j \right)
         - \IAAAzN (\pp + \Hperp - \Iperp)
                   \left( (\pe + \Uconf - \Hconf) \hci_j \right)
   \right)
\nonumber \\ & &
 + \DDEMI \frac{1}{N} \gamma_{i k} \IAB (\nabla^2 + K (N - 1)) \vi_j
 + \IAB \nabla_j \nabla_{(i} \vi_{k)} + \IAA \nabla_j \dot{\TT{E}}_{i k}
\nonumber \\ & &
 - [j \leftrightarrow k]
\nonumber \\ & &
 + (C_{0 i 0 k} \nabla_j + C_{0 i j 0} \nabla_k) \ABBE
 - (C_{0 i \finN k} \nabla_j + C_{0 i j \finN} \nabla_k) \ANEE
\nonumber \\ & &
 + C_{l i j k} \left(\nabla^l \dot E - \IBzA \vi^l \right) , \\
\delta C_{0 i 0 \finN} & = &
 - \frac{N - 1}{N} \IBB \nabla_i
   \left(\pp (\Phi + \Psi) + (\Iperp - \Hperp) (\hpp + \Psi) \right)
 + \frac{N - 1}{N} \IBN
   \left(\TDEMI \pezB + \Uconf - \Hconf \right) \nabla_i \vp
\nonumber \\ & &
 - \DDEMI \frac{1}{N} \IBBNzA (\nabla^2 + K(N - 1) ) \hci_i
\nonumber \\ & &
 + \DDEMI \frac{N - 1}{N}
   \left(  \IAB (\pe + \Hconf - \Uconf)
                \left( (\pp + \Iperp - \Hperp) \vi_i \right)
         - \INA (\pe + \Hconf - \Iconf)
                \left( (\pe + \Uconf - \Hconf) \hci_i \right) \right)
\nonumber \\ & &
 - C_{0 \finN 0 \finN} \nabla_i \ANEE
 + C_{j \finN 0 i} \left(\dot E^j - \IBzA \vi^j \right)
 + C_{0 j 0 i} \left(E^j{}' - \INzA \hic^j \right) , \\
\delta C_{0 i\finN j} & = &
 - \DDEMI \IBN 
   \left(\nabla_{i j} - \frac{1}{N} \nabla^2 \gamma_{i j} \right) \vp
 - \frac{1}{N} \IAAzBN C_{0 \finN 0 \finN} \gamma_{i j} \vp
\nonumber \\ & &
 + \DDEMI \IAB (\pp + \Hperp - \Iperp) \nabla_{(i} \vi_{j)}
 + \DDEMI \INA (\pe + \Hconf - \Uconf) \nabla_{(i} \hci_{j)}
\nonumber \\ & &
 + \IAA \left(  (\TDEMI \pp + \Hperp - \Iperp) \pe
              + (\TDEMI \pe + \Hconf - \Uconf) \pp \right) \TT{E}_{i j}
\nonumber \\ & &
 + (\dot C_{0 i \finN j} + C_{0 i \finN j} \pe + C_{0 i 0 j} \pp) \ABBE
 - (C'_{0 i \finN j} + C_{0 i \finN j} \pp + C_{\finN i \finN j} \pe) \ANEE
\nonumber \\ & &
 + C_{0 k \finN j} (E^k_i - \delta^k_i \Phi)
 + C_{0 i \finN k} (E^k_j - \delta^k_j \Phi) , \\
\delta C_{i j k l} & = &
 - \frac{1}{N} \IAA 
   \left(\nabla_{i k} - \frac{1}{N} \nabla^2 \gamma_{i k} \right)
   \gamma_{j l} (2 \Phi + \Psi - \hpp)
\nonumber \\ & &
 - \frac{1}{N^2 (N + 1)} \IAA \gamma_{i k} \gamma_{j l}
   \left(\nabla^2 + K N \right) (2 \Phi + \Psi - \hpp)
\nonumber \\ & &
 + C_{i j k l} (\hpp - \Psi)
 + \IAA \left(  \nabla_{k i} + K \gamma_{k i}
              - \frac{1}{N} \gamma_{k i} (\nabla^2 - 2 K) \right)
   \TT{E}_{j l}
\nonumber \\ & &
 - \frac{1}{N (N + 1)} \gamma_{i k} \gamma_{j l}
   \left(\IAAAAzBB \XZ + \IAAAAzNN \XP + \IAAAAzBN \VP \right)
\nonumber \\ & &
 + \IAAAAzBB \frac{1}{N} \gamma_{i k} \EZA_{j l}
 - \IAAAAzNN \frac{1}{N} \gamma_{i k} \EPA_{j l}
\nonumber \\ & &
 - [i \leftrightarrow j]
 + [ik \leftrightarrow jl]
 - [k \leftrightarrow l]
\nonumber \\ & &
 + C_{m j k l} (E^m_i - \delta^m_i \Phi)
 + C_{i m k l} (E^m_j - \delta^m_j \Phi)
 + C_{i j m l} (E^m_k - \delta^m_k \Phi)
 + C_{i j k m} (E^m_l - \delta^m_l \Phi)
\nonumber \\ & &
 + \dot C_{i j k l} \ABBE - C'_{i j k l} \ANEE , \\
\delta C_{i j 0 \finN} & = &
   \nabla_i \left(  \IAB (\pp + \Iperp - \Hperp) \vi_j
                  - \INA (\pe + \Uconf - \Hconf) \hci_j \right)
 - [i \leftrightarrow j] , \\
\delta C_{\finN i j k} & = &
 - \frac{1}{N} \gamma_{i k} \nabla_j
   \left(  \IAA \pp (\Phi + \Psi)
         + \IAA (\Iperp - \Hperp) (\Psi + \hpp)
         - \INAAzB (\TDEMI \pe + \Uconf - \Hconf) \vp
   \right)
\nonumber \\ & &
 + \DDEMI \frac{1}{N} \gamma_{i k}
   \left(  \IAAAzB (\pe + \Hconf - \Uconf)
                   \left( (\pp + \Iperp - \Hperp) \vi_j \right)
         - \INAAAzBB (\pe + \Hconf - \Iconf)
                     \left( (\pe + \Uconf - \Hconf) \hci_j \right) \right)
\nonumber \\ & &
 + \DDEMI \frac{1}{N} \gamma_{i k} \INA (\nabla^2 + K (N - 1)) \hci_j
 + \INA \nabla_j \nabla_{(i} \hci_{k)} + \IAA \nabla_j \TT{E}{}'_{i k}
\nonumber \\ & &
 - [j \leftrightarrow k]
\nonumber \\ & &
 + (C_{\finN i 0 k} \nabla_j + C_{\finN i j 0} \nabla_k) \ABBE
 - (C_{\finN i \finN k} \nabla_j + C_{\finN i j \finN} \nabla_k) \ANEE
\nonumber \\ & &
 + C_{l i j k} \left(\nabla^l E' - \INzA \hic^l \right) , \\
\delta C_{0 \finN 0 \finN} & = &
   \IBBNNzAA \frac{N - 1}{N (N + 1)} \left(\nabla^2 + K N \right)
   (2 \Phi + \Psi - \hpp)
\nonumber \\ & &
 + \frac{N - 1}{N + 1}
   \left(\INN \XZ + \IBB \XP + \IBN \VP \right)
\nonumber \\ & &
 + (\dot C_{0 \finN 0 \finN} + 2 C_{0 \finN 0 \finN} \pe) \ABBE
 - (C'_{0 \finN 0 \finN} + 2 C_{0 \finN 0 \finN} \pp) \ANEE , \\
\label{cloclo}
\delta C_{\finN 0 \finN i} & = &
   \frac{N - 1}{N} \INN \nabla_i
   \left(  \pe(\Phi - \hpp) + (\Hconf - \Uconf) (\Psi + \hpp)
         + (\TDEMI \pp + \Iperp - \Hperp) \vp
  \right)
\nonumber \\ & &
 + \DDEMI \frac{1}{N} \IBNNzA (\nabla^2 + K (N - 1))  \vi_i
\nonumber \\ & &
 - \DDEMI \frac{N - 1}{N}
   \left(  \IAB (\pp + \Hperp - \Uperp)
                \left( (\pp + \Iperp -\Hperp) \vi_i \right)
         - \INA (\pp + \Hperp - \Iperp)
                \left( (\pe + \Uconf - \Hconf) \hci_i \right) \right)
\nonumber \\ & &
 + C_{\finN 0 \finN 0} \nabla_i \ABBE
 + C_{j 0 \finN i} \left(E^j{}' - \INzA \hic^j \right)
 + C_{\finN j \finN i} \left(\dot E^j - \IBzA \vi^j \right) , \\
\delta C_{\finN i \finN j} & = &
 - \INN \left(\nabla_{i j} - \frac{1}{N} \nabla^2 \gamma_{i j} \right)
   \left(\frac{N - 1}{N} (\hpp - \Phi) + \frac{1}{N} (\Phi + \Psi) \right)
\nonumber \\ & &
 + \INN \frac{N - 1}{N^2 (N + 1)} \gamma_{i j}
   \left(\nabla^2 + K N \right) (2 \Phi + \Psi - \hpp)
\nonumber \\ & &
 - C_{\finN i \finN j} (\Psi + \hpp)
 - \INN \frac{1}{N} (\nabla^2 - 2 K) \TT{E}_{i j}
\nonumber \\ & &
 + \frac{N - 1}{N (N + 1)} \gamma_{i j}
   \left(\INNAAzBB \XZ + \IAA \XP + \INAAzB \VP \right)
\nonumber \\ & &
 + \INNAAzBB \left(\frac{1}{N} \EZA_{i j} + \EZB_{i j} \right)
 + \IAA \left(\frac{N - 1}{N} \EPA_{i j} + \EPB_{i j} \right)
\nonumber \\ & &
 + (\dot C_{\finN i \finN j} + 2 C_{\finN i \finN j} \pe) \ABBE
 - (C'_{\finN i \finN j} + 2 C_{\finN i \finN j} \pp) \ANEE
\nonumber \\ & &
 + C_{\finN k \finN j} (E^k_i - \delta^k_i \Phi)
 + C_{\finN i \finN k} (E^k_j - \delta^k_j \Phi) .
\end{eqnarray}

\APPsection{Perturbed matter content}
\label{sec_app_pert_bulk_matt}

\subAPPsection{Unit vectors}

\begin{eqnarray}
\delta u^\alpha & = &
\left(- \IzB A, {\bf 0}, 0 \right) , \\
\delta u_\alpha & = &
\left(\IB A, \IA B_i, \IN \Bperp \right) , \\
\delta n^\alpha & = &
\left(0, {\bf 0}, - \IzN \epp \right) , \\
\delta n_\alpha & = &
\left(- \IB \Bperp, - \IA \epi_i, - \IN \epp \right) , \\
\delta U^\alpha & = &
\left(\IzB \GAM (\BET \www - A - \BET \Bperp), \IzA \RVD{v}^i,
      \IzN \GAM (\www + \BET \epp) \right) , \\
\delta U_\alpha & = &
\left(  \IB \GAM (\BET \www + A),
      - \IA (\RVU{v}_i - \GAM B_i - \BGM \epi_i) ,
      - \IN \GAM (\www - \Bperp - \BET \epp) \right) , \\
\delta N^\alpha & = &
\left(\IzB \GAM (- \www + \BET A + \Bperp), \IzA \RVD{\dni}^i,
      - \IzN \GAM (\epp + \BET \www) \right) , \\
\delta N_\alpha & = &
\left(- \IB \GAM (\www + \BET A),
      - \IA (\RVU{\dni}_i + \BGM B_i + \GAM \epi_i) ,
        \IN \GAM (\BET \www - \epp - \BET \Bperp) \right) ,
\end{eqnarray}
with
\begin{eqnarray}
u_\mu U^\mu & = & \GAM , \\
\delta (u_\mu U^\mu) & = & \delta \GAM , \\
\www & = & \frac{\delta \GAM}{\BGM}
          =   \frac{\delta (\BGM)}{\GAM} , \\
\frac{\www}{\BET}
 & = &   \frac{\delta \BET}{\BET}
       + \frac{\delta \GAM}{\GAM}
   =     \frac{\delta (\BGM)}{\BGM} , \\
\GAM \www
 & = &   \delta (\BGM) , \\
\RVU{v}_i & = & \nabla_i \RVD{v} + \VVRVU{v}_i , \\
\RVU{\dni}_i & = & \nabla_i \RVD{\dni} + \VVRVU{\dni}_i .
\end{eqnarray}
Since $n^\alpha + \delta n^\alpha$ is not orthogonal to $u^\alpha +
\delta u^\alpha$, one does not have $\delta (n_\mu U^\mu) = \delta
(\BGM)$, but rather $\delta (n_\mu U^\mu) = \delta (\BGM) + (U^\mu
u_\mu) \delta(n^\nu u_\nu)$.

\subAPPsection{Gauge transformation}

\begin{eqnarray}
\RVD{v} & \to & \RVD{v}  - \IAzB \GAM \dot L - \IAzN \BGM L' , \\
\VVRVU{v}_i
 & \to &   \VVRVU{v}_i
         - \IAzB \GAM \dot{\VV{L}}_i - \IAzN \BGM {\VV{L}}_i{}' , \\
\RVD{\dni} & \to & \RVD{\dni} + \IAzB \BGM \dot L + \IAzN \GAM L' , \\
\VVRVU{\dni}_i
 & \to &   \VVRVU{\dni}_i
         + \IAzB \BGM \dot{\VV{L}}_i + \IAzN \GAM {\VV{L}}_i{}' , \\
\www
 & \to &   \www + \frac{\dot \GAM}{\BGM} T
                   + \frac{\GAM'}{\BGM} \Lperp
                      - \INzB \dot \Lperp , \\
\label{lie_scal_bulk}
\REST{\delta X} & \to & \REST{\delta X} + \REST{\dot{X}} T
                                        + \REST{X}' \Lperp ,
\end{eqnarray}
where $\REST{X}$ is any ($N+2$)-scalar quantity (density
$\REST{\rho}$, pressure $\REST{P}$, etc). Note: $\gamma$, $\beta$,
$\rho$, $\PRESY$, $F$ are {\em not} scalars as they are defined
through the vector fields $u^\alpha$, $n^\alpha$ which specifically
depend on a coordinate choice.

\subAPPsection{Gauge invariant quantities}

\begin{eqnarray}
\label{gi_v_scal}
\RVD{v}^\sharp & = & \RVD{v} + \IAzB \GAM \dot E + \IAzN \BGM E' , \\
\label{gi_v_vect}
\VVRVU{v}_i{}^\sharp
 & = & \VVRVU{v}_i + \IAzB \GAM \dot{\VV{E}}_i + \IAzN \BGM
 {\VV{E}}_i{}' , \\
\RVD{\dni}^\sharp & = & \RVD{\dni} - \IAzB \BGM \dot E - \IAzN \GAM E' , \\
\VVRVU{\dni}_i{}^\sharp
 & = &   \VVRVU{\dni}_i
       - \IAzB \BGM \dot{\VV{E}}_i - \IAzN \GAM {\VV{E}}_i{}' , \\
\www^\sharp
 & = &   \www
       - \frac{\dot \GAM}{\BGM} \ABBE
       + \frac{\gamma'}{\BGM} \ANEE
       - \INzB \pe \ANEE 
\nonumber \\
 & = &   \www
       - \frac{\pe (\BGM)}{\GAM} \ABBE
       + \frac{(\BGM)}{\GAM}' \ANEE
       - \INzB \pe \ANEE , \\
\REST{\delta X}^\sharp & = & \REST{\delta X} - \REST{\dot{X}} \ABBE
                                             + \REST{X}' \ANEE .
\end{eqnarray}
It is useful to define
\begin{eqnarray}
v^\sharp & \equiv &   \GAM \RVD{v}^\sharp
                    + \BGM \RVD{\dni}^\sharp , \\
\VV{v}{}^\sharp_i & \equiv &   \GAM \VVRVU{v}_i{}^\sharp
                             + \BGM \VVRVU{\dni}_i{}^\sharp , \\
\dni^\sharp & \equiv &   \BGM \RVD{v}^\sharp
                       + \GAM \RVD{\dni}^\sharp , \\
\VV{\dni}{}^\sharp_i & \equiv &   \BGM \VVRVU{v}_i{}^\sharp
                                + \GAM \VVRVU{\dni}_i{}^\sharp .
\end{eqnarray}

\subAPPsection{Stress-energy tensor}

\begin{equation}
\delta T_{\alpha \beta}
 =   \delta ((\REST{P} + \REST{\rho}) U_\alpha U_\beta )
   - \delta ((\REST{P} - \REST{\PRESY}) N_\alpha N_\beta)
   - \delta (\REST{P} g_{\alpha \beta}) + \pi_{\alpha \beta} .
\end{equation}
\begin{eqnarray}
\pi_{\alpha \mu} U^\mu & = & 0 , \\
\pi_{\alpha \mu} N^\mu & = & 0 , \\
\pi_{0 \alpha} & = & 0 , \\
\pi_{\finN \alpha} & = & 0 , \\
\pi_{i j} & = & \IAA \Pi_{i j}
            =   \IAA \left(  \left(  \nabla_{i j}
                                   - \frac{1}{N} \nabla^2 \gamma_{ij}
                             \right) \Pi
                           + \nabla_{(i} \VV{\Pi}_{j)}
                           + \TT{\Pi}_{i j} \right) ,
\end{eqnarray}
\begin{eqnarray}
\delta T_{0 0} & = &
   \IBB \GMM (  \delta \REST{\rho} + \BTT \delta \REST{\PRESY}
               + 2 A (\REST{\rho} + \BTT \REST{\PRESY})
 + 2 \BET (\REST{\PRESY} + \REST{\rho}) \www)
\nonumber \\ & = &
   \IBB \left(  \delta \rho^\sharp + 2 \rho \Psi \right)
\nonumber \\ & &
 + (\dot T_{0 0} + 2 T_{0 0} \pe) \ABBE
 - (T'_{0 0} + 2 T_{0 \finN} \pe) \ANEE , \\
\delta T_{0 i} & = &
 - \IAB ( (\REST{P} + \REST{\rho}) \GAM \RVU{v}_i
 -(\REST{\PRESY} - \REST{P}) \BGM \RVU{\dni}_i
 -   \GMM (\REST{\rho} + \BTT \REST{\PRESY}) B_i
              - (\REST{\rho} + \REST{\PRESY}) \epi_i)
\nonumber \\ & = &
 - \IAB \left( (P + \rho) v^\sharp_i - F \dni^\sharp_i \right)
 + \IAzB T_{0 0} \vi_i - \IAzN T_{0 \finN} \hci_i
\nonumber \\ & &
 + T_{0 0} \nabla_i \ABBE
 - T_{0 \finN} \nabla_i \ANEE
 + T_{i j} \dot E^j , \\
\delta T_{i j} & = &
\IAA (  \delta \REST{P} \gamma_{i j} + \Pi_{i j}
      + 2 \REST{P} (C \gamma_{i j} +  E_{i j}))
\nonumber \\ & = &
   \IAA \gamma_{i j} \delta P^\sharp + \IAA \Pi_{i j}
\nonumber \\ & &
 + \dot T_{i j} \ABBE
 - T'_{i j} \ANEE
 + 2 T_{k (i} (E^k_{j)} - \delta^k_{j)} \Phi) , \\
\delta T_{0 \finN} & = &
 - \IBN \GMM \left(  \BET (\delta \REST{\PRESY} + \delta \REST{\rho})
              + (\REST{\PRESY} + \REST{\rho})  (1 + \BTT) \www
              + \BET (\REST{\PRESY} + \REST{\rho})  (A - \epp)
              -  (\REST{\rho} + \BTT \REST{\PRESY}) \Bperp \right)
\nonumber \\ & = &
 - \IBN \left(  \delta F^\sharp + F (\Psi - \hpp) - \rho \vp \right)
\nonumber \\ & &
 + (\dot T_{0 \finN} + T_{0 \finN} \pe + T_{0 0} \pp) \ABBE
 - (   T'_{0 \finN} + T_{0 \finN} \pp
    + T_{\finN \finN} \pe) \ANEE , \\
\delta T_{i \finN} & = &
\INA \left(  \GAM (\REST{P} - \REST{\PRESY}) \RVU{\dni}_i
           + \BGM (\REST{P} + \REST{\rho}) \RVU{v}_i
           - \GMM (\REST{\PRESY} + \BTT \REST{\rho}) \epi_i
           - \BGG (\REST{\rho} + \REST{\PRESY}) B_i \right) 
\nonumber \\ & = &
   \INA \left(F v^\sharp_i + (P - \PRESY) \dni^\sharp_i \right)
 + \IAzB T_{0 \finN} \vi_i - \IAzN T_{\finN \finN} \hci_i
\nonumber \\ & &
 + T_{0 \finN} \nabla_i \ABBE
 - T_{\finN \finN} \nabla_i \ANEE
 + T_{i j} E^j{}' , \\
\delta T_{\finN \finN} & = &
   \INN \GMM \left(   (  \delta \REST{\PRESY} + \BTT \delta \REST{\rho})
              + 2 \BET (\REST{\PRESY} + \REST{\rho}) \www
              - 2  (\REST{\PRESY} + \BTT \REST{\rho}) \epp
              - 2 \BET (\REST{\PRESY} + \REST{\rho}) \Bperp \right) 
\nonumber \\ & = &
   \INN \left(  \delta \PRESY^\sharp - 2 \PRESY \hpp - 2 F \vp \right) 
\nonumber \\ &   &
 + (\dot T_{\finN \finN} + 2 T_{0 \finN} \pp) \ABBE
 - (T'_{\finN \finN} + 2 T_{\finN \finN} \pp) \ANEE .
\end{eqnarray}
\begin{eqnarray}
\delta T^{0 0} & = &
   \IzBB \left(  \delta \rho^\sharp - 2 \rho \Psi - 2 F \vp \right)
\nonumber \\ & &
 + (\dot T^{0 0} - 2 T^{0 0} \pe - 2 T^{0 \finN} \pp) \ABBE
 - T^{0 0}{}'  \ANEE , \\
\delta T^{0 i} & = &
   \IzAB \left( (P + \rho) v_\sharp^i - F \dni_\sharp^i \right)
 - \IAzB T^{i j} \vi_j
\nonumber \\ & &
 - T^{i j} \nabla_j \ABBE
 - (T^{0 0} \pe + T^{0 \finN} \pp) E^i , \\
\delta T^{i j} & = &
   \IzAA \left(\delta P^\sharp \gamma^{i j} + \Pi^{i j} \right)
\nonumber \\ & &
 + \dot T^{i j} \ABBE
 - T^{i j}{}' \ANEE
 - 2 T^{k (i} (E_k^{j)} - \delta_k^{j)} \Phi) , \\
\delta T^{0 \finN} & = &
   \IzBN \left(  \delta F^\sharp + F (\hpp - \Psi)
               - \PRESY \vp \right)
\nonumber \\ & &
 + (\dot T^{0 \finN} - T^{0 \finN} \pe - T^{\finN \finN} \pp) \ABBE
 - (   T^{0 \finN}{}' - T^{0 \finN} \pp - T^{0 0} \pe) \ANEE , \\
\delta T^{i \finN} & = &
   \IzNA \left( F v_\sharp^i + (P - \PRESY) \dni_\sharp^i \right)
 + \IAzN T^{i j} \hci_j
\nonumber \\ & &
 + T^{i j} \nabla_j \ANEE
 - (T^{0 \finN} \pe + T^{\finN \finN} \pp) E^i , \\
\delta T^{\finN \finN} & = &
 \IzNN \left(  \delta \PRESY^\sharp + 2 \PRESY \hpp \right)
\nonumber \\ & &
 + \dot T^{\finN \finN} \ABBE
 - (T^{\finN \finN}{}' - 2 T^{\finN \finN} \pp - 2 T^{0 \finN} \pe ) \ANEE .
\end{eqnarray}

\subAPPsection{An example: a scalar field}
\label{ssec_app_sf}

\begin{eqnarray}
\REST{\rho} & = & \DDEMI D_\mu \phi D^\mu \phi + V 
\nonumber \\
            & = & \DDEMI \left(  \frac{\dot \phi^2}{\IBB}
                               - \frac{\phi'{}^2}{\INN} \right) + V , \\
\REST{P} & = & \DDEMI D_\mu \phi D^\mu \phi - V 
\nonumber \\
         & = & \DDEMI \left(  \frac{\dot \phi^2}{\IBB}
                               - \frac{\phi'{}^2}{\INN} \right) - V , \\
\REST{\PRESY} & = & \REST{P} , \\
U_\alpha & = & \frac{D_\alpha \phi}{\pm \sqrt{D_\mu \phi D^\mu \phi}} , \\
\GAM & = & \frac{1}
                {\displaystyle
                 \sqrt{1 - \IBBzNN \frac{\phi'{}^2}{\dot \phi^2}}} , \\
\BGM & = & - \IBzN \frac{\phi'}{\dot \phi}
             \frac{1}
                  {\displaystyle
                   \sqrt{1 - \IBBzNN \frac{\phi'{}^2}{\dot \phi^2}}} , \\
\BET & = & - \IBzN \frac{\phi'}{\dot \phi} ,
\end{eqnarray}
with the $\pm$ sign determined by the condition $U_0 \geq 0$.
\begin{eqnarray}
\rho & = & \DDEMI \left(  \frac{\dot \phi^2}{\IBB}
                        + \frac{\phi'{}^2}{\INN} \right) + V , \\
P & = & \DDEMI \left(  \frac{\dot \phi^2}{\IBB}
                     - \frac{\phi'{}^2}{\INN} \right) - V , \\
\PRESY & = & \DDEMI \left(  \frac{\dot \phi^2}{\IBB}
                          + \frac{\phi'{}^2}{\INN} \right) - V , \\
F & = & - \frac{\dot \phi}{\IB} \frac{\phi'}{\IN} .
\end{eqnarray}
\begin{eqnarray}
\REST{\delta \rho}^\sharp & = &
   \frac{\dot \phi \dot{\delta \phi}{}^\sharp}{\IBB}
 - \frac{\phi' \delta \phi^\sharp{}'}{\INN}
 - \frac{\dot \phi^2}{\IBB} \Psi - \frac{\phi'{}^2}{\INN} \hpp
 + \frac{\dot \phi}{\IB} \frac{\phi'}{\IN} \vp
 + \frac{\ddd V}{\ddd \phi} \delta \phi^\sharp , \\
\REST{\delta P}^\sharp & = &
   \frac{\dot \phi \dot{\delta \phi}{}^\sharp}{\IBB}
 - \frac{\phi' \delta \phi^\sharp{}'}{\INN}
 - \frac{\dot \phi^2}{\IBB} \Psi - \frac{\phi'{}^2}{\INN} \hpp
 + \frac{\dot \phi}{\IB} \frac{\phi'}{\IN} \vp
 - \frac{\ddd V}{\ddd \phi} \delta \phi^\sharp , \\
\delta \rho^\sharp & = &
   (P + \rho)
   \left(\frac{\dot{\delta \phi}{}^\sharp}{\dot \phi} - \Psi \right)
 + (\PRESY - P) \left(\frac{\delta \phi^\sharp{}'}{\phi'} + \hpp\right)
 + F \vp
 + \frac{\ddd V}{\ddd \phi} \delta \phi^\sharp , \\
\delta \PRESY^\sharp & = &
   (P + \rho)
   \left(\frac{\dot{\delta \phi}{}^\sharp}{\dot \phi} - \Psi \right)
 + (\PRESY - P) \left(\frac{\delta \phi^\sharp{}'}{\phi'} + \hpp\right)
 + F \vp
 - \frac{\ddd V}{\ddd \phi} \delta \phi^\sharp , \\
\delta P^\sharp & = &
   (P + \rho) \left(\frac{\dot{\delta \phi}{}^\sharp}{\dot \phi} - \Psi \right)
 - (\PRESY - P) \left(\frac{\delta \phi^\sharp{}'}{\phi'} + \hpp\right)
 - F \vp
 - \frac{\ddd V}{\ddd \phi} \delta \phi^\sharp , \\
\delta F^\sharp & = &
   (P + \rho) \vp
 + F \left(  \frac{\dot{\delta \phi}{}^\sharp}{\dot \phi} - \Psi
           + \frac{\delta \phi^\sharp{}'}{\phi'} + \hpp\right) ,
\end{eqnarray}
\begin{eqnarray}
\IA \RVD{v}^\sharp & = &
 - \IB
   \frac{\displaystyle \frac{\delta \phi^\sharp}{\dot \phi}}
        {\displaystyle
         \sqrt{1 - \IBBzNN \frac{\phi'{}^2}{\dot \phi^2}}} , \\
\VVRVU{v}_i{}^\sharp & = &
   \GAM \vi_i + \BGM \hci_i , \\
\IA \RVD{\dni}^\sharp & = &
 - \BET \IA \RVD{v}^\sharp , \\
\VVRVU{\dni}_i{}^\sharp & = &
 - \GAM \hci_i - \BGM \vi_i , \\
\www & = &
   \BGG \left(  A + \epp
              + \frac{\delta \phi'}{\phi'}
              - \frac{\dot{\delta \phi}}{\dot \phi} \right)
 + \GMM \Bperp , \\
\www^\sharp & = &
   \BGG \left(  \Psi + \hpp
              + \frac{\delta \phi^\sharp{}'}{\phi'}
              - \frac{\dot{\delta \phi}{}^\sharp}{\dot \phi} \right)
 + \GMM \vp , \\
\pi_{i j} & = & 0 .
\end{eqnarray}
Here, the components $\RVD{\dni}^i$ are arbitrary as the eigenvalues
$\REST{P}$ and $\REST{\PRESY}$ are degenerate. The expression chosen
here is purely for convenience in order to simplify the following
untilded components.
\begin{eqnarray}
\IA v^\sharp & = &
 - \IB \frac{\delta \phi^\sharp}{\dot \phi} , \\
\VV{v}{}^\sharp_i & = & \vi_i , \\
\IA \dni^\sharp & = & 0 , \\
\VV{\dni}{}^\sharp_i & = & - \hci_i .
\end{eqnarray}

\subAPPsection{Interaction term}

\begin{eqnarray}
\delta Q^0  & = &
   \IzB \left(  \delta \Gamma^\sharp - \Gamma \Psi - D \vp \right)
\nonumber \\ & &
 + (\dot Q^0 - Q^0 \pe - Q^\perp \pp) \ABBE - Q^0{}' \ANEE , \\
\delta Q^i & = &
   \IzA \QINT^i_\sharp
 - (Q^0 \pe + Q^\perp \pp) E^i , \\
\delta Q^\perp & = &
   \IzN \left(  \delta D^\sharp + D \hpp \right)
\nonumber \\ & &
 + \dot Q^\perp \ABBE - (Q^\perp{}' - Q^\perp \pp - Q^0 \pe) \ANEE ,
\end{eqnarray}
with
\begin{equation}
\QINT^\sharp_i = \nabla_i \QINT^\sharp + \VV{\QINT}^\sharp_i .
\end{equation}

\subAPPsection{Conservation equations}

\begin{eqnarray}
   (\pezB + N \HcB + 2 \UcB) (\delta \rho^\sharp - F \vp)
 + N \HcB \delta P^\sharp
 + \UcB (\delta \PRESY^\sharp - \delta \rho^\sharp)
\nonumber \\
 + (\ppzN + N \HpN + 2 \IpN) \left(  \delta F^\sharp + F (\Psi + \hpp) \right)
\nonumber \\
 + \DzAA \left( (P + \rho) \IA v^\sharp - F \IA \dni^\sharp \right)
\nonumber \\
 - N (P + \rho) \pezB \Phi - (\rho + \PRESY) \pezB \hpp - F \pezB \vp
 + F \ppzN (\Psi - \hpp - N \Phi)
 & = & \delta \Gamma^\sharp + \Gamma \Psi , \\
   \left(\pezB + N \HcB + \UcB \right)
        \left((P + \rho) \IA v^\sharp - F \IA \dni^\sharp \right)
\nonumber \\
 + \left(\ppzN + N \HpN + \IpN \right)
   \left(F \IA v^\sharp + (P - \PRESY) \IA \dni^\sharp \right)
\nonumber \\
 + \delta P^\sharp + \frac{N - 1}{N} (\DzAA + N \KzAA) \IAA \Pi
 + (P + \rho) \Psi + (\PRESY - P) \hpp + F \vp
 & = & \IA \QINT^\sharp , \\
   (\pezB + N \HcB + 2 \UcB)
   \left(  \delta F^\sharp - F (\Psi + \hpp) - (\rho + \PRESY) \vp \right)
\nonumber \\
 + (\ppzN + N \HpN + 2 \IpN)
   (\delta \PRESY^\sharp - F \vp)
 - N \HpN \delta P^\sharp - \IpN (\delta \PRESY^\sharp - \delta \rho^\sharp)
\nonumber \\
 + \DzAA \left( F \IA v^\sharp + (P - \PRESY) \IA \dni^\sharp \right)
\nonumber \\
 + F \pezB (\Psi - \hpp - N \Phi)
 + N (P - \PRESY) \ppzN \Phi + (\rho + \PRESY) \ppzN \Psi + F \ppzN \vp
 & = & \delta D^\sharp - D \hpp - \Gamma \vp  ,
\end{eqnarray}
\begin{eqnarray}
   \left(\pezB + (N + 1) \HcB + \UcB \right)
   \left(  (P + \rho) (\VV{v}^\sharp_i - \vi_i)
         - F (\VV{\dni}^\sharp_i + \hci_i) \right)
\nonumber \\
 + \left(\ppzN + (N + 1) \HpN + \IpN \right)
   \left(  F (\VV{v}^\sharp_i - \vi_i)
         + (P - \PRESY) (\VV{\dni}^\sharp_i + \hci_i) \right)
\nonumber \\
 + \DDEMI \left(\DzAA + (N - 1) \KzAA\right) \IA \VV{\Pi}_i
 & = & \VV{\QINT}^\sharp_i - D \hci_i - \Gamma \vi_i .
\end{eqnarray}

\subAPPsection{Einstein equations}

\begin{eqnarray}
\label{pert_ein_first}
  \DzAA ((N - 1) \Phi + \hpp) + \KzAA N (N - 1) \Phi
\nonumber \\
 - N \left(  \left((N - 1) \HcB^2 + 2 \HcB \UcB \right) \Psi
           + \HcB \pezB \hpp
           + \left((N - 1) \HcB + \UcB \right) \pezB \Phi \right)
\nonumber \\
 - N \left(\HpN \ppzN + (N + 1) \HpN^2 \right) \hpp
 - 2 N \hpp \ppzN \HpN
\nonumber \\
 + N (\ppzN + (N + 1) \HpN ) \ppzN \Phi
\nonumber \\
 + N \left(\ppzN + N \HpN + \IpN \right) (\HcB \vp)
 & = & \KAPPAN \left(\delta \rho^\sharp - F \vp \right) , \\
   \DDEMI \left(\ppzN + (N - 2) \HpN + 2 \IpN \right) \vp
\nonumber \\
 -   (\UcB + (N - 1) \HcB) \Psi
        -  (\pezB + \UcB - \HcB) \hpp - (N - 1) \pezB \Phi
 & = & \KAPPAN \left( (P + \rho) \IA v^\sharp - F \IA \dni^\sharp \right) , \\
 - \KzAA (N - 1) (N - 2) \Phi
\nonumber \\
 + \left(2 (\UcB + (N - 1) \HcB) (\pezB + \UcB) + N (N - 1) \HcB^2 \right) \Psi
\nonumber \\
 + 2 \Psi \pezB (\UcB + (N - 1) \HcB)
\nonumber \\
 + \left(2 (\IpN + (N - 1) \HpN) (\ppzN + \IpN) + N (N - 1) \HpN^2 \right) \hpp
\nonumber \\
 + 2 \hpp \ppzN (\IpN + (N - 1) \HpN)
\nonumber \\
 + (\pezB + \UcB) \pezB (\hpp + (N - 1) \Phi)
\nonumber \\
 - (\UcB + (N - 1) \HcB) \pezB (\Psi - \hpp - N \Phi)
\nonumber \\
 + (\ppzN + \IpN) \ppzN (\Psi - (N - 1) \Phi)
\nonumber \\
 + (\IpN + (N - 1) \HpN) \ppzN (\Psi - \hpp - N \Phi)
\nonumber \\
 - \TDEMI (\ppzN \pezB + \pezB \ppzN + \IpN \pezB + \UcB \ppzN) \vp
\nonumber \\
 - \left(  (\UcB + (N - 1) \HcB) \ppzN 
         + (\IpN + (N - 1) \HpN) \pezB \right) \vp
\nonumber \\
 - \vp \left(  (\ppzN + \IpN) (\UcB + (N - 1) \HcB)
             + (\pezB + \UcB) (\IpN + (N - 1) \HpN) \right)
\nonumber \\
 - N \left(\UcB \pezB - \IpN \ppzN \right) \Phi
 - N (N - 1) \HcB \HpN \vp
 & = & \KAPPAN \left(\delta P^\sharp + \frac{N - 1}{N} \nabla^2 \Pi
               \right) , \\
N \Phi - (\Phi + \Psi) + (\hpp - \Phi)
 & = & \KAPPAN \IAA \Pi , \\
 - N \left(  (\pezB \ppzN + (\HpN - \IpN) \pezB + \HcB \ppzN) \Phi
           + \HcB \ppzN \Psi - \HpN \pezB \hpp \right)
\nonumber \\
 - \DDEMI \DzAA \vp
 - N \vp (\pezB \HcB  + \HcB^2 - \HcB \UcB) 
 & = & \KAPPAN \left(  \delta F^\sharp + F (\Psi - \hpp) \right) , \\
   \DDEMI \left(\pezB + (N - 2) \HcB + 2 \UcB \right) \vp
\nonumber \\
 -  (\ppzN + \IpN - \HpN) \Psi
         - ((N - 1) \HpN + \IpN) \hpp
         + (N - 1) \ppzN \Phi
 & = & \KAPPAN \left(F \IA v^\sharp + (P - \PRESY) \IA \dni^\sharp \right) , \\
 - \left(\DzAA ((N - 1) \Phi - \Psi) + \KzAA N (N - 1) \Phi \right)
\nonumber \\
 + N \left(\HcB \pezB + (N + 1) \HcB^2 \right) \Psi
 + 2 N \Psi \pezB \HcB
\nonumber \\
 + N \left(\pezB + (N + 1) \HcB \right) \pezB \Phi
\nonumber \\
 + N \left(  \left(\HpN \ppzN \right) \Psi
           + \left((N - 1) \HpN^2 + 2 \HpN \IpN \right) \hpp
           - \left((N - 1) \HpN \ppzN + \IpN \ppzN \right) \Phi \right)
\nonumber \\
 - N \left(\pezB + N \HcB + \UcB \right) (\HpN \vp)
 & = & \KAPPAN \left(\delta \PRESY^\sharp - F \vp \right) ,
\end{eqnarray}
\begin{eqnarray}
 - \DDEMI (\DzAA +  (N - 1)\KzAA) \vi_i
\nonumber \\
 - \DDEMI \left(\ppzN + (N + 1) \HpN \right)
          \left( (\ppzN + \IpN - \HpN) \vi_i \right)
\nonumber \\
 + \DDEMI \left(\ppzN + (N + 1) \HpN \right)
          \left( (\pezB + \UcB - \HcB) \hci_i \right)
 & = & \KAPPAN \left(  (P + \rho) (\VV{v}^\sharp_i - \vi_i)
                     - F (\VV{\dni}^\sharp_i + \hci_i) \right) , \\
   \left(\pezB + (N - 1) \HcB + \UcB \right) \vi_i
 - \left(\ppzN + (N - 1) \HpN + \IpN \right) \hci_i
 & = & \KAPPAN \IA \VV{\Pi}_i , \\
   \DDEMI (\DzAA + (N - 1) \KzAA) \hci_i
\nonumber \\
 + \DDEMI \left(\pezB + (N + 1) \HcB \right)
          \left( (\ppzN + \IpN - \HpN) \vi_i \right)
\nonumber \\
 - \DDEMI \left(\pezB + (N + 1) \HcB \right)
          \left( (\pezB + \UcB - \HcB) \hci_i \right)
 & = & \KAPPAN \left(  F (\VV{v}^\sharp_i - \vi_i)
                     + (P - \PRESY) (\VV{\dni}^\sharp_i + \hci_i ) \right) ,
\end{eqnarray}
\begin{eqnarray}
\label{pert_ein_last}
 - (\DzAA - 2 \KzAA) \TT{E}_{i j}
 + \left(\pezB + N \HcB + \UcB \right) \pezB \TT{E}_{i j}
 - \left(\ppzN + N \HpN + \IpN \right) \ppzN \TT{E}_{i j}
 & = & \KAPPAN \TT{\Pi}_{i j}  .
\end{eqnarray}

\APPsection{Perturbed brane-related quantities}
\label{sec_app_pert_brane}

\subAPPsection{Brane position}

\begin{eqnarray}
X^0 & = & \sigma^0 + \ZETA^0 (\sigma^a) , \\
X^i & = & \sigma^i + \ZETA^i (\sigma^a) , \\
X^\finN & = & \POSBRN + \UPS (\sigma^a) .
\end{eqnarray}
Under an infinitesimal coordinate change, the brane position
$\UPS$ transforms into
\begin{equation}
\UPS \to \UPS - \Lperp .
\end{equation}

\subAPPsection{Normal vector to the brane}
\label{ssec_app_pert_brane2}

\begin{eqnarray}
(\NORMVEC_\mu+\DNORMVEC_\mu)
    \frac{\partial X^\mu}{\partial \sigma_a} & = & 0 , \\
\DNORMVEC^\alpha
 & = & \left(  \IzB \left(\Bperp - \INzB \dot \UPS \right),
               \IzA \left(\INzA \nabla^i \UPS - \eip^i \right),
             - \IzN \epp \right) , \\
\DNORMVEC_\alpha
 & = & \left(- \IB \INzB \dot \UPS, - \IA \INzA \nabla_i \UPS, - \IN \epp
       \right) .
\end{eqnarray}
Since $\NORMVEC^\alpha$ plays the same role for the brane as
$N^\alpha$ for a bulk component, this means that formally, the
quantities $\www$, $\dni_i$ can be defined for the brane (we will note
them $\BRN{\www}$ and $\BRN{\dni}_i$ respectively) at $\fd =
\POSBRN$,
\begin{eqnarray}
\BRN{\www} & = & \INzB \dot \UPS , \\
\BRN{\dni}_i + \epi_i & = & \INzA \nabla_i \UPS , \\
\end{eqnarray}
or, equivalently,
\begin{eqnarray}
\BRN{\www}^\sharp & = & \INzB \dot \UPS^\sharp , \\
\BRN{\dni}^\sharp & = & \INzA \UPS^\sharp , \\
\BRN{\VV{\dni}}^\sharp_i & = & - \hci_i .
\end{eqnarray}

\subAPPsection{Induced metric}

\begin{eqnarray}
\DFFF_{0 0} & = & 2 \IBB (A + \Iperp \UPS) , \\
\DFFF_{0 i} & = & \IAB B_i , \\
\DFFF_{i j}
& = & - 2 \IAA (C + \Hperp \UPS) \gamma_{i j} - 2 \IAA E_{i j} , \\
\DFFF_{0 \finN}
& = & \IBN \left(\Bperp - \INzB \dot \UPS \right) , \\
\DFFF_{i \finN}
& = & \INA \left(\epi_i - \INzA \nabla_i \UPS \right) , \\
\DFFF_{\finN \finN} & = & 0 .
\end{eqnarray}

\subAPPsection{First Israel condition}

\begin{eqnarray}
\DISC{A} + \DISC{\Iperp} \UPS & = & 0 , \\
\DISC{B_i} & = & 0 , \\
\DISC{C} + \DISC{\Hperp} \UPS & = & 0 , \\
\DISC{E_i} & = & 0 , \\
\DISC{\TT{E}_{i j}} & = & 0 , \\
\DISC{\ABBE} & = & 0 .
\end{eqnarray}
Or, equivalently
\begin{eqnarray}
\DISC{\Psi} & = & - \DISC{\Iperp \UPS - \Iperp \ANEE} , \\
\DISC{\Phi} & = & \DISC{\Hperp \UPS - \Hperp \ANEE} , \\
\DISC{\vi_i} & = & 0 , \\
\DISC{\TT{E}_{i j}} & = & 0 .
\end{eqnarray}
With,
\begin{eqnarray}
\BRNPsi & = & A - (\pe + \Iconf) \ABBE + \Iperp \UPS
          =   \Psi + \Iperp \left(\UPS - \ANEE \right) , \\
\BRNPhi & = & - C + \Hconf \ABBE - \Hperp \UPS
          =   \Phi - \Hperp \left(\UPS - \ANEE \right) ,
\end{eqnarray}
one has
\begin{eqnarray}
\DFFF_{0 0}
& = & 2 \IBB \BRNPsi + (\dot{\FFF}_{0 0} + 2 \FFF_{0 0} \pe) \ABBE , \\
\DFFF_{0 i} & = & - \IBzA \FFF_{i j} \vi^j
                  + \FFF_{0 0} \nabla_i \ABBE + \FFF_{i j} \dot E^j , \\
\DFFF_{i j}
& = &   2 \FFF_{k (i} \left(E_{j)}^k - \delta^k_{j)} \BRNPhi \right) ,
\end{eqnarray}

\begin{equation}
\epp \to \epp - \Uconf T - \IzN (\IN \Lperp)' ,
\end{equation}
therefore
\begin{eqnarray}
\DISC{\Lperp} & = & 0 , \\
\DISC{\IN \Lperp} & = & 0 .
\end{eqnarray}
Then,
\begin{eqnarray}
\DISC{\IN} \neq 0 \Rightarrow \Lperp (\fd = \POSBRN) & = & 0 , \\
\DISC{\IN \UPS} & = & 0 .
\end{eqnarray}

\begin{eqnarray}
\IN \Bperp - \INNzB \UPS & \to & \IN \Bperp - \INNzB \UPS + \IB T' , \\
\IN \epi_i - \INNzA \nabla_i \UPS
 & \to & \IN \epi_i - \INNzA \nabla_i \UPS - \IA L'_i ,
\end{eqnarray}
and
\begin{eqnarray}
\DISC{\IN \Bperp} - \DISC{\INNzB} \dot \UPS
   =   \DISC{\IN \Bperp}
 & = & 0 , \\
\DISC{\IN \epi_i} - \DISC{\INNzA} \nabla_i \UPS
   =   \DISC{\IN \epi_i} \UPS
 & = & 0 ,
\end{eqnarray}
therefore
\begin{eqnarray}
\DISC{T'} & = & 0 , \\
\DISC{L^i{}'} & = & 0 .
\end{eqnarray}

\subAPPsection{New brane-related gauge invariant quantities}

At the brane position (or on both sides of the brane),
\begin{eqnarray}
\UPS^\sharp & \equiv & \UPS - \ANEE , \\
\BRNvp & \equiv & \Bperp - \IBzN \pp \ABBE - \INzB \pe \UPS
         =        \vp - \INzB \pe \UPS^\sharp , \\
\BRNhpp & \equiv & \epp +\Uconf \ABBE - \Uperp \UPS
          =        \hpp - \IzN \pp (\IN \UPS^\sharp) , \\
\BRNPsi' & \equiv & A' - \pp (\pe + \Iconf) \ABBE + \Iperp' \UPS
           =        \Psi' + \Iperp' \UPS^\sharp - \Iperp \ANEE' , \\
\BRNPhi' & \equiv & - C' + (\Hconf \pp + \Hconf') \ABBE - \Hperp' \UPS
           =        \Psi' - \Hperp' \UPS^\sharp + \Hperp \ANEE' , \\
\ppzN \BRNPsi & \equiv & \IzN \BRNPsi' , \\
\ppzN \BRNPhi & \equiv & \IzN \BRNPhi' .
\end{eqnarray}
\begin{eqnarray}
\CONT{\IN \UPS^\sharp} & \equiv & \BRUPS , \\
\DISC{\IN \UPS^\sharp} & \equiv & - \BRNXI .
\end{eqnarray}
\begin{eqnarray}
\DISC{\Psi}
 & = & - \DISC{\frac{\Iperp}{\IN}} \BRUPS
       + \CONT{\frac{\Iperp}{\IN}} \BRNXI , \\
\DISC{\Phi}
 & = &   \DISC{\frac{\Hperp}{\IN}} \BRUPS
       - \CONT{\frac{\Hperp}{\IN}} \BRNXI , \\
\DISC{\vp}
 & = &   \DISC{\BRNvp}
       - \IzB \left(\pe - \CONT{\Uconf} \right) \BRNXI
       - \IzB \DISC{\Uconf} \BRUPS , \\
\DISC{\hpp}
 & = &   \DISC{\BRNhpp}
       + \DISC{\frac{\pp}{\IN} \BRUPS} ,
\end{eqnarray}
\begin{eqnarray}
\CONT{\Psi}
 & = &   \BRNPsi
       - \CONT{\frac{\Iperp}{\IN}} \BRUPS
       + \frac{1}{4} \DISC{\frac{\Iperp}{\IN}} \BRNXI , \\
\CONT{\Phi}
 & = &   \BRNPhi
       + \CONT{\frac{\Hperp}{\IN}} \BRUPS
       - \frac{1}{4} \DISC{\frac{\Hperp}{\IN}} \BRNXI , \\
\CONT{\vp}
 & = &   \CONT{\BRNvp}
       + \IzB \left(\pe - \CONT{\Uconf} \right) \BRUPS
       + \frac{1}{4} \IzB \DISC{\Uconf} \BRNXI , \\
\CONT{\hpp}
 & = &   \CONT{\BRNhpp}
       + \CONT{\frac{\pp}{\IN} \BRUPS} ,
\end{eqnarray}
where the terms $\pp \BRUPS$ are defined by setting $\UPS$ constant.

\subAPPsection{Extrinsic curvature}
\label{ssec_app_pert_cex}

\begin{eqnarray}
\DCEX_{0 0} & = &
 \IB (\pe + \Uconf) \left(\vp - \INzB \dot \UPS + \INzB \pe \ANEE \right)
 - \IBBzN (\Psi' + \Iperp \hpp)
 + 2 \CEX_{0 0} \Psi + \CEX'_{0 0} \UPS
\nonumber \\ & &
 + (\dot{\CEX}_{0 0} + 2 \CEX_{0 0} \pe) \ABBE
 - \CEX'_{0 0} \ANEE
\nonumber \\ & = &
 \IB (\pe + \Uconf) \BRNvp
 - \IBBzN (\BRNPsi' + \Iperp \BRN{\hpp})
 + 2 \CEX_{0 0} \BRNPsi
 + (\dot{\CEX}_{0 0} + 2 \CEX_{0 0} \pe) \ABBE , \\
\DCEX_{0 i} & = &
   \DDEMI \IB \vp
 - (\pe - \Hconf) \left(\IN \UPS - \IN \ANEE \right)
\nonumber \\ & &
 + \DDEMI \IA (\pe + \Uconf - \Hconf) \hci_i
 - \DDEMI \IABzN (\pp + \Hperp - \Iperp) \vi_i + \IAzB \CEX_{0 0} \vi_i
\nonumber \\ & &
 + \CEX_{0 0} \nabla_i \ABBE
 + \CEX_{i j} \dot E^j
\nonumber \\ & = &
   \DDEMI \IB \BRNvp
 - (\TDEMI \pe + \TDEMI \Uconf - \Hconf) \left(\IN \UPS^\sharp \right)
 + \DDEMI \IA (\pe + \Uconf - \Hconf) \hci_i
 - \DDEMI \IABzN (\pp + \Hperp - \Iperp) \vi_i
\nonumber \\ & &
 + \IAzB \CEX_{0 0} \vi_i
 + \CEX_{0 0} \nabla_i \ABBE
 + \CEX_{i j} \dot E^j , \\
\DCEX_{i j} & = &
   \IAAzB \gamma_{i j} \Hconf
   \left(\INzB \dot \UPS - \INzB \pe \ANEE - \vp \right)
 - \IAAzN \gamma_{i j} (\Phi' - \Hperp \hpp)
\nonumber \\ & &
 - \nabla_{i j} \left(\IN \UPS - \IN \ANEE \right)
 + \IA \nabla_{(i} \hci_{j)}
 + \IAAzN \TT{E}'_{i j}
 + \CEX'_{i j} \UPS
\nonumber \\ & &
 + \dot{\CEX}_{i j} \ABBE
 - \CEX'_{i j} \ANEE
 + 2 \CEX_{k (i} (E^k_{j)} - \delta^k_{j)} \Phi)
\nonumber \\ & = &
 - \IAAzB \gamma_{i j} \Hconf \BRNvp
 - \IAAzN \gamma_{i j} (\BRNPhi' - \Hperp \BRNhpp)
 - \nabla_{i j} (\IN \UPS^\sharp)
 + \IA \nabla_{(i} \hci_{j)}
 + \IAAzN \TT{E}'_{i j}
\nonumber \\ & &
 + \dot{\CEX}_{i j} \ABBE
 + 2 \CEX_{k (i} (E^k_{j)} - \delta^k_{j)} \BRNPhi) , \\
\DCEX_{0 \finN} & = &
   \CEX_0^0 \DFFF_{0 \finN} , \\
\DCEX_{i \finN} & = &
   \CEX_i^j \DFFF_{j \finN} , \\
\DCEX_{\finN\ finN} & = & 0 .
\end{eqnarray}

\subAPPsection{{\protect $\STT_{\alpha \beta}$} tensor}
\label{ssec_app_pert_stt}

The perturbed stress-energy tensor on the brane is given in
Eqns~(\ref{tensorbranebeg}--\ref{tensorbraneend}).  Formally, for any
quantity $\BRN{X}$ defined on the brane, the quantity $\DIR \BRN{X}$
is a bulk scalar.  The quantity $\DIR$ is also a scalar quantity of
the bulk since one is allowed to consider the case $\BRN{X} = {\rm
constant}$. Its perturbation is
\begin{equation}
\delta \DIR = \DIR \epp - \UPS (\pp + \Uperp) \DIR .
\end{equation}
Note that this derivation is a bit formal: since $\epp$ and $\BRNhpp$
can be discontinuous, this expression and the next one are
ill-defined, even in the case where $\IN$ is continuous. But again,
all the pathological terms cancel each other when one writes the
Einstein equations, so that we consider that this is not a serious
problem. Using the formula~(\ref{lie_scal_bulk}), it is possible to
build the gauge invariant counterparts of both $\delta \DIR$ and
$\BRN{\delta X}$:
\begin{eqnarray}
\delta \DIR^\sharp & = & \DIR \BRNhpp - \DIR' \UPS^\sharp , \\
\BRN{\delta X}^\sharp & = & \BRN{\delta X} - \dot {\BRN{X}} \ABBE .
\end{eqnarray}
This last quantity is invariant under any infinitesimal
reparametrization of the $\sigma^a$. Equivalently, one has,
using~(\ref{gi_v_scal},\ref{gi_v_vect}), as well as the fact that
$\GAM = 1$ for the brane,
\begin{eqnarray}
\BRN{v}^\sharp & = & \BRN{v} + \IAzB \dot E , \\
\BRN{\VV{v}}^\sharp_i
 & = & \BRN{\VV{v}}_i + \IAzB \dot{\VV{E}}_i .
\end{eqnarray}
With these definitions,
\begin{eqnarray}
\DSTT_{0 0} & = &
   \IBB \left(  \frac{N - 1}{N} \BRN{\delta \rho}^\sharp
              + \BRN{\delta P}^\sharp \right)
\nonumber \\ & &
 + 2 \STT_{0 0} \BRNPsi
 + (\dot{\STT}{}_{0 0} + 2 \STT_{0 0} \pe) \ABBE , \\
\DSTT_{0 i} & = &
 - \IAB (\BRN{P} + \BRN{\rho}) \BRN{v}^\sharp_i
\nonumber \\ & &
 + \IAzB \STT_{0 0} \vi_i
 + \STT_{0 0} \nabla_i \ABBE
 + \STT_{i j} \dot E^j , \\
\DSTT_{i j} & = &
   \IAA \BRN{\Pi}_{i j}
 + \IAA \frac{1}{N} \BRN{\delta \rho}^\sharp \gamma_{i j}
\nonumber \\ & &
 + \dot{\STT}{}_{i j} \ABBE
 + 2 \STT_{k (i} (E^k_{j)} - \delta^k_{j)} \BRNPhi) , \\
\DSTT_{0 \finN} & = &
   \STT_0^0 \DFFF_{0 \finN} , \\
\DSTT_{i \finN} & = &
   \STT_i^j \DFFF_{j \finN} , \\
\DSTT_{\finN \finN} & = & 0 .
\end{eqnarray}

\subAPPsection{Second Israel condition}

\begin{eqnarray}
 - \DzAA \BRNXI
 + N \DISC{\ppzN  \BRNPhi - \HpN \BRNhpp}
 + N \HcB \DISC{\BRNvp}
 & = & \KAPPAN \BRN{\delta \rho}^\sharp , \\
 \DDEMI \DISC{\BRNvp - (\pezB + \UcB - 2 \HcB) (\IN \UPS^\sharp)}
 & = & \KAPPAN (\BRN{P} + \BRN{\rho}) \IA \BRN{v}^\sharp , \\
   \DISC{\ppzN \BRNPsi + \IpN \BRNhpp}
 - \DISC{\left(\pezB + \UcB \right) \BRNvp}
 & = & \KAPPAN \left(  \BRN{\delta P}^\sharp
                    + \frac{N - 1}{N} \BRN{\delta \rho}^\sharp \right) , \\
 - \BRNXI
 & = & \KAPPAN \IAA \BRN{\Pi} ,
\end{eqnarray}
\begin{eqnarray}
 - \DDEMI
   \DISC{\ppzN + \IpN - \HpN} \vi_i
 + \DDEMI \DISC{(\pezB + \UcB - \HcB) \hci_i}
 & = & \KAPPAN
       (\BRN{P} + \BRN{\rho}) (\BRN{v}^\sharp_i - \vi_i), \\
 - \DISC{\hci_i}
 & = & \KAPPAN \IA \BRN{\VV{\Pi}}{}_i ,
\end{eqnarray}
\begin{eqnarray}
 - \DISC{\ppzN \TT{E}_{i j}}
 & = & \KAPPAN \BRN{\TT{\Pi}}{}_{i j} .
\end{eqnarray}

\subAPPsection{Projected Weyl tensor}

As this quantity is defined on the brane, it is more convenient to
express it in term of the brane-related (underlined) metric
perturbations instead of the bulk-related (non underlined) metric
perturbations as it was the case for the Weyl tensor.
\begin{eqnarray}
\BRN{\delta \WEYL}^\sharp
 & \equiv &   \frac{2}{N} (\DzAA + N \KzAA ) \BRNPhi
            + \frac{1}{N} \DzAA (\BRNPsi - \BRNhpp)
\nonumber \\ & &
            + \left(\pezB + \UcB \right)
              \left(\pezB \BRNPhi + \HcB \BRNPsi \right)
            - \left(\pezB + 2 \UcB - \HcB \right)
              \left(\pezB \BRNhpp + \UcB \BRNPsi \right)
\nonumber \\ & &
            - \left(\ppzN + \IpN \right)
              \left(  \ppzN \BRNPhi - \HpN \BRNhpp + \HcB \BRNvp
                    + \frac{1}{N} \DzAA (\IN \UPS^\sharp) \right)
\nonumber \\ & &
            - \left(\ppzN + 2 \IpN - \HpN \right)
              \left(  \ppzN \BRNPsi + \IpN \BRNhpp
                    - \left(\pezB + \UcB \right) \BRNvp \right)
\nonumber \\ & &
            - \BRNPsi \pezB \left(\UcB - \HcB \right)
            - \left(\BRNhpp \ppzN - \BRNvp \pezB \right)
              \left(\IpN - \HpN \right) , \\
\DWN^v
 & \equiv &   \frac{N - 1}{N}
              \left(- \left(\HcB \BRNPsi + \pezB \BRNPhi \right)
                    + \left(\pezB \BRNhpp + \UcB \BRNPsi \right)
                    + \left(\UcB - \HcB \right) \BRNhpp
                    + \left(\HpN - \IpN \right) \BRNvp
\right. \nonumber \\ & & \left. \qquad \qquad
                    - \TDEMI \ppzN
                      \left(  \BRNvp
                            - \left(\pezB + \UcB - 2 \HcB \right)
                              (\IN \UPS^\sharp) \right) \right) , \\
\DWN^\Pi
 & \equiv & \frac{1}{N}
            \left(  (N - 2) \BRNPhi - \BRNPsi
                  - (N - 1) \BRNhpp
                  + ((N - 2) \HpN + \IpN - (N - 1) \ppzN) (\IN \UPS^\sharp)
            \right) ,
\end{eqnarray}
\begin{eqnarray}
\VV{\DWN}{}^v_i
 & \equiv & - \DDEMI \frac{1}{N} (\DzAA + (N - 1)\KzAA ) \vi_i
            + \DDEMI \frac{N - 1}{N}
              \left( \ppzN + \HpN \right)
              \left(  \left(\ppzN + \IpN - \HpN \right) \vi_i
                    - \left(\pezB + \UcB - \HcB \right) \hci_i \right) , \\
\VV{\DWN}{}^\Pi_i
 & \equiv & \frac{1}{N}
            \left(  \left(\pezB + (N - 1) \left(\HcB - \UcB \right)
                    \right) \vi_i
                  + \left((N - 1) \ppzN + \left(\HpN - \IpN \right)
                    \right) \hci_i \right) ,
\end{eqnarray}
\begin{eqnarray}
\TT{\DWN}{}^\Pi_{i j}
 & \equiv & \frac{1}{N}
            \left(  \left(  \pezB + N \HcB - (N - 1) \UcB \right) \pezB
                  - \left(\DzAA - 2 \KzAA \right)
                  + \left(  (N - 1) \ppzN + N \HpN - \IpN \right) \ppzN
            \right) \TT{E}_{i j} ,
\end{eqnarray}
\begin{eqnarray}
\DWN_{0 0} & = &
   \frac{N - 1}{N + 1} \IBB \BRN{\delta \WEYL}^\sharp
 + 2 \WN_{0 0} \BRNPsi
\nonumber \\ & &
 + (\dot{\WN}_{0 0} + 2 \WN_{0 0} \pe) \ABBE , \\
\DWN_{0 i} & = &
 - \IB \nabla_i \DWN^v - \IAB \VV{\DWN}{}^v_i
\nonumber \\ & &
 + \WN_{0 0} \nabla_i \ABBE
 + \WN_{i j} \dot E^j , \\
\DWN_{i j} & = &
   \frac{N - 1}{N (N + 1)} \IAA \gamma_{i j} \BRN{\delta \WEYL}^\sharp
\nonumber \\ & &
 + \left(\nabla_{i j} - \frac{1}{N} \nabla^2 \gamma_{i j} \right) \DWN^\Pi
 + \IA \nabla_{(i} \VV{\DWN}{}^\Pi_{j)}
 + \IAA \TT{\DWN}{}^\Pi_{i j}
\nonumber \\ & &
 + \dot{\WN}_{i j} \ABBE
 + 2 \WN_{k(i} (E^k_{j)} - \delta^k_{j)} \BRNPhi) , \\
\DWN_{0 \finN} & = &
   \WN^0_0 \DFFF_{0 \finN} , \\
\DWN_{i \finN} & = &
   \WN^j_i \DFFF_{j \finN} , \\
\DWN_{\finN \finN} & = & 0 .
\end{eqnarray}

\APPsection{Brane point of view, perturbed case}
\label{sec_app_pert_brane_pov}

\subAPPsection{New Einstein equations}
\label{ssec_app_pert_newee}

We rewrite the perturbed Einstein near the brane, that is near
$\fd = \POSBRN + \UPS$.  We first define some new bulk
matter content perturbations by
\begin{eqnarray}
\UDT{\delta \rho}^\sharp
 & =  &   \delta \rho^\sharp 
        + \rho' \UPS^\sharp
        - 2 F \INzB \dot \UPS^\sharp , \\
\UDT{\delta P}^\sharp
 & = &   \delta P^\sharp
       + P' \UPS^\sharp , \\
\UDT{\delta F}^\sharp
 & = &   \delta F^\sharp
       + F' \UPS^\sharp
       - (\rho + \PRESY) \INzB \dot \UPS^\sharp , \\
\UDT{\delta \PRESY}^\sharp
 & = &   \delta \PRESY^\sharp 
       + \PRESY' \UPS^\sharp
       - 2 F \INzB \dot \UPS^\sharp , \\
\IA \UDT{\dni}^\sharp
 & = &   \IA \dni^\sharp
       - \IN \UPS^\sharp , \\
\UDT{\delta \Gamma}^\sharp
 & = &   \delta \Gamma^\sharp
       + \Gamma' \UPS^\sharp
       - D \INzB \dot \UPS^\sharp
       + F \frac{\nabla^2}{\IAA} \IN \UPS^\sharp , \\
\IA \UDT{\QINT}^\sharp
 & = &   \IA \QINT^\sharp
       + D \IN \UPS^\sharp , \\
\UDT{\delta D}^\sharp
 & = &   \delta D^\sharp - \Gamma \vp
       + D' \UPS^\sharp
       + (\PRESY - P) \frac{\nabla^2}{\IAA} \IN \UPS^\sharp .
\end{eqnarray}
The terms proportional to $\UPS^\sharp$ come from the fact that we are
considering the bulk perturbations at $\fd = \POSBRN + \UPS$ instead
of $\fd = \POSBRN$. The other terms come from the fact that the brane
is not at rest with respect to the bulk coordinate system, and are a
mere consequence of a Lorentz boost with velocity $v^\perp = \INzB
\dot \UPS$ along the $\fd$ axis. Going from the non underlined (bulk)
metric perturbation to the underlined (brane-related) metric
perturbations, the Einstein equation simplify a little bit to
\begin{eqnarray}
\label{patho00}
   (N - 1) \left(\DzAA + N \KzAA  \right) \BRNPhi
 - N (N - 1) \HcB
   \left(\HcB \BRNPsi + \pezB \BRNPhi \right)
\nonumber \\
 - N \UcB
   \left(\HcB \BRNPsi + \pezB \BRNPhi \right)
 - N \HcB
   \left(\pezB \BRNhpp + \UcB \BRNPsi \right)
\nonumber \\
 + \DzAA \BRNhpp
 - N \BRNhpp \ppzN \CONT{\HpN}
 + N \BRNvp \pezB \HpN
\nonumber \\
 + N
   \left(\ppzN + (N + 1) \HpN \right)
   \left(  \ppzN \BRNPhi - \HpN \BRNhpp
         + \HcB \BRNvp
         + \frac{1}{N} \DzAA \IN \UPS^\sharp
   \right)
 & = & \KAPPAN \left(  \DIR \BRN{\delta \rho}^\sharp_\SBRN
                     + \UDT{\delta \rho}^\sharp_\SBLK \right) , \\
 - (N - 1) \left(\HcB \BRNPsi + \pezB \BRNPhi \right)
\nonumber \\
 - \left(\pezB \BRNhpp + \UcB \BRNPsi \right)
 - \left(\UcB - \HcB \right) \BRNhpp
 - \left(\HpN - \IpN \right) \BRNvp
\nonumber \\
 + \DDEMI \left(  \ppzN + N \HpN \right)
          \left(\BRNvp - \left(  \pezB + \UcB
                               - 2 \HcB \right)
                         (\IN \UPS^\sharp) \right)
 & = & \KAPPAN \left(  \DIR (\BRN{P}_\SBRN + \BRN{\rho}_\SBRN)
                            \IA \BRN{v}_\SBRN^\sharp
\right. \nonumber \\ & & \left. \quad
                     + (P_\SBLK + \rho_\SBLK) \IA v^\sharp_\SBLK
                     - F_\SBLK \IA \UDT{\dni}^\sharp_\SBLK
               \right) , \\
\label{pathoij}
 - (N - 2) (N - 1) \KzAA \BRNPhi
\nonumber \\
 + (N - 1) \left(\pezB + N \HcB \right)
   \left(\HcB \BRNPsi + \pezB \BRNPhi \right)
 + (N - 1) \BRNPsi \pezB \HcB
\nonumber \\
 + (N - 1) \UcB
   \left(\HcB \BRNPsi + \pezB \BRNPhi \right)
 + \BRNPsi \pezB \UcB
\nonumber \\
 + \left(\pezB + 2 \UcB + (N - 1) \HcB
   \right)
   \left(\pezB \BRNhpp + \UcB \BRNPsi \right)
\nonumber \\
 - (N - 1) \left(  \ppzN
                 + \IpN + N \HpN \right)
   \left(  \ppzN \BRNPhi - \HpN \BRNhpp
         + \HcB \BRNvp
   \right)
\nonumber \\
 + \left(  \ppzN
         + 2 \IpN + (N - 1) \HpN \right)
   \left(  \ppzN \BRNPsi + \IpN \BRNhpp - \left(\pezB + \UcB \right) \BRNvp
   \right)
\nonumber \\
 + \BRNhpp \ppzN
   \CONT{\IpN + (N - 1) \HpN}
 - \BRNvp \pezB
   \left(\IpN + (N - 1) \HpN \right)
 & = & \KAPPAN \left(  \DIR \BRN{\delta P}^\sharp_\SBRN
                     + \frac{N - 1}{N} \DIR \nabla^2 \Pi_\SBRN
\right. \\ & & \left. \qquad
                     + \UDT{\delta P}^\sharp_\SBLK
                     + \frac{N - 1}{N} \nabla^2 \Pi_\SBLK  \right) , \\
   (N - 2) \BRNPhi - \BRNPsi
\nonumber \\
 + \BRNhpp
 + \left(  \ppzN
         + (N - 2) \HpN + \IpN
   \right) (\IN \UPS^\sharp)
 & = & \KAPPAN \left(\DIR \IAA \BRN{\Pi}_\SBRN + \IAA \Pi_\SBLK \right) ,  \\
 - N \left(\pezB + \HcB \right)
     \left(  \ppzN \BRNPhi - \HpN \BRNhpp + \HcB \BRNvp
           + \frac{1}{N} \DzAA (\IN \UPS^\sharp)
     \right)
\nonumber \\
 - N \HcB
     \left(  \ppzN \BRNPsi + \IpN \BRNhpp
           - \left(\pezB + \UcB \right) \BRNvp \right)
\nonumber \\
 - \DDEMI \DzAA
   \left(  \BRNvp
         - \left(  \pezB + \UcB
                 - 2 \HcB \right) (\IN \UPS^\sharp)
   \right)
 - N \left(\HpN - \IpN \right)
     \pezB \BRNPhi
 & = & \KAPPAN \left(  \UDT{\delta F}^\sharp_\SBLK + F_\SBLK \BRNPsi
               \right) , \\
 - \DDEMI \left(\pezB + N \HcB \right)
   \left(\BRNvp - \left(  \pezB + \UcB
                        - 2 \HcB \right) (\IN \UPS^\sharp)
   \right)
\nonumber \\
 - \left(   \IzN \BRNPsi' + \IpN \BRNhpp
          - \left(\pezB + \UcB\right) \BRNvp
   \right)
\nonumber \\
 + (N - 1)
   \left(  \IzN \BRNPhi' - \HpN \BRNhpp
         + \HcB \BRNvp
         + \frac{1}{N} \DzAA (\IN \UPS^\sharp) \right)
\nonumber \\
 + \left(\HpN - \IpN \right) \BRNPsi
 - \frac{N - 1}{N} \left(\DzAA + N \KzAA \right)
   (\IN \UPS^\sharp)
 & = & \KAPPAN \left(  F_\SBLK \IA v^\sharp_\SBLK
                     + (P_\SBLK - \PRESY_\SBLK) \IA \UDT{\dni}^\sharp_\SBLK
               \right) , \\
 - (N - 1) \left( \DzAA + N \KzAA  \right) \BRNPhi
 + \DzAA \BRNPsi
\nonumber \\
 + N \BRNPsi \pezB \HcB
 + N \left(\pezB + (N + 1) \HcB \right)
   \left( \pezB \BRNPhi + \HcB \BRNPsi \right)
\nonumber \\
 + N \HpN
     \left(  \ppzN \BRNPsi + \IpN \BRNhpp
           - \left(\pezB + \UcB \right) \BRNvp
     \right)
\nonumber \\
 - N \left((N - 1) \HpN + \IpN \right)
   \left(  \ppzN \BRNPhi - \HpN \BRNhpp
           + \HcB \BRNvp
           + \frac{1}{N} \DzAA (\IN \UPS^\sharp)
   \right)
 & = & \KAPPAN \UDT{\delta \PRESY}^\sharp_\SBLK ,
\end{eqnarray}
\begin{eqnarray}
 - \DDEMI \left(\DzAA + (N-1) \KzAA  \right) \vi_i
\nonumber \\
 - \DDEMI \left(\ppzN + (N + 1) \HpN \right)
          \left( \left(\ppzN + \IpN - \HpN\
          \right) \vi_i \right)
\nonumber \\
 + \DDEMI \left(\ppzN + (N + 1) \HpN \right)
          \left( \left(\pezB + \UcB - \HcB
          \right) \hci_i \right)
 & = & \KAPPAN \left(  \DIR (\BRN{P}_\SBRN + \BRN{\rho}_\SBRN)
                       (\BRN{\VV{v}}_{\SBRN\;i}^\sharp - \vi_i)
\right. \nonumber \\ & & \left. \quad
                     + (P_\SBLK + \rho_\SBLK)
                       (\VV{v}^\sharp_{\SBLK\;i} - \vi_i)
                     - F_\SBLK (\VV{\dni}^\sharp_{\SBLK\;i} + \hci_i)
               \right) , \\
   \left(\pezB + (N - 1) \HcB \right) \vi_i
 - \left(  \ppzN + (N - 1) \HpN \right) \hci_i
\nonumber \\
 + \UcB \vi_i - \IpN \hci_i
 & = & \KAPPAN \left(  \DIR \IA \VV{\BRN{\Pi}}_i^\SBRN
                     + \IA \VV{\Pi}_i^\SBLK \right) , \\
   \DDEMI \left(\DzAA +(N-1) \KzAA \right) \hci_i
\nonumber \\
 + \DDEMI \left(\pezB + (N + 1) \HcB \right)
          \left( \left(\ppzN + \IpN - \HpN \right) \vi_i \right)
\nonumber \\
 - \DDEMI \left(\pezB + (N + 1) \HcB \right)
          \left( \left(\pezB + \UcB - \HcB \right) \hci_i \right)
 & = & \KAPPAN \left(  F_\SBLK (\VV{v}^\sharp_{\SBLK\;i} - \vi_i)
                     + (P_\SBLK - \PRESY_\SBLK)
                       (\VV{\dni}^\sharp_{\SBLK\;i} + \hci_i )
               \right) ,
\end{eqnarray}
\begin{eqnarray}
   \left(\pezB + N \HcB \right)
   \pezB \TT{E}_{i j}
 - \left(\DzAA - 2 \KzAA \right) \TT{E}_{i j}
\nonumber \\
 + \UcB \pezB \TT{E}_{i j}
 - \left(  \ppzN
         + N \HpN + \IpN \right)
         \left(\ppzN \TT{E}_{i j} \right)
 & = & \KAPPAN \left(  \DIR \TT{\BRN{\Pi}}_{i j}^\SBRN
                     + \TT{\Pi}_{i j}^\SBLK \right) .
\end{eqnarray}
(A sum on all the brane and bulk species is implicitly assumed on
the right hand side of these equations.) One can easily check that
the singular part of the above equation reduces to the second
Israel condition.

\subAPPsection{Sail equation}

In the following a sum on all the brane and bulk species is
implicitly assumed.
\begin{eqnarray}
   N \CONT{\HpN}\BRN{\delta P}^\sharp - \CONT{\IpN} \BRN{\delta \rho}^\sharp
\nonumber \\
 - N \BRN{P}
     \CONT{  \ppzN \BRNPhi - \HpN \BRNhpp + \HcB \BRNvp
           + \frac{1}{N} \DzAA \BRUPS }
\nonumber \\
 - \BRN{\rho}
   \CONT{  \ppzN \BRNPsi + \IpN \BRNhpp - \left(\pezB + \UcB \right) \BRNvp }
 & = & \DISC{  \UDT{\delta \PRESY}^\sharp} ,
\end{eqnarray}
\begin{eqnarray}
\label{sail2}
 - \pezB^2 (\BRN{\rho} \BRUPS)
 - N \HcB \pezB \left(2 \BRN{\rho} \BRUPS + \BRN{P} \BRUPS \right)
 - \DzAA \BRN{P} \BRUPS
 + \KzAA (N - 1) \BRN{\rho} \BRUPS
\nonumber \\
 - N \left(\BRN{P} + \frac{N - 1}{N} \BRN{\rho} \right)
   \BRUPS \left(2 \pezB \HcB + (N + 1) \HcB^2 \right)
\nonumber \\
 - (\BRN{P} + \BRN{\rho}) \BRUPS
   \left(  N \CONT{\HpN} \CONT{\HpN - \IpN}
         + \frac{\KAPPAN}{4} \BRN{\rho} (\BRN{P} + \BRN{\rho}) \right)
\nonumber \\
 + \BRUPS \left(  N \CONT{\HpN} \DISC{\PRESY - P}
                  + \CONT{\IpN} \DISC{\PRESY + \rho} \right)
 & = &   \DISC{  \delta \PRESY^\sharp}
       - N \CONT{\HpN}\BRN{\delta P}^\sharp
       + \CONT{\IpN} \BRN{\delta \rho}^\sharp
\nonumber \\ & &
       + N \BRN{P} \CONT{  \ppzN \Phi - \HpN \hpp + \HcB \vp }
\nonumber \\ & &
       + \BRN{\rho}
         \CONT{  \ppzN \Psi + \IpN \hpp - \left(\pezB + \UcB \right) \vp }
\nonumber \\ & &
       + (2 \pezB + N \HcB) (\CONT{F} \BRNXI)
       - \BRNXI \pezB \CONT{F}
\nonumber \\ & &
       + N \CONT{\HpN} \BRNXI
         \left(  \frac{\KAPPAN}{4} (\BRN{P} + \BRN{\rho})^2
               + \CONT{\PRESY - P} \right)
\nonumber \\  & &
       + \CONT{\IpN} \BRNXI
         \left(- \frac{\KAPPAN}{4} (\BRN{P} + \BRN{\rho}) \BRN{\rho}
               + \CONT{\PRESY + \rho} \right) .
\end{eqnarray}

\subAPPsection{Perturbed conservation equation}
\label{ssec_app_pert_cons}

They transform into
\begin{eqnarray}
   \left(\pezB + N \HcB + \UcB \right)
   \left(\UDT{\delta \rho}^\sharp - F \BRNvp \right)
 + N \HcB \UDT{\delta P}^\sharp
 + \UcB \left(\UDT{\delta \PRESY}^\sharp - F \BRNvp \right)
\nonumber \\
 + \left(\ppzN + N \HpN + 2 \IpN \right)
   \left(\UDT{\delta F}^\sharp + F (\BRNPsi + \BRNhpp) \right)
\nonumber \\
 + \DzAA \left(  \IA (P + \rho) v^\sharp - \IA F \dni^\sharp \right)
\nonumber \\
 - N (P + \rho) \pezB \BRNPhi
 - (\rho + \PRESY) \pezB \BRNhpp
 - F \pezB \BRNvp
 + F \ppzN (\BRNPsi - \BRNhpp - N \BRNPhi)
 & = & \UDT{\delta \Gamma}^\sharp + \Gamma \BRNPsi , \\
   \left(  \pezB + N \HcB
         + \UcB\right)
   \left(\IA (P + \rho) v^\sharp - \IA F \UDT{\dni}^\sharp \right)
\nonumber \\
 + \left(  \ppzN + N \HpN
         + \IpN \right)
       \left(\IA F v^\sharp + \IA (P - \PRESY) \UDT{\dni}^\sharp \right)
\nonumber \\
 + \UDT{\delta P}^\sharp
 + \frac{N - 1}{N} \left(\DzAA + N \KzAA \right) \IAA \Pi
 + (P + \rho) \BRNPsi + (\PRESY - P) \BRNhpp + F \BRNvp
 & = & \IA \UDT{\QINT}^\sharp , \\
   \left(\pezB + N \HcB + 2 \UcB \right)
   \left(  \UDT{\delta F}^\sharp - F (\BRNPsi + \BRNhpp)
         - (\rho + \PRESY) \BRNvp \right)
\nonumber \\
 + \left(\ppzN + N \HpN + 2 \IpN \right)
   \left(\UDT{\delta \PRESY}^\sharp - F \BRNvp \right)
 - N \HpN \UDT{\delta P}^\sharp
 - \IpN (\UDT{\delta \PRESY}^\sharp - \UDT{\delta \rho}^\sharp)
\nonumber \\
 + \DzAA
   \left(\IA F v^\sharp + \IA (P - \PRESY) \dni^\sharp \right)
\nonumber \\
 + F \pezB (\BRNPsi - \BRNhpp - N \BRNPhi)
 + N (P - \PRESY) \ppzN \BRNPhi
 + (\rho + \PRESY) \ppzN \BRNPsi
 + F \ppzN \BRNvp
 & = & \UDT{\delta D}^\sharp - D \BRNhpp ,
\end{eqnarray}
\begin{eqnarray}
  \left(\pezB + (N + 1) \HcB + \UcB
  \right)
  \left(  (P + \rho) (\VV{v}^\sharp_i - \vi_i)
              - F (\VV{\dni}^\sharp_i + \hci_i)
  \right)
\nonumber \\
 + \left(\ppzN + (N + 1) \HpN + \IpN
   \right)
   \left(  F (\VV{v}^\sharp_i - \vi_i)
         + (P - \PRESY) (\VV{\dni}^\sharp_i + \hci_i) \right)
\nonumber \\
 + \DDEMI \left(\DzAA + (N - 1) \KzAA \right) \IA \VV{\Pi}_i
 & = & \VV{\QINT}^\sharp_i - D \hci_i - \Gamma \vi_i .
\end{eqnarray}
For the brane components, they are obtained by considering the
discontinuity of the $\{0 \SfinN\}$, $\{i \SfinN\}$ components of
Einstein equation or by taking the singular part of the above
equations.
\begin{eqnarray}
   \pezB \BRN{\delta \rho}^\sharp_\SBRN
 + N \HcB (  \BRN{\delta \rho}^\sharp_\SBRN
                        + \BRN{\delta P}^\sharp_\SBRN)
\nonumber \\
 + (\BRN{P}_\SBRN + \BRN{\rho}_\SBRN)
   \DzAA \IA \BRN{v}^\sharp_\SBRN
 - N (\BRN{P}_\SBRN + \BRN{\rho}_\SBRN) \pezB \BRNPhi
 & = & \BRN{\delta \Gamma}^\sharp_\SBRN + \Gamma_\SBRN \BRNPsi , \\
\sum_\SBRN \BRN{\delta \Gamma}^\sharp_\SBRN + \Gamma_\SBRN \BRNPsi
 & = & - \sum_\SBLK \DISC{\UDT{\delta F}^\sharp_\SBLK + F_\SBLK \BRNPsi} , \\
   \left(  \pezB + N \HcB \right)
   \left((\BRN{P}_\SBRN + \BRN{\rho}_\SBRN) \IA \BRN{v}^\sharp_\SBRN \right)
 + \BRN{\delta P}^\sharp_\SBRN
\nonumber \\
 + \frac{N - 1}{N} \left(\DzAA + N \KzAA \right) \IAA \BRN{\Pi}_\SBRN
 + (\BRN{P}_\SBRN + \BRN{\rho}_\SBRN) \BRNPsi
 & = & \IA \QINT^\sharp_\SBRN , \\
\sum_\SBRN \IA \QINT^\sharp_\SBRN
 & = & - \sum_\SBLK \DISC{  F_\SBLK \IA v^\sharp_\SBLK
                          + (P_\SBLK - \PRESY_\SBLK)
                            \IA \UDT{\dni}^\sharp_\SBLK} , \\
  \left(\pezB + (N + 1) \HcB \right)
  \left(  (\BRN{P}_\SBRN + \BRN{\rho}_\SBRN)
          (\VV{\BRN{v}}^{\SBRN\;\sharp}_i - \vi_i)  \right)
\nonumber \\
 + \DDEMI \left(\DzAA + (N - 1) \KzAA \right) \IA \VV{\BRN{\Pi}}^\SBRN_i
 & = & \VV{\QINT}^{\SBRN\;\sharp}_i - \BRN{\Gamma}_\SBRN \vi_i , \\
\sum_\SBRN \VV{\QINT}^{\SBRN\;\sharp}_i - \BRN{\Gamma}_\SBRN \vi_i
 & = & - \sum_\SBLK \DISC{  F_\SBLK (\VV{v}^{\SBLK\;\sharp}_i - \vi_i)
                          + (P_\SBLK - \PRESY_\SBLK)
                            (\VV{\dni}^{\SBLK\;\sharp}_i + \hci_i) } .
\end{eqnarray}

\subAPPsection{Einstein equations using the Weyl tensor}

\begin{eqnarray}
   (N - 1) \left(\DzAA + N \KzAA \right) \BRNPhi
\nonumber \\
 - N (N - 1) \HcB
   \left(\HcB \BRNPsi + \pezB \BRNPhi \right)
 & = &   \frac{1}{4} \frac{N - 1}{N} \KAPPAN^2
         \left( \sum_\SBRN \BRN{\rho}_\SBRN \right)
         \sum_\SBRN \BRN{\delta \rho}_\SBRN^\sharp
\nonumber \\ & &
       - (N - 1) \CONT{\HpN}
         \CONT{  N \ppzN \BRNPhi- N \HpN \BRNhpp + N \HcB \BRNvp
               + \DzAA \IN \UPS^\sharp  }
\nonumber \\ & &
       + \frac{N - 1}{N + 1} \KAPPAN
         \sum_\SBLK \CONT{  \UDT{\delta P}^\sharp_\SBLK
                          + \UDT{\delta \rho}^\sharp_\SBLK
                          - \UDT{\delta \PRESY}^\sharp_\SBLK }
       + \frac{N - 1}{N + 1} \CONT{\BRN{\delta \WEYL}^\sharp} , \\
 - (N - 1) \left(\HcB \BRNPsi + \pezB \BRNPhi \right)
 & = &   \frac{1}{4} \frac{N - 1}{N} \KAPPAN^2
         \left( \sum_\SBRN \BRN{\rho}_\SBRN \right)
         \sum_\SBRN \left((\BRN{P}_\SBRN + \BRN{\rho}_\SBRN)
                          \IA v_\SBRN^\sharp \right)
\nonumber \\ & &
       - \frac{N - 1}{2} \CONT{\HpN}
         \CONT{\BRNvp - (\pezB + \UcB - 2 \HcB) (\IN \UPS^\sharp)}
\nonumber \\ & &
       + \frac{N - 1}{N} \KAPPAN
         \sum_\SBLK \CONT{  (P_\SBLK + \rho_\SBLK) \IA v_\SBLK^\sharp
                          - F_\SBLK \IA \UDT{\dni}^\sharp_\SBLK }
       + \CONT{\DWN^v} , \\
   \frac{N - 1}{N} \left(  \DzAA \BRNPsi
                         - (N - 2) (\DzAA + N \KzAA) \BRNPhi \right)
\nonumber \\
 + (N - 1) \left(\pezB + N \HcB \right)
   \left(\HcB \BRNPsi + \pezB \BRNPhi \right)
\nonumber \\
 + (N - 1) \BRNPsi \pezB \HcB
 & = &   \frac{1}{4} \frac{N - 1}{N} \KAPPAN^2
         \left( \sum_\SBRN \BRN{\rho}_\SBRN \right)
         \sum_\SBRN \BRN{\delta P}_\SBRN^\sharp
\nonumber \\ & &
       - (N - 1) \CONT{\HpN}
         \CONT{  \ppzN \BRNPsi + \IpN \BRNhpp - (\pezB + \UcB) \BRNvp }
\nonumber \\ & &
       + (N - 1) \CONT{(N - 2) \HpN + \IpN}
         \CONT{  \ppzN \BRNPhi - \HpN \BRNhpp + \HcB \BRNvp
               + \frac{1}{N} \DzAA \IN \UPS^\sharp }
\nonumber \\ & &
       + \frac{N - 1}{N (N + 1)} \KAPPAN
         \sum_\SBLK \CONT{  \UDT{\delta P}^\sharp_\SBLK
                          + \UDT{\delta \rho}^\sharp_\SBLK
                          + N \UDT{\delta \PRESY}^\sharp_\SBLK }
\nonumber \\ & &
       + \frac{1}{4} \frac{N - 1}{N} \KAPPAN^2
         \left( \sum_\SBRN (\BRN{P}_\SBRN + \BRN{\rho}_\SBRN) \right)
         \sum_\SBRN \BRN{\delta \rho}_\SBRN^\sharp
       + \frac{N - 1}{N (N + 1)} \CONT{\BRN{\delta \WEYL}^\sharp} , \\
(N - 2) \BRNPhi - \BRNPsi
 & = &   \frac{1}{4} \frac{N - 1}{N} \KAPPAN^2
         \left(\sum_\SBRN \BRN{\rho}_\SBRN \right)
         \IAA \sum_\SBRN \BRN{\Pi}_\SBRN
\nonumber \\ & &
       - \CONT{(N - 2) \HpN + \IpN}
         \BRUPS
\nonumber \\ & &
       - \frac{1}{4} \KAPPAN^2
         \left(\sum_\SBRN (\BRN{P}_\SBRN + \BRN{\rho}_\SBRN)
         \right) \IAA \sum_\SBRN \BRN{\Pi}_\SBRN
\nonumber \\ & &
       + \frac{N - 1}{N} \KAPPAN \IAA \sum_\SBLK \CONT{\Pi_\SBLK}
       + \CONT{\DWN^\Pi} ,
\end{eqnarray}
\begin{eqnarray}
 - \DDEMI \left(\DzAA + (N - 1) \KzAA \right) \vi_i
 & = &   \frac{1}{4} \frac{N - 1}{N} \KAPPAN^2
         \left( \sum_\SBRN \BRN{\rho}_\SBRN \right)
         \sum_\SBRN \left((\BRN{P}_\SBRN + \BRN{\rho}_\SBRN)
                          (\VV{v}_i^{\SBRN\;\sharp} - \vi_i) \right)
\nonumber \\ & &
       + \frac{N - 1}{2} \CONT{\HpN}
         \CONT{  (\ppzN + \IpN - \HpN) \vi_i
               - (\pezB + \UcB - \HcB) \hci_i}
\nonumber \\ & &
       + \frac{N - 1}{N} \KAPPAN
         \sum_\SBLK \CONT{  (P_\SBLK + \rho_\SBLK)
                            (\VV{v}^{\SBLK\;\sharp}_i - \vi_i)
                          - F_\SBLK (\VV{\dni}^{\SBLK\;\sharp}_i + \hci_i) }
       + \CONT{\VV{\DWN}{}^v_i} , \\
\left(\pezB + (N - 1) \HcB \right) \vi_i
 & = &   \frac{1}{4} \frac{N - 1}{N} \KAPPAN^2
         \left( \sum_\SBRN \BRN{\rho}_\SBRN \right)
         \IA \sum_\SBRN \BRN{\VV{\Pi}}^\SBRN_i
\nonumber \\ & &
       + \CONT{(N - 2) \HpN + \IpN}
         \CONT{\hci_i}
\nonumber \\ & &
       - \frac{1}{4} \KAPPAN^2
         \left( \sum_\SBRN (\BRN{P}_\SBRN + \BRN{\rho}_\SBRN) \right)
         \IA \sum_\SBRN \BRN{\VV{\Pi}}^\SBRN_i
\nonumber \\ & &
       + \frac{N - 1}{N} \KAPPAN \IA \sum_\SBLK \CONT{\VV{\Pi}^\SBLK_i}
       + \CONT{\VV{\DWN}{}^\Pi_i} ,
\end{eqnarray}
\begin{eqnarray}
   \left(\pezB + N \HcB \right)
   \pezB \TT{E}_{i j}
 - \left(\DzAA - 2 \KzAA \right) \TT{E}_{i j}
 & = &   \frac{1}{4} \frac{N - 1}{N} \KAPPAN^2
         \left( \sum_\SBRN \BRN{\rho}_\SBRN \right)
         \sum_\SBRN \BRN{\TT{\Pi}}{}^\SBRN_{i j}
\nonumber \\ & &
       + \CONT{(N - 2) \HpN + \IpN}
         \CONT{\ppzN \TT{E}_{i j}}
\nonumber \\ & &
       - \frac{1}{4} \KAPPAN^2
         \left( \sum_\SBRN (\BRN{P}_\SBRN + \BRN{\rho}_\SBRN) \right)
         \sum_\SBRN \BRN{\TT{\Pi}}{}^\SBRN_{i j}
\nonumber \\ & &
       + \frac{N - 1}{N} \KAPPAN \sum_\SBLK \CONT{\TT{\Pi}{}^\SBLK_{i j}}
       + \CONT{\TT{\DWN}{}^\Pi_{i j}} .
\end{eqnarray}

\subAPPsection{Relationship between {\protect $\CONT{\DCEX_{\alpha
\beta}}$} and {\protect $\DISC{\DWN_{\alpha \beta}}$}}
\label{ssec_app_pffffff}

\begin{eqnarray}
 - \BRN{\rho}_\SBRN \CONT{  \ppzN \BRNPhi - \HpN \BRNhpp + \HcB \BRNvp
                          + \frac{1}{N} \DzAA \BRUPS}
 & = & - \BRN{\delta \rho}^\sharp_\SBRN \CONT{\HpN}
\nonumber \\ & &
       + \frac{1}{N + 1}
         \left(  \DISC{  \UDT{\delta P}^\sharp_\SBLK
                       + \UDT{\delta \rho}^\sharp_\SBLK
                       - \UDT{\delta \PRESY}^\sharp_\SBLK}
                       + \frac{1}{\KAPPAN} \DISC{\BRN{\delta \WEYL}^\sharp}
         \right) , \\
 - (N \BRN{P}_\SBRN + \BRN{\rho}_\SBRN)
   \CONT{  \ppzN \BRNPhi - \HpN \BRNhpp
         + \HcB \BRNvp + \frac{1}{N} \DzAA \BRUPS}
\nonumber \\
 - \BRN{\rho}_\SBRN \CONT{  \ppzN \BRNPsi + \IpN \BRNhpp
                          - \left(\pezB + \UcB \right) \BRNvp}
 & = & - (N \BRN{\delta P}^\sharp_\SBRN + \BRN{\delta \rho}^\sharp_\SBRN)
         \CONT{\HpN}
       + \BRN{\delta \rho}^\sharp_\SBRN \CONT{\IpN}
\nonumber \\ & &
       + \frac{1}{N + 1}
         \DISC{  \UDT{\delta P}^\sharp_\SBLK
               + \UDT{\delta \rho}^\sharp_\SBLK
               + N \UDT{\delta \PRESY}^\sharp_\SBLK }
\nonumber \\ & &
       + \frac{1}{N + 1}
         \frac{1}{\KAPPAN} \DISC{\BRN{\delta \WEYL}^\sharp} , \\
 - \BRN{\rho}_\SBRN
   \CONT{  \TDEMI \BRNvp
         - \TDEMI \left(\pezB + \UcB - 2 \HcB \right) \IN \UPS^\sharp}
 & = & - N \CONT{\HpN}
         (\BRN{P}_\SBRN + \BRN{\rho}_\SBRN) \IA \BRN{v}^\sharp_\SBRN
\nonumber \\ & &
       + \DISC{  (P_\SBLK + \rho_\SBLK) \IA v_\SBLK^\sharp
               - F_\SBLK \IA \UDT{\dni}^\sharp_\SBLK}
\nonumber \\ & &
       + \frac{N}{N - 1} \frac{1}{\KAPPAN} \DISC{\DWN^v} , \\
(N \BRN{P}_\SBRN + \BRN{\rho}_\SBRN) \BRUPS
 & = & - N \CONT{(N - 2) \HpN + \IpN} \IAA \BRN{\Pi}_\SBRN
\nonumber \\ & &
       + (N - 1) \IAA \DISC{\Pi_\SBLK}
       + N \frac{1}{\KAPPAN} \DISC{\DWN^\Pi} ,
\end{eqnarray}
\begin{eqnarray}
 - \BRN{\rho}_\SBRN
   \CONT{  \TDEMI \left(\pezB + \UcB - \HcB \right) \hci_i
         - \TDEMI \left(\ppzN + \IpN - \HpN \right) \vi_i}
 & = & - N \CONT{\HpN}
         (\BRN{P}_\SBRN + \BRN{\rho}_\SBRN) (\BRN{v}^\sharp_{\SBRN\;i} - \vi_i)
\nonumber \\ & &
       + \DISC{  (P_\SBLK + \rho_\SBLK) (\VV{v}^\sharp_{\SBLK\;i} - \vi_i)
               - F_\SBLK (\VV{\dni}^\sharp_{\SBLK\;i} + \hci_i)}
\nonumber \\ & &
       + \frac{N}{N - 1} \frac{1}{\KAPPAN} \DISC{\VV{\DWN}{}^v_i} , \\
 - (N \BRN{P}_\SBRN + \BRN{\rho}_\SBRN) \CONT{\hci_i}
 & = & - N \CONT{(N - 2) \HpN + \IpN} \IAA \BRN{\VV{\Pi}}{}^\SBRN_i
\nonumber \\ & &
       + (N - 1) \IA \DISC{\VV{\Pi}{}_i^\SBLK}
       + N \frac{1}{\KAPPAN} \DISC{\VV{\DWN}^\Pi_i} ,
\end{eqnarray}
\begin{eqnarray}
 - (N \BRN{P}_\SBRN + \BRN{\rho}_\SBRN) \CONT{\ppzN \TT{E}_{i j}}
 & = & - N \CONT{(N - 2) \HpN + \IpN} \BRN{\TT{\Pi}}{}_{i j}^\SBRN
\nonumber \\ & &
       + (N - 1) \DISC{\TT{\Pi}{}^\SBLK_{i j}}
       + N \frac{1}{\KAPPAN} \DISC{\TT{\DWN}{}^\Pi_{i j}} .
\end{eqnarray}


\begin{thebibliography}{100}

\bibitem{Polch}
J.~Polchinski, {\it String Theory} (Cambridge Universty Press,
Cambridge, England, 1998), volumes 1 and 2.

\bibitem{HW1}
P.~Horava and E.~Witten, Nucl.\ Phys.\ {\bf B640}, 506 (1996).

\bibitem{HW2}
P.~Horava and E.~Witten, Nucl.\ Phys.\ {\bf B675}, 94 (1996).




\bibitem{ebp1}
K.~Akama, {\it Pregeometry}, Lecture Notes in Physics {\bf 176},
edited by K.~Kikkawa, N.~Nakanishi, and H.~Nariai (Springer-Verlag,
Berlin, 1982), p.~267.

\bibitem{ebp2}
V.A.~Rubakov and M.E.~Shaposhnikov, \pl B {\bf 125}, 136 (1983).

\bibitem{ebp3}
M.~Visser, \pl B {\bf 159}, 22 (1985).

\bibitem{ebp4}
E.J.~Squires, \pl B {\bf 167}, 286 (1986).

\bibitem{ebp5}
N.~Arkani-Hamed, S.~Dimopolous, and G.~Dvali, \pl B {\bf 429}, 263
(1998).

\bibitem{ebp6}
I.~Antoniadis, N.~Arkani-Hamed, S.~Dimopolous, and G.~Dvali, \pl B
{\bf 436}, 257 (1998).

\bibitem{RS1}
L.~Randall and R.~Sundrum, \prl {\bf 83}, 3370 (1999).

\bibitem{RS2}
L.~Randall and R.~Sundrum, \prl {\bf 83}, 4670 (1999).

\bibitem{Bine1}
P.~Bin\'etruy, C.~Deffayet, and D.~Langlois, Nucl.\ Phys.\ {\bf B565},
269 (2000).


\bibitem{Bine2}
P.~Bin\'etruy, C.~Deffayet, U.~Ellwanger, and D.~Langlois, \pl B {\bf
477}, 285 (1999).


\bibitem{Muko}
S.~Mukohyama, \pl B {\bf 473}, 241-245 (2000).


\bibitem{gcc1}
J.M.~Cline, C.~Grojean, and G.~Servant, \prl {\bf 83}, 4245 (1999).

\bibitem{gcc2}
C.~Csaki, M.~Graesser, C.~Kolda, and J.~Terning, \pl B {\bf 462}, 34
(1999).

\bibitem{gcc3}
E.E.~Flannagan, S.-H.~Tye, and I.~Wasserman, \prd {\bf 62}, 044039
(2000).

\bibitem{Lorenzo}
J.~Lesgourgues, S.~Pastor, M.~Peloso, and L.~Sorbo, \pl B {\bf 489},
411 (2000).


\bibitem{Deruelle}
N.~Deruelle, T.~Dole\v{z}el, and J.~Katz, \prd {\bf 63}, 083513 (2001).

\bibitem{Deruelle2}
N.~Deruelle and J.~Katz, \prd {\bf 64}, 083515 (2001).

\bibitem{Deruelle3}
N.~Deruelle and T.~Dole\v{z}el, \prd {\bf 64}, 103506 (2001).

\bibitem{Tanaka}
J.~Garriga and T.~Tanaka, \prl {\bf 84}, 2778-2781 (2000).

\bibitem{GKR}
S.B.~Giddings, E.~Katz, and L.~Randall, JHEP {\bf 0003}, 023 (2000).

\bibitem{fields}
S.~Kachru, M.~Schulz, E.~Silverstein, \prd {\bf 62}, 045021 (2000).

\bibitem{bulksf}
P.F.~Gonzalez-Diaz \pl B {\bf 481}, 353 (2000).


\bibitem{DDPV00}
A.-C. Davis,  I.R. Vernon, S. Davis, and W.B. Perkins,
\pl B {\bf 504}, 254 (2001).

\bibitem{perkins00}
W.B. Perkins, \pl B {\bf 504}, 28 (2001).

\bibitem{Carter1}
B.~Carter, Int.\ J.\ Theor.\ Phys.\ {\bf 40}, 2099 (2001).

\bibitem{Carter2}
R.A.~Battye and B.~Carter, \pl B {\bf B50}, 331 (2001).

\bibitem{JP1}
B.~Carter and J.-P.~Uzan, Nucl.\ Phys.\ {\bf B606}, 45 (2001).

\bibitem{JP2}
R.A.~Battye, B.~Carter, A.~Mennim, and J.-P.~Uzan, \prd {\bf 64},
124007 (2001).

\bibitem{JP3}
B.~Carter, J.-P.~Uzan, R.A.~Battye, and A.~Mennim, Class.\ Quantum
Grav.\ {\bf 18}, 4871 (2001).



\bibitem{Fterm}
C.~van de Bruck, M.~Dorca, C.J.A.P.~Martins, and M.~Parry, \pl B {\bf
495}, 183 (2000).


\bibitem{SMS}
T.~Shiromizu, K.~Maeda, and M.~Sasaki, \prd {\bf 62}, 024012 (2000).

\bibitem{Roy}
R.~Maartens, \prd {\bf 62}, 084023 (2000).


\bibitem{Carsten}
C.~van de Bruck, M.~Dorca, R.~Brandenberger, and A.~Lukas, \prd {\bf
62}, 123515 (2000).

\bibitem{bp1}
H.~Kodama, A.~Ishibashi, and O.~Seto, \prd {\bf 62}, 064022 (2000).

\bibitem{bp2a}
S.~Mukohyama, \prd {\bf 62}, 084015 (2000).

\bibitem{bp2b}
S.~Mukohyama, Class.\ Quantum Grav.\ {\bf 17}, 4777 (2000).

\bibitem{bp2c}
S.~Mukohyama, \prd {\bf 64}, 064006 (2001).


\bibitem{bp4}
D.~Langlois, \prd {\bf 62}, 126012 (2000).

\bibitem{bp5}
D.~Langlois, \prl {\bf 86}, 2212 (2001).

\bibitem{bp6}
A.~Neronov and I.~Sachs, \pl B {\bf 513}, 173 (2001).

\bibitem{bp7}
C.~van de Bruck and M.~Dorca, hep-th/0012073.

\bibitem{bp8}
H.~Bridgman, K.~Malik, and D.~Wands, \prd {\bf 65}, 043502 (2002).

\bibitem{bp9}
K.~Koyama and J.~Soda, \prd {\bf 62}, 123502 (2000).

\bibitem{bp10}
K.~Koyama and J.~Soda, \prd {\bf 65}, 023514 (2002).

\bibitem{Bozza}
V.~Bozza, M.~Gasperini, and G.~Veneziano, Nucl.\ Phys.\ {\bf B619},
191 (2001).

\bibitem{seto}
O.~Seto and H.~Kodama, \prd {\bf 63}, 123506 (2001).

\bibitem{giovannini}
M.~Giovannini, \prd {\bf 65}, 064008 (2002).

\bibitem{BarrowRoy}
J.D.~Barrow and R.~Maartens, gr-qc/0108073.



\bibitem{Khoury}
J.~Khoury and R.-J.~Zhang, hep-th/0203274.

\bibitem{Anisotropicstress}
M.~Dorca and C.~van de Bruck, Nucl.\ Phys.\ {\bf B605}, 215 (2001).

\bibitem{Dick}
R.~Dick, \pl B {\bf491}, 333 (2000).

\bibitem{Christos}
P.~Bowcock, C.~Charmousis, and R.~Gregory, Class.\ Quantum Grav.\ {\bf
17}, 4745 (2000).

\bibitem{Kraus}
P.~Kraus, JHEP {\bf 9912}, 011 (1999).

\bibitem{Ida}
D.~Ida, JHEP {\bf 0009}, 014 (2000).

\bibitem{Mukoya}
S.~Mukohyama, T.~Shiromizu, and K.~Maeda, \prd {\bf 62}, 024028
(2000).

\bibitem{radion}
C.~Charmousis, R.~Gregory, and V.~Rubakov, \prd {\bf 62}, 067505 (2000).

\bibitem{Bar}
J.M.~Bardeen, \prd {\bf 22}, 1882 (1980).

\bibitem{KS}
H.~Kodama and M.~Sasaki, Prog.\ Theor.\ Phys.\ {\bf 78}, 1 (1984).

\bibitem{STW}
J.M.~Stewart and M.~Walker, Proc.\ Roy.\ Soc.\ London, {\bf A341}
(1974).

\bibitem{gravitruc1}
H.A.~Bridgman, K.A.~Malik, and D.~Wands, \prd {\bf 63}, 084012 (2001).

\bibitem{gravitruc2}
Y.S.~Myung, hep-th/0009117

\bibitem{Ijc}
W.~Israel, Nuovo Cimento {\bf B44}, 1 (1966); Erratum: {\it ibid.}
{\bf B48}, 463 (1967).

\bibitem{Wald}
R.M.~Wald, {\it General Relativity}, (The University of Chicago Press,
Chicago, 1984).


\bibitem{induced}
B.~Carter and P.~Peter, \pl B {466}, 41 (1999).

\bibitem{petrov}
A.Z.~Petrov, {\it Einstein Spaces}, (Pergamon Press, Oxford, 1969).

\bibitem{fundform}
B.~Carter, \pl B {\bf 228}, 446 (1989).

\bibitem{MTW}
W.~Misner, K.S.~Thorne, and J.A.~Wheeler, {\it Gravitation}, (Freeman, New
York, 1970).



\end{thebibliography}
\end{document}